\documentclass[a4paper,15pt]{article}

\RequirePackage{amsthm,amsmath,amsfonts,amssymb}
\RequirePackage[numbers]{natbib}
\RequirePackage[colorlinks,citecolor=blue,urlcolor=blue,linkcolor=red]{hyperref}
\RequirePackage{graphicx}
\usepackage{authblk}

\usepackage{geometry}
\newtheorem{theorem}{Theorem}[section]

\newtheorem{prop}[theorem]{Proposition}

\newtheorem{defi}[theorem]{Definition}
\newtheorem{ass}{Assumption}
\newtheorem{rem}{Remark}

\usepackage[utf8]{inputenc} %
\usepackage[T1]{fontenc}    %
\usepackage{hyperref}       %
\usepackage{url}            %
\usepackage{booktabs}       %
\usepackage{amsfonts}       %
\usepackage{nicefrac}       %
\usepackage{microtype}      %
\usepackage[linesnumbered,ruled]{algorithm2e}

\usepackage{amsmath}
\usepackage{amsfonts}
\usepackage{mathtools}
\usepackage{amssymb}
\usepackage{amsthm}
\usepackage{bbm}
\usepackage{hyperref}
\usepackage{enumitem}
\usepackage{etoolbox} %
\usepackage{array}
\usepackage{extpfeil}
\usepackage{placeins}
\usepackage{bbm}

\usepackage[dvipsnames, table]{xcolor}

\usepackage{caption}
\graphicspath {{./figures/}}

\renewcommand\epsilon{\varepsilon}
\renewcommand\subset{\subseteq}

\newcommand\N{\mathbb{N}}

\newcommand\R{\mathbb{R}}

\renewcommand\P{\mathbb{P}}
\newcommand\E{\mathbb{E}}
\newcommand\var{\operatorname{Var}}
\newcommand\cov{\operatorname{Cov}}

\newcommand\PP{\mathcal{P}}

\newcommand\FF{\mathcal{F}}
\newcommand\GG{\mathcal{G}}
\newcommand\II{\mathcal{I}}
\newcommand\HH{\mathcal{H}}
\newcommand\MM{\mathcal{M}}

\newcommand\fs{f_{\diamond}}

\newcommand\gs{g_{\diamond}}

\newcommand\bes{b_{\diamond}}
\newcommand\beb{\bar{b}}

\newcommand{\define}[4]{\expandafter#1\csname#3#4\endcsname{#2{#4}}}
\forcsvlist{\define{\newcommand}{\mathcal}{c}}{A,B,C,D,E,F,G,H,I,J,K,L,P,R,M,N,U,S,T,W,V,X,Y,Z}
\forcsvlist{\define{\newcommand}{\mathbb}{b}}{A,B,C,D,E,F,G,H,I,J,K,L,P,Q,M,N,Z,T,X,Y}
\forcsvlist{\define{\newcommand}{\mathbf}{f}}{A,B,C,D,E,F,G,H,I,J,K,L,M,N,O,P,Q,R,S,T,U,W,V,X,Y,Z}

\renewcommand{\phi}{\varphi}
\newcommand{\ep}{\epsilon}
\usepackage{tcolorbox}
\newtcbox{\mybox}{nobeforeafter,colframe=black!50,colback=white,boxrule=1.8pt,arc=5pt,
  boxsep=0pt,left=6pt,right=6pt,top=6pt,bottom=6pt,tcbox raise base}

\newcommand{\B}[1]{\mathbf{#1}}
\newcommand{\indep}{\perp \!\!\! \perp}
\newcommand{\independent}{\perp \!\!\! \perp}
\newcommand{\given}{\, \vert \,}
\newcommand{\st}{\, : \,}
\newcommand{\iid}{\overset{\text{iid}}{\sim}}
\DeclarePairedDelimiterX{\card}[1]{\lvert}{\rvert}{#1}
\DeclarePairedDelimiterX{\norm}[1]{\lVert}{\rVert}{#1}
\DeclarePairedDelimiterX{\abs}[1]{\lvert}{\rvert}{#1}
\DeclareMathOperator*{\argmin}{argmin}

\newcommand{\supp}{\mathrm{supp}}
\newcommand{\eqas}{\ensuremath{\stackrel{\text{a.s.}}{=}}}
\newcommand{\eqd}{\ensuremath{\stackrel{\text{d}}{=}}}
\DeclareSymbolFont{largesymbolsA}{U}{txexa}{m}{n}
\DeclareMathSymbol{\varprod}{\mathop}{largesymbolsA}{16} %
\DeclareMathOperator{\sign}{sign}

\newcommand\todo[1]{{\color{brown}todo: #1}}
\newcommand\Rune[1]{{\color{green!70!black}Rune: #1}}

\newcommand\Rm[1]{{\color{lightgray}Remove: #1}}
\newcommand\Niklas[1]{{\color{Fuchsia}Niklas: #1}}
\newcommand\Martin[1]{{\color{PineGreen}Martin: #1}}
\newcommand\Nicola[1]{{\color{RubineRed}Nicola: #1}}

\title{
\bf
A causal framework for distribution generalization
}
\author{Rune Christiansen$^\flat$
\quad Niklas Pfister$^\flat$ \quad Martin Emil Jakobsen$^{\flat}$ \\ Nicola Gnecco$^{\sharp}$ \quad Jonas Peters$^\flat$}
\affil{
\normalsize{
\textit{$^\flat$University of Copenhagen, Denmark}\\
\textit{\{krunechristiansen,
np,
m.jakobsen,
jonas.peters\}@math.ku.dk} \vspace{0.2cm}\\ 
\textit{$^\sharp$University of Geneva, Switzerland}\\ 
\textit{nicola.gnecco@unige.ch} \vspace{-0.1cm}\\ 
}
}
\date{\today}

\begin{document}

\maketitle

\begin{abstract}
  We consider the problem of predicting a 
  response $Y$ from a set of covariates $X$
  when test and training distributions differ. 
  Since such differences may 
have causal explanations,
we
consider
test distributions that 
emerge from interventions in a structural causal model,
and 
focus 
on 
minimizing
the worst-case risk.
 Causal 
 regression
  models, which regress the
  response on its direct causes, 
  remain 
  unchanged
  under
  arbitrary interventions on the covariates, but they are not always optimal in the above sense.
  For example, for linear models
and bounded
  interventions,  
  alternative solutions
  have been shown to be minimax prediction optimal.
We introduce the formal framework 
of distribution generalization 
that allows us to analyze the above problem
in partially observed nonlinear models
for both direct interventions on $X$ and 
  interventions that occur indirectly via exogenous variables $A$.
It takes into account that, in practice, 
minimax solutions need to be identified from data.  
Our framework allows us to characterize under which 
class of interventions
the causal function is minimax optimal.  
  We prove sufficient conditions for distribution 
  generalization and present corresponding impossibility results.
We propose a
  practical method, 
  NILE, that 
achieves distribution   
  generalization
  in a nonlinear IV
  setting
  with linear extrapolation. 
  We prove consistency and present empirical results.
\end{abstract}

\section{Introduction} \label{sec:intro}
Large-scale learning systems, particularly those focusing on
prediction tasks, have been successfully applied in various domains of
application. 
Since inference is usually done during training time,
any difference between training and test distribution poses a
challenge for prediction methods \cite{Quionero2009, Pan2010,
  Csurka2017, Arjovsky2019}.  Dealing with these differences
is of great importance in 
several
fields such as 
environmental sciences,
where methods need to extrapolate
both in space and time.
Tackling this problem requires restrictions on how the distributions may
differ, since, clearly, generalization becomes impossible if the test
distribution may be arbitrary.  Given a response $Y$ and some
covariates~$X$, several existing procedures aim to find a 
minimax
function
$f$ which minimizes the worst-case risk
$\sup_{P \in \mathcal{N}} \mathbb{E}_P [(Y - f(X))^2]$ across
distributions contained in a small neighborhood $\mathcal{N}$ of the
training distribution. The neighborhood~$\mathcal{N}$ should be
representative of the difference between the training and test
distributions, and often mathematical tractability is taken into
account, too \citep{abadeh2015distributionally, sinha2017certifying}.
A typical approach is to define a $\rho$-ball of distributions
$\cN_\rho(P_0) := \{P: D(P, P_0) \leq \rho\}$ around the (empirical)
training distribution $P_0$, with respect to some divergence measure
$D$,
such
as the Kullback-Leibler divergence
\citep{bagnell2005robust, hu2013kullback}.  
While 
some
divergence
functions only consider distributions with the same support as $P_0$,
the Wasserstein distance allows %
for a neighborhood of distributions around $P_0$ with possibly
different supports \citep{abadeh2015distributionally,
  sinha2017certifying,esfahani2018data,blanchet2019data}.

  In our
analysis, we do not start from a divergence measure, but instead
model the difference between training and test distribution using the 
concept of 
interventions \citep{pearl2009causality, Peters2017book}.
We believe that for many problems this provides a useful description of distributional changes.
We will see that, depending on the considered setup, this approach allows to find models that perform well even on test distributions which would be considered far away from the training distribution in any commonly used metric.
For this class of distributions, causal regression models 
appear naturally because of the following
well-known observation.
A prediction
model, which uses only the direct causes of the response $Y$ as
covariates, is 
invariant under interventions on variables
other than $Y$:
the conditional distribution of $Y$ given its causes does not change
(this principle is known, e.g., as invariance, autonomy or modularity)
\citep{Aldrich1989, Haavelmo1944, pearl2009causality}.  Such a causal regression %
model yields the minimal worst-case risk when considering
all interventions on variables other than $Y$
\citep[e.g.,][Theorem~1,
Appendix]{Rojas2016}. It has therefore been suggested to use causal
models in problems of distributional shifts
\citep{Scholkopf2012, Rojas2016, HeinzeDeml17, Magliacane2018,
  Meinshausen2018, Arjovsky2019, pfister2019stabilizing}.
In practice, however, not all relevant causal variables might be observed.
One may further argue that
causal methods are too
conservative in that the interventions 
which induce the test distributions may not
be arbitrarily strong.
Instead,
methods which 
focus on a
trade-off between predictability and
causality
have been proposed for linear models
\citep{rothenhausler2018anchor, Pfister2019pnas}, see also Section~\ref{sec:existingmethods}.
Anchor regression \citep{rothenhausler2018anchor} is shown to be predictive optimal under a set of bounded interventions.

In this work, we introduce the general framework of distribution
generalization, which permits a unifying perspective on the potentials
and limitations of applying causal concepts to the problem of generalizing 
regression models from training to test distribution.
In particular, we use it to characterize the
relationship between a minimax optimal solution and the causal
function, and to classify settings under which the minimax solution is
identifiable from the 
training
distribution.

\subsection{Further related work}

The field of
distributional robustness or out-of-distribution generalization
aims to develop procedures that are robust to 
changes between training and test distribution. 
This problem has been actively studied from an empirical perspective in
machine learning research, for example, in image
classification by using
adversarial attacks,
where small 
digital 
 \citep{goodfellow2014explaining}
or
physical
\citep{evtimov2017robust}
perturbations of pictures 
can deteriorate 
the performance of a model. Arguably, 
these procedures are not yet fully understood theoretically. A more
theoretical perspective 
is given by the previously mentioned
minimization of a worst-case risk across distributions contained in a
neighborhood of the training distribution,
in our case, distributions generated by interventions.

  Our framework includes the problems of multi-task
learning, domain generalization and transfer learning
\citep{Baxter2000, Candela2009, Caruana1997, Mansour2009} (see
Section~\ref{sec:intro_focus} for more details), with a focus on
minimizing the worst-case risk.
In settings of covariate shift
\citep[e.g.,][]{Shimodaira2000, Sugiyama2005, Sugiyama2008}, one
usually assumes that the training and test distribution of the
covariates are different, while the conditional distribution of the
response given the covariates remains invariant
\citep{daume2006domain, bickel2009discriminative, David10,
  muandet2013domain}.  Sometimes, it is additionally assumed that the
support of the training distribution covers that of the test
distribution \citep{Shimodaira2000}.  In this work, the
conditional distribution of the response given the covariates is
allowed to change between interventions, due to the existence of 
hidden confounders,
and we consider settings where the test
observations lie outside the training support.

  Data augmentation methods have become successful techniques,
  e.g.\ in image classification, to adapt prediction procedures
  to such types of distribution shifts. These methods %
  increase the diversity of the training data
by changing the geometry and the color of the images (e.g., by
rotation, cropping or changing saturation) \citep{zhang2017mixup,
  shorten2019survey}. This allows the user to create models that
generalize better to unseen
environments %
\citep[e.g.,][]{volpi2018adversarial}.  We view these approaches as 
ways to enlarge the support of the covariates, which, as our results
show, comes with theoretical advantages, see
Section~\ref{sec:generalizability}.  

Minimizing the worst-case risk is considered in robust methods \citep{Ghaoui03,
  Kim2005}, too.  It can also be formulated in terms of minimizing the
regret in a multi-armed bandit problem \citep{lai1985asymptotically,
  auer2002finite, bartlett2008high}.  In that setting, the agent can
choose
the distribution which generates the data.
In our setting, though, we do not assume to have 
control over the interventions, 
and, hence, neither over the distribution of the sampled data.

\subsection{Contribution and structure}
This work contains four main contributions:
(1) 
A novel framework for analyzing the problem of generalization from 
training to test distribution, 
using the notion of distribution generalization (Section~\ref{sec:framework}).
(2) Results elucidating the relationship between a causal function and a minimax solution (Section~\ref{sec:robustness}).
(3) Sufficient conditions which ensure distribution
generalization, along with corresponding impossibility results (Section~\ref{sec:generalizability}).
(4) A practical method, called NILE (`Non-linear Intervention-robust Linear Extrapolator'), which 
learns a minimax solution from i.i.d.\ observational data (Section~\ref{sec:learning}). 

Our framework describes how structural causal models can be used as 
technical devices for modeling
 plausible test distributions. It further allows us to formally 
define distribution generalization, which describes the ability
to identify
generalizing
regression models (i.e., minimax solutions) from the observational distribution.
While it is well known that
the causal function is minimax optimal under the set of all
interventions on the covariates \citep[e.g.,][]{Rojas2016}, we extend
this result in several ways, for example, by allowing for hidden variables and
by characterizing more general sets of interventions under which the causal
function is minimax optimal.
We further derive conditions on the
model class, the observational distribution and the family of
interventions under which distribution generalization is
possible, and present impossibility results proving the 
necessity of some of these conditions. 
For example, we show that 
strong assumptions on the 
functional relationship between $X$ and $Y$ 
are 
needed whenever the interventions extend the
training support of $X$. %
An example of such an assumption is to consider the class of 
differentiable functions that
linearly extrapolate outside
the support of $X$. For that model class, we propose the explicit method NILE, 
which obtains distribution generalization by exploiting a nonlinear instrumental variables setup.
We show that our method learns a
minimax solution which corresponds to the causal
function.  
We prove consistency and compare our algorithm to
state-of-the art approaches empirically.

We believe that our results 
shed
some light on the potential merits of 
using causal concepts in 
the context 
of generalization. 
The framework allows us to make first steps 
towards answering
when it can be beneficial to use non-causal functions for prediction under interventions,
and 
what might happen under misspecification of the intervention class.
Our results also formalize in which sense
  methods that generalize
  in the linear case -- such as IV and anchor regression \citep{rothenhausler2018anchor} --
can be extended to nonlinear settings. %
Further, 
our framework 
implies impossibility statements for multi-task learning that relate to existing 
results
\citep{David10}.

Our code
  is available as an \texttt{R}-package at
  \url{https://runesen.github.io/NILE}; scripts generating all our
  figures and results can be found at the same url. Additional
  supporting material is given in the online appendix.
Appendix~\ref{sec:causal_relations_X} shows how to represent several
causal models
in our framework. Appendix~\ref{sec:IVconditions} summarizes %
existing results on identifiability in IV models.
Appendix~\ref{sec:test_statistic} provides details on the test
statistic that we use for NILE.
Appendix~\ref{sec:additional_experiments} contains additional
experiments. All proofs are provided in Appendix~\ref{app:proofs}.

\section{Framework} \label{sec:framework}
For a real-valued response 
 $Y\in\R$ and predictors $X \in \R^d$,
we consider the problem
of
identifying
a regression function that works
well not only on the training data, but also under 
perturbed distributions that we will model by
interventions.

\subsection{Modeling intervention-induced distributions}\label{sec:modeling_int_ind_distr}
We
require a model that is able to 
model 
an observational 
distribution of $(X,Y)$ 
(as training distribution)
and the distribution of $(X,Y)$
under a
class of interventions on (parts of) $X$ (as test distribution).
We will do so by means of a
structural causal model (SCM) \citep{Bollen1989, pearl2009causality}.
More precisely, denoting by $H \in \R^q$ some additional (unobserved) 
variables, we consider the~SCM
\begin{equation}
  \label{eq:SCMmodelreduced}
    H \coloneqq \ep_H, \,\,\,
    X \coloneqq h_2(H, \epsilon_X), \,\,\,
    Y \coloneqq f(X) + h_1(H, \epsilon_Y),
\end{equation}
where the assignments for $H$, $X$ and $Y$ consist of $q$, $d$ and $1$
coordinate(s), respectively.
Here, $f$, $h_1$ and $h_2$ are measurable functions, and the innovation
terms $\ep_X$, $\ep_Y$ and $\ep_H$ are independent vectors with
possibly dependent coordinates. 
Two comments
are in order. First, the joint distribution of $(X, Y)$ is constrained only by requiring
that $X$ and $h_1(H, \ep_Y)$ enter the assignment for $Y$
additively. This constraint affects the allowed conditional
distributions of $Y$ given $X$, but does not make any restriction on
the marginal distributions of either $X$ or $Y$. 
Second, we only use the above SCM as a technical device for
  modeling training and test distributions, by considering
  interventions on $X$ or $A$ (introduced in Section~\ref{sec:interventions}), for which we are
  analyzing the predictive performance of different models -- similarly
  to how one could have considered a ball around the training
  distribution. 
	We therefore only require the SCM to 
	correctly (a) model the 
	training-distribution, 
	and (b) induce the
	test-distributions through interventions.
  Any other
  causal implications of the SCM, such as causal orderings between
  variables, causal effects or counterfactual statements, are not
  assumed to be correctly specified.
  As such, our framework includes a wide range of cases, 
  including situations where training and test distribution come from interventions in 
  an SCM with a different structure than \eqref{eq:SCMmodelreduced}, where, 
  for example, some of the variables in $X$ are not ancestors but descendants of $Y$.
  To see whether our framework applies, one needs to check if
  the considered training and test distributions can be equivalently expressed as 
  interventions in a model of our form. If the structure of the true 
  data generating SCM is known, this can be done 
  by directly 
  transforming the SCM and the interventions. The following remark shows 
  an example of such a transformation
  and may be
interesting to readers with a special interest in causality.  It can
be skipped at first reading.

\begin{rem}[Transforming causal models] \label{rem:model}
Assume that the training distribution is induced by the following SCM
\begin{equation*}
 		X_1 \coloneqq \ep_1, \quad
        X_2 \coloneqq k(Y) + \epsilon_2, \quad
        Y \coloneqq f(X_1) + \epsilon_3,
\end{equation*}
with $(\epsilon_1,\epsilon_2, \ep_3)\sim Q$, and that we consider 
test distributions arising from shift interventions on $X_2$. This 
set of training and test distributions can be equivalently modeled 
by 
the 
reduced SCM
\begin{equation*}
H \coloneqq \ep_3, \quad
        X \coloneqq h_2(H,(\ep_1,\ep_2)), \quad
        Y \coloneqq f(X_1) + H,
\end{equation*}
with $(\epsilon_1,\epsilon_2, \ep_3)\sim Q$, and where
$h_2$ is defined by $h_2(H,(\ep_1, \ep_2))\coloneqq(\ep_1, k(f(\ep_1)+H) + \ep_2)$. 
Both SCMs induce the same observational distribution over $(X_1,X_2, Y)$ and
shift interventions on $X_2$ in the original SCM correspond to shift interventions on 
$X = (X_1, X_2)$ in the reduced SCM (where only the second coordinate is shifted). 
Our framework can then be used, for example, to give sufficient conditions under 
which generalization (formally defined below) is possible, see Proposition~\ref{prop:genX_intra}~and~\ref{prop:genX_extra}.
	It is not always possible to transform an SCM into our reduced form, and it might also happen that 
	the transformed interventions are not covered by our framework.
For example, we do not allow for direct interventions on $Y$ in the original model. In 
other cases, where the original SCM may contain additional hidden variables, even
interventions on (parts of) $X$ in the original SCM may translate into interventions
on $H$ in the reduced SCM, and are therefore not covered. 
	Details and a more general treatment %
	are provided in
	Appendix~\ref{sec:causal_relations_X}. 
\end{rem}

  Sometimes, the vector of covariates $X$
  contains variables, which are independent of $H$, that enter into the assignments of the other
  covariates additively and cannot be used
  for the prediction (e.g., because they are not observed during
  testing).
	If such covariates exist, 
	it can be useful to explicitly distinguish them from the 
	remaining predictors.
We will denote them by $A$ and call them exogenous variables. 
Such
variables are interesting for several reasons.
(i) 
We will see that in general, 
interventions on $A$ lead to intervention distributions with desirable properties for distribution generalization, see Section~\ref{sec:int_onA}.
(ii) 
Some of our results rely on 
the function $f$ being identifiable from the observational distribution, see Assumption~\ref{ass:identify_f} below.
The variables $A$ can be used to state explicit conditions for identifiability.
Under additional assumptions, for example,
they 
can be used as
instrumental variables \citep[e.g.,][]{Bowden1985, greene2003econometric}, 
a well-established tool for 
recovering $f$
from the observational distribution of 
$(X,Y,A)$.
  (iii) The variable $A$ can be used to model a 
  covariate that is not observed under testing. 
  It can also be used to index tasks (which we discuss at the end of Section~\ref{sec:intro_focus}). 
In the remainder of this work, we therefore 
consider a slightly larger class of SCMs that also includes exogenous variables $A$. 
It contains the SCM \eqref{eq:SCMmodelreduced} as a special 
case.\footnote{This follows from choosing $A$ as an independent noise variable and a constant $g$.}
We derive results for settings with and without exogenous variables $A$.

\subsection{Model}
Formally, we consider a response $Y \in \R^1$, covariates $X\in \R^d$,
exogenous variables $A\in \R^r$, and unobserved variables
$H \in \R^q$.  Let further $\FF\subseteq\{f:\R^d\rightarrow\R\}$,
$\GG\subseteq\{g:\R^r\rightarrow\R^d\}$,
$\HH_1\subseteq\{h_1:\R^{q+1}\rightarrow\R\}$ and
$\HH_2\subseteq\{h_2:\R^{q+d}\rightarrow\R^d\}$ be fixed sets of
measurable functions. Moreover, let $\mathcal{Q}$ be a collection of
probability distributions on $\R^{d+1+r+q}$, such that for all
$Q\in\mathcal{Q}$ it holds that if
$(\epsilon_X,\epsilon_Y, \ep_A, \ep_H)\sim Q$, then
$\epsilon_X,\epsilon_Y, \ep_A$ and $\ep_H$ are jointly independent, and
for all $h_1\in\HH_1$ and $h_2\in\HH_2$ it holds that
$\xi_Y := h_1(\ep_H, \epsilon_Y)$ and
$\xi_X := h_2(\ep_H, \epsilon_X)$ have mean zero.\footnote{ This can
  be assumed w.l.o.g.\ if $\FF$ and $\GG$ are closed
  under addition and scalar multiplication, and contain the
  constant~function.}  Let
$\mathcal{M}\coloneqq
\FF\times\GG\times\HH_1\times\HH_2\times\mathcal{Q}$ denote the model
class. Every model $M=(f, g, h_1, h_2, Q)\in \cM$ then specifies an
SCM by\footnote{ For an appropriate choice of $h_2$, the model
  includes settings in which (parts of) $A$ directly influence
  $Y$.}
\begin{minipage}{0.45\columnwidth}
  \begin{align*}
    A &\coloneqq \ep_A 	\\
    H &\coloneqq \ep_H 	\\
    X &\coloneqq g(A) + h_2(H, \epsilon_X)\\
    Y &\coloneqq f(X) + h_1(H, \epsilon_Y)
  \end{align*}
\end{minipage}%
\begin{minipage}{0.55\columnwidth}
\includegraphics[width=.9\textwidth]{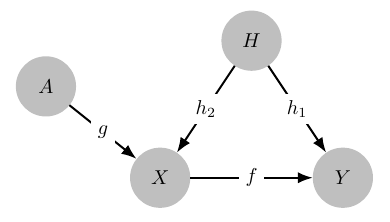}
%
\end{minipage}

\noindent with $(\epsilon_X,\epsilon_Y, \ep_A, \ep_H)\sim Q$, 
where the assignments for $A$, $H$ , $X$ and $Y$ consist of 
$r$, $q$, $d$ and~$1$ coordinate(s), respectively.
For each model $M = (f,g,h_1,h_2,Q) \in\mathcal{M}$, we refer to 
$f$ as the
\textit{causal function} (for the pair $(X,Y)$), 
and denote by $\P_M$ the joint distribution over the observed variables $(X, Y, A)$.
We assume that this distribution has finite second moments.
If no exogenous
variables $A$ exist, one can think of the function $g$ as being 
constant. A model $M$ that correctly models 
the training and test distributions will be referred to as the `true model'. 

\subsection{Interventions}\label{sec:interventions}
Each SCM $M\in\mathcal{M}$ can now be modified by the concept of
interventions \citep[e.g.,][]{pearl2009causality,Peters2017book}. An
intervention corresponds to replacing one or more of the structural
assignments of the SCM 
(see Section~\ref{sec:types_of_interventions} for details). 
For example, we intervene on some of the covariates
$X$ by replacing the corresponding assignments with, e.g.,
a Gaussian random vector that is independent of the other noise variables.
Importantly, an
intervention on some of the variables does not change the assignment
of any other variable.  In particular, an intervention on $X$ does not
change the conditional distribution of $Y$, given $X$ and $H$ (this is
an instance of the invariance property mentioned in
Section~\ref{sec:intro}) 
but it may change the conditional distribution of $Y$, given $X$.  

The problems addressed in this work require us to simultaneously consider 
several different SCMs that are all subject to the same (set of) interventions.
Formally, we therefore regard an intervention $i$
as a mapping from the model class $\MM$ into a (possibly larger) set of SCMs,
which takes as input a model $M \in \MM$ and outputs another model $M(i)$
over variables $(X^i, A^i, Y^i, H^i)$, the intervened model.
We do not %
need to assume
that the intervened
  model $M(i)$ belongs to the model class $\cM$, 
  but we require that $M(i)$ induces a joint distribution over
  $(X^i,Y^i,A^i,H^i)$\footnote{If the context does not allow for any
    ambiguity, we omit the superscript $i$.} with finite second
  moments.
  We denote the 
  corresponding distribution over the observed $(X^i, Y^i, A^i)$ by $\P_{M(i)}$, 
   and 
   use $\II$ for a collection of interventions.
  In our work, the test distributions are modeled as distributions
  generated by these types of intervened models, and the set
  $\mathcal{I}$ therefore indexes 
  the set of 
  test distributions. We
 will be 
  interested in the mean squared prediction error on each test
  distribution $i$, formally written as $\E_{M(i)}[(Y-f(X))^2]$.
(In this work, we consider a univariate $Y$, but writing
$\E [\| Y - \fs(X) \|_{\R^d}^2] = \sum_{j=1}^d \E [(Y_j - f_{\diamond, j}(X))^2]$, 
most 
our results extend straight-forwardly to a $d$-dimensional response.)

The support of random variables under interventions will play
an important role for the analysis of distribution generalization. 
Throughout this paper, $\supp^{M}(Z)$ denotes the support
of the random variable $Z\in\{A, X, H, Y\}$ under the distribution
induced by the SCM $M\in\mathcal{M}$. 
Moreover,
$\supp_{\II}^{M}(Z)$ denotes the union of $\supp^{M(i)}(Z)$ over all
interventions $i\in\II$. We call a collection of interventions on $Z$ 
\textit{support-reducing} (w.r.t.\ $M$) if $\supp_\cI^M(Z)\subseteq \supp^M(Z)$ and
\textit{support-extending} (w.r.t.\ $M$)
 if
$\supp_\cI^M(Z)\not \subseteq \supp^M(Z)$. 
Whenever it is clear from the context which model is considered, we
may drop the indication of $M$ altogether and simply write $\supp(Z)$.

\subsection{Distribution generalization} \label{sec:intro_focus}
Let $\MM$ be a fixed model class, 
let $M = (f,g,h_1, h_2, Q) \in \MM$
and let $\II$ be a class of interventions.
In this work, we aim to find a function $f^*:\R^d\rightarrow\R$, 
such that the predictive model $\hat Y = f^*(X)$ has low worst-case
risk over all test distributions induced by the interventions $\cI$ in model $M$.
We therefore consider, for the true $M$, the optimization problem 
\begin{equation} \label{eq:minimax_problem}
\argmin_{\fs \in \FF} \sup_{i \in \II} \E_{M(i)} \big[ (Y - \fs(X))^2 \big],
\end{equation}
where $\E_{M(i)}$ is the expectation 
in the intervened model
$M(i)$. In general, this optimization problem is neither guaranteed to
have a solution, 
nor is the solution, if it exists, ensured
to be unique. 
Whenever a solution $f^*$ to~\eqref{eq:minimax_problem} exists, we refer to it as a
\textit{minimax solution} (for model $M$ w.r.t.\ ($\mathcal{F},\II$)).

Depending on the model class $\MM$, there may be several models $\tilde{M} \in \MM$ that induce the observational distribution $\P_M$, that is, the same distribution over the observed variables $A$, $X$ and $Y$, but do not agree with $M$ on all intervention distributions %
induced by 
$\cI$. 
Thus, each such model induces a potentially different minimax problem with different solutions. 
Given knowledge only of
$\P_M$, it is therefore generally not possible to identify a solution
to \eqref{eq:minimax_problem}.
In this paper, we study conditions
on $\MM$, $\P_M$ and $\II$, under which this becomes possible. More
precisely, we aim to characterize under which conditions $(\P_M, \mathcal{M})$
admits distribution generalization to $\mathcal{I}$.

\begin{defi}[Distribution generalization]
  \label{defi:general}
  $(\P_{M},\mathcal{M})$ is said to 
  \emph{admit distribution generalization to} $\II$, 
  or simply to \emph{admit generalization to} $\II$,
  if for every
  $\epsilon > 0$ there exists a function $f^{*}_\varepsilon\in\mathcal{F}$ such
  that, for all models $\tilde{M}\in\mathcal{M}$ with
  $\P_{\tilde{M}}=\P_{M}$, it holds that %
  \begin{align} \label{eq:def_generalization}
  \begin{split}
    &\ \left \vert \sup_{i\in\II}\E_{\tilde{M}(i)}\big[(Y-f_\varepsilon^*(X))^2\big] \right.
    \left.- \inf_{\fs\in\FF}\,
      \sup_{i\in\II}\E_{\tilde{M}(i)}\big[(Y-\fs(X))^2\big] \right \vert \leq \epsilon.
  \end{split}
  \end{align}
\end{defi}
Distribution generalization does not require the existence of a minimax solution in
$\FF$ (which would 
require
further assumptions on the function class $\FF$)
and instead focuses on whether an approximate solution can be
identified based only on the observational distribution $\P_{M}$.
If, however, there exists a function $f^* \in \FF$ which, for every
$\tilde{M}\in\MM$ with $\P_{\tilde{M}} = \P_M$, is a minimax solution for $\tilde{M}$
w.r.t.\ $(\FF, \II)$, then, in particular,
$(\P_{M},\mathcal{M})$ admits generalization to $\II$.

Our framework also includes several settings of multitask learning
(MTL) 
  and domain adaptation \citep{Candela2009}, where one often assumes to observe different training tasks. 
In MTL,
one is then interested in using the 
different tasks to improve the predictive performance on either one or all training tasks -- this is often referred to as asymmetric and symmetric MTL, respectively.
In our framework, 
such a setup
can be modeled using a categorical
variable $X$. 
If, however, 
one is interested in predicting on an unseen task
or if one does not know which of the observed tasks the new test data come from, one may instead use a categorical $A$ 
with support-extending or support-reducing interventions,
  respectively.

\section{Minimax solutions and the causal function} \label{sec:robustness}
To address the question of distribution generalization, we first 
study properties of the minimax optimization problem \eqref{eq:minimax_problem}. 
In the simplest case, where
$\II$ consists only of the trivial intervention, that
is, $\P_M = \P_{M(i)}$, we are looking for the best predictor on the
observational distribution.  In that case, the minimax solution is 
attained at
any conditional mean function,
$f^*: x\mapsto\E[Y\vert X=x]$ (provided that 
$f^*\in\mathcal{F}$).  For larger classes of interventions, however, the
conditional mean may become sub-optimal in terms of prediction.
To see this, it is instructive to decompose the risk under 
an intervention.
Since the structural assignment for $Y$ remains unchanged 
for all interventions that we consider in this work, it holds
  for all $\fs \in \FF$ and all interventions $i$ on either $A$ or $X$ that
\begin{align*}
	\E_{M(i)}[(Y - \fs(X))^2] = &\, \E_{M(i)}[(f(X) - \fs(X))^2]+\E_{M}[\xi_Y^2] +2\E_{M(i)}[\xi_Y(f(X)-\fs(X))].
	\end{align*}
  Here, the middle term does not depend on $i$ since
  $\xi_Y = h_1(H, \ep_Y)$ remains fixed. 
  We call the intervention $i$
 \begin{align*}
 \begin{array}{ll}
 &\quad\text{\emph{confounding-removing}}\phantom{ii}\qquad
	\begin{array}{ll}
	\text{if for all models } M \in \mathcal{M} \text{ it holds that }\\
	 X \indep H, \text{ under } M(i).
	\end{array}
 \end{array}
\end{align*} 
  For such an intervention, we have that $\xi_Y\indep X$ under $\P_{M(i)}$, and hence, since 
  $\E_{M}[\xi_Y] = 0$, the last term in the above equation
  vanishes.
  Therefore, if 
  $\II$ consists only of confounding-removing
  interventions, the causal function is a solution to the
  minimax
  problem \eqref{eq:minimax_problem}. 
The
following proposition shows that an even stronger statement holds:
The causal function is already a minimax solution if $\II$ contains
at least one confounding-removing intervention on~$X$.

\begin{prop}[Confounding-removing interventions on $X$]
  \label{prop:minimax_equal_causal}
  Let $\II$ be a set of interventions on $X$ or $A$ 
such that
  there
  exists at least one $i\in\II$ 
  that is confounding-removing.
  Then, the minimal worst-case risk is attained at a
  confounding-removing intervention, and the causal function $f$ is a
  minimax solution. %
\end{prop}
We now prove that, 
in a linear setting, 
the causal
function is also minimax optimal
if the
interventions create unbounded variability in all directions of the covariance matrix of $X$.
\begin{prop}[Unbounded interventions on $X$ with linear $\mathcal{F}$]
  \label{prop:shift_interventions}
  Let $\FF$ be the class of all linear functions, and let $\II$ be a
  set of interventions on $X$ or $A$ s.t.\
  $\sup_{i\in \cI} \lambda_{\min}\big(\E_{M(i)}\big[XX^\top\big]\big) =\infty$,
  where $\lambda_{\min}$ denotes the smallest eigenvalue.
  Then, the
  causal function $f$ is the unique minimax solution. 
\end{prop}
The unbounded eigenvalue condition above is satisfied 
if $\II$ is the set of all shift interventions on $X$.
These interventions, formally defined in Section~\ref{subsec:intallowgen},
appear in linear IV models and recently gained further attention in the causal community \citep{rothenhausler2018anchor, Sani2020}. 
The proposition above considers a linear function class $\mathcal{F}$; in this way, shift interventions are related to linear models.

Even if the 
causal function $f$ does not 
solve the minimax problem~\eqref{eq:minimax_problem},
the difference
between the minimax solution and the causal function
cannot
be arbitrarily large.  
The following proposition shows that the worst-case 
$L_2$-distance between $f$ and any function $\fs$ that performs better than $f$ (in terms of worst-case risk) can be bounded by a term which is related to the strength of the confounding. 
\begin{prop}[Difference between causal function and minimax solution]
  \label{prop:difference_to_causal_function}
  Let $\II$ be a set of interventions on $X$ or $A$. Then, for any
  function $\fs\in\FF$ which satisfies that
  \begin{equation*}
    \sup_{i\in\II}\E_{M(i)}[(Y-\fs(X))^2]\leq\sup_{i\in\II}\E_{M(i)}[(Y-f(X))^2],
  \end{equation*}
  it holds that
  \begin{equation*}
    \sup_{i\in\II}\E_{M(i)}[(f(X)-\fs(X))^2]\leq 4\var_M[\xi_Y].
  \end{equation*}
\end{prop}
Even though the difference can be bounded, it may be non-zero, and one may %
benefit
from choosing a function that differs from the causal function $f$.
This choice, however, comes at a cost: it relies on the fact that we know the class of interventions $\mathcal{I}$. 
In general, being a minimax solution is not entirely
robust with respect to misspecification of $\mathcal{I}$. 
  In particular, if the 
set $\cI_2$ of interventions
  describing the test distributions is misspecified by a set
  $\cI_1\neq\cI_2$, then the considered minimax solution with respect
  to $\cI_1$ may perform worse than the causal function on the test
  distributions. 

\begin{prop}[Properties of the minimax solution under mis-specified
  interventions]
  \label{prop:misspecification_minimax}
  Let $\II_1$ and $\II_2$ be any two sets of interventions on
  $X$, and let $f_1^*\in\mathcal{F}$ be a minimax solution w.r.t.\
  $\II_1$.  Then, if $\II_2\subseteq\II_1$, it holds that
  \begin{equation*}
    \sup_{i\in\II_2}\E_{M(i)}\big[(Y-f_1^*(X))^2\big] \leq \sup_{i\in\II_2}\E_{M(i)}\big[(Y-f(X))^2\big].
  \end{equation*}
  If $\II_2\not\subseteq\II_1$, however, it can happen (even if
  $\mathcal{F}$ is linear) that
  \begin{equation*}
    \sup_{i\in\II_2}\E_{M(i)}\big[(Y-f_1^*(X))^2\big] > \sup_{i\in\II_2}\E_{M(i)}\big[(Y-f(X))^2\big].
  \end{equation*}
\end{prop}
 The second part of 
 the proposition should be understood as 
 a non-robustness
 property of non-causal minimax solutions.
 Improvements on the causal 
 function are possible 
in situations, where 
one has reasons to believe that 
the test distributions do not stem from 
a set of interventions that is much larger than the specified set.

\section{Distribution generalization}\label{sec:generalizability}
As described in Section~\ref{sec:intro_focus},
we consider a fixed model class $\mathcal{M}$ containing the true (but unknown) 
model $M$, and let $\mathcal{I}$ be a class of interventions.
By definition, the optimizer of the minimax problem~\eqref{eq:minimax_problem}
depends on the true model $M$.
Section~\ref{sec:robustness} relates 
this optimizer
to the causal function $f$, 
whose knowledge, too, requires knowing $M$.
In practice, however, we do not have access to the true model $M$, but only to its observational distribution $\P_M$. 
This motivates the notion of distribution generalization, see~\eqref{eq:def_generalization}. 
In words, it states that approximate minimax solutions (which depend on the intervention distributions $\P_{M(i)}$, $i \in \II$)
are identified from the observational distribution $\P_M$. This holds true, in particular, if the 
intervention distributions themselves are identified from $\P_M$.
\begin{prop}[Sufficient conditions for distribution generalization]
  \label{prop:suff_general}
  Assume that for all $\tilde{M} \in \MM$ it holds that
  \begin{equation*}
  \P_{\tilde{M}} = \P_M \quad \Rightarrow \quad \P^{(X,Y)}_{\tilde{M}(i)} = \P^{(X,Y)}_{M(i)} \quad \forall i \in \II,
  \end{equation*}
 where $\P^{(X,Y)}_{M(i)}$ 
 is the joint distribution of $(X, Y)$ under
  $M(i)$. Then, $(\P_M, \MM)$ admits generalization to~$\II$.
\end{prop}
Proposition~\ref{prop:suff_general} provides verifiable conditions for
distribution generalization, and 
can be used to prove
possibility statements. 
 It is, however, not a necessary condition.
Indeed, we will see that, under certain types of interventions, 
distribution generalization becomes possible even in cases
  where the interventional marginal of~$X$ is not identified.

In this section, we study conditions on $\MM$, $\P_M$ and $\II$ which ensure 
generalization, and present 
corresponding impossibility results proving the necessity of some of these conditions. 
Two aspects will be of central importance. The first is related
to causal identifiability, i.e., whether the causal function $f$ is sufficiently identified from the observational 
distribution $\P_M$ (Section~\ref{sec:causal_identifiability}). The other aspect is related to the 
types of interventions (Section~\ref{sec:types_of_interventions}). 
  We consider interventions on $X$ in
  Section~\ref{sec:gen_int_on_X} and interventions on $A$ in
  Section~\ref{sec:int_onA}.  
  Parts of our results are summarized in Table~\ref{tab:generalizability}.
\begin{table}
  \centering
   {%
   \small
  \renewcommand{\arraystretch}{1.15}
 \begin{tabular}{>{\centering\arraybackslash}p{3cm}>{\centering\arraybackslash}p{2.5cm}>{\centering\arraybackslash}p{3.5cm}|>{\centering\arraybackslash}p{3cm}}
    \toprule    
    intervention on & $\supp_{\II}(X)$ & assumptions & result  \\
    \toprule
    $X$ (well-behaved) & $ \subseteq \supp(X)$ & Assumption~\ref{ass:identify_f} & Proposition~\ref{prop:genX_intra} \\
     $X$ (well-behaved) & $ \not \subseteq \supp(X)$ &Assumptions~\ref{ass:identify_f}~and~\ref{ass:gen_f} & Proposition~\ref{prop:genX_extra}\\
    $A$ & $ \subseteq \supp(X)$ & Assumptions~\ref{ass:identify_f}~and~\ref{ass:identify_g} & Proposition~\ref{prop:genA}\\
    $A$ & $\not \subseteq \supp(X)$ & Assumptions~\ref{ass:identify_f},~\ref{ass:gen_f}~and~\ref{ass:identify_g} & Proposition~\ref{prop:genA}\\
    \arrayrulecolor{black}\bottomrule
  \end{tabular}
  }
  \caption{Summary of conditions under which generalization is
    possible. Corresponding impossibility results are shown in
    Propositions~\ref{prop:impossibility_interpolation},~\ref{prop:impossibility_extrapolation} and~\ref{prop:impossibility_intA}.}
  \label{tab:generalizability}
\end{table}

\subsection{Identifiability of the causal function} \label{sec:causal_identifiability}
For specific types of interventions, the causal function $f$ is itself a minimax solution,
see Propositions~\ref{prop:minimax_equal_causal}~and~\ref{prop:shift_interventions}. 
If, in addition,
these interventions are 
support-reducing, 
generalization is directly implied by the following assumption.

\begin{ass}[Identifiability of $f$ on the support of $X$]
  \label{ass:identify_f}
  For all
  $\tilde{M}=(\tilde{f}, \dots) \in \mathcal{M}$ with $\P_{\tilde{M}} = \P_{M}$, it holds
  that $\tilde{f}(x) = f(x)$ for all $x \in \supp(X)$.
\end{ass}

Assumption~\ref{ass:identify_f} will play a central role in proving distribution generalization
even in situations where the causal function is not a minimax solution.
We 
use it as a starting point for 
most
of our results. 
The assumption
is
violated, for example, in a linear Gaussian setting with a single covariate $X$ (without $A$). Here, in general, we cannot identify $f$ and distribution generalization does not hold. 
Assumption~\ref{ass:identify_f}, however, is not necessary for  generalization.
In Section, \ref{sec:int_onA} we discuss 
a linear setting
where 
distribution generalization is possible, even if Assumption~\ref{ass:identify_f} does not hold.

The question of causal identifiability
has received a lot of attention in the literature. In linear instrumental
variables settings, for example, one assumes that the functions $f$ and
$g$ are linear and 
identifiability follows if
the product moment between $A$ and $X$ has rank at
least the dimension of $X$
\cite[e.g.,][]{wooldridge2010econometric}.
In
linear non-Gaussian models, %
one can identify the function
$f$ even 
if there are no instruments
\citep{Hoyer2008b}. %
For nonlinear models, restricted SCMs can be
exploited, too. In that case, Assumption~\ref{ass:identify_f} holds
under regularity conditions if $h_1(H, \epsilon_Y)$ is independent of
$X$ \citep{Zhang2009, Peters2014jmlr, Peters2017book}
and first attempts 
have been made
to extend such results to
non-trivial confounding cases 
\citep{Janzing2009uai}.  
The nonlinear IV setting \citep[e.g.,][]{amemiya1974nonlinear,newey2013nonparametric,newey2003instrumental}
is discussed in more detail in 
Appendix~\ref{sec:IVconditions}, where we give a brief overview of
identifiability
results for linear, parametric and non-parametric function classes. 
Assumption~\ref{ass:identify_f} states that $f$ is identifiable, 
even on $\P_M$-null sets, which 
is usually achieved by placing further 
constraints on the function class, such as smoothness.
Even though this issue seems 
technical, it becomes important when considering hard interventions
that
set $X$ to a fixed value, for example.

\subsection{Types of interventions} \label{sec:types_of_interventions}
Whether distribution generalization is admitted depends on
the intervention class $\mathcal{I}$.
In this work,
we only consider interventions on the covariates $X$ and $A$. 
Each of these types of interventions can be characterized by a measurable 
function $\psi^i$,
which determines the structural assignment of the intervened variable, and a (possibly degenerate) random vector $I^i$,
which serves as an independent noise innovation. More formally, 
for an intervention on $X$, the pair $(\psi^i, I^i)$ defines the intervention which
maps the input model $M=(f,g,h_1,h_2,Q)\in \cM$ to the intervened model $M(i)$ given by the assignments
\begin{align*}
  A^i := \ep_A^i, \quad H^i := \ep_H^i, \quad
  X^i := \psi^i(g, h_2, A^i,H^i, \ep^i_X ,I^i), \quad
  Y^i := f(X^i) + h_1(H^i,\ep_Y^i).
\end{align*}
Similarly, for an intervention on $A$, 
$(\psi^i, I^i)$ specifies the intervention which outputs %
\begin{align*}
  A^i := \psi^i(I^i, \ep_A^i), \quad  H^i := \ep_H^i, \quad
  X^i := g(A^i) + h_2(H^i, \ep_X^i), \quad
  Y^i := f(X^i) + h_1(H^i,\ep_Y^i).
\end{align*}
In both cases, $(\ep_X^i, \ep_Y^i, \ep_A^i,\ep_H^i) \sim Q$ and $I^i \indep (\ep_X^i, \ep_Y^i, \ep_A^i,\ep_H^i)$.
We will see below that this
class of interventions is rather flexible.
It does, however, not allow for arbitrary manipulations of $M$. 
For example, it does not allow for changes in the structural assignments for $Y$ or $H$, 
or for the noise variable $\ep_Y^i$ to enter the assignment of the intervened variable. 
As the following section highlights, further constraints on the types of interventions are 
necessary to ensure distribution generalization.

\subsubsection{Impossibility of generalization without constraints on the interventions}

Let $\mathcal{Q}$ be a class of product 
distributions on $\R^4$, such that for all $Q \in \mathcal{Q}$, 
the coordinates of $Q$ are non-degenerate, zero-mean with finite second moment. 
Let $\MM$ be the class of all models of the form 
\begin{equation*}
A\coloneqq \epsilon_A, \quad %
H\coloneqq\sigma\epsilon_H, \quad %
X\coloneqq \gamma A + \epsilon_X + \tfrac{1}{\sigma}H, \quad %
Y\coloneqq \beta X + \epsilon_Y + \tfrac{1}{\sigma}H,
\end{equation*}
with $\gamma, \beta \in \R$, $\sigma > 0$ and $(\epsilon_A, \epsilon_X,\epsilon_Y,\epsilon_H) \sim Q \in \mathcal{Q}$. 
Assume that $\P_M$ is induced by some model $M = M(\gamma, \beta, \sigma, Q)$ from the above model class
(here, we slightly adapt the notation from Section~\ref{sec:framework}). 
The following proposition shows that, without constraining the set of interventions $\II$, 
distribution generalization is not always ensured.
 
\begin{prop}[Impossibility of generalization without constraining the class of interventions]
    \label{prop:impossibility_interpolation}
     Assume that $\MM$ is given as defined above, 
     let $\II \subseteq \R_{>0}$ be 
     a 
     compact, non-empty
    set and define the interventions on $X$
    by $\psi^i(g, h_2, A^i, H^i, \epsilon_X^i, I^i) = iH$, for
    $i \in \II$.
    Then, $(\P_M, \cM)$ does not admit generalization to $\II$
    (even if Assumption~\ref{ass:identify_f} is satisfied).
In addition,
    any prediction model other than the causal model
    may perform arbitrarily bad under the interventions $\II$. 
    That is, for any $b \neq \beta$ and any $c > 0$,
    there exists a model $\tilde{M}\in\cM$
    with $\P_{\tilde{M}} = \P_{M}$, such that
    \begin{equation*}
      \abs[\Big]{\sup_{i\in\II}\E_{\tilde{M}(i)}\big[(Y-bX)^2\big]
      - \inf_{\bes \in \R}\,
      \sup_{i\in\II} \E_{\tilde{M}(i)}\big[(Y-\bes X)^2\big]} \geq c.
    \end{equation*}
\end{prop}
We now give some intuition about the above result.
By definition, distribution generalization is ensured if there exist prediction functions
that are (approximately) minimax optimal for all models which induce the same observational
distribution as $M$. Since, in the above example, the distribution of $(X,Y,A)$ does not depend on $\sigma$, 
this includes all models of the form $M_{\tilde{\sigma}} = M(\gamma, \beta, \tilde{\sigma}, Q)$
for some $\tilde{\sigma} > 0$. However, while agreeing on the observational distribution, each of these 
models induces fundamentally different intervention distributions
(under $M_{\tilde{\sigma}}(i)$, $(X,Y)$ is equal in distribution to $(i \epsilon_H, (\beta i + \frac{1}{\tilde{\sigma}}) \epsilon_H)$)
and results in different (approximate) minimax solutions. 
Below, we introduce two types of interventions which ensure distribution generalization 
in a wide range of settings by constraining the influence of $H$ on $X$.

\subsubsection{Interventions which allow for generalization} \label{subsec:intallowgen}
In Section~\ref{sec:robustness}, we already introduced confounding-removing interventions, 
which break the dependence between $X$ and $H$.
For an intervention set $\II$ which contains at least one confounding-removing intervention, 
the causal function $f$ is always a minimax solution (see Proposition~\ref{prop:minimax_equal_causal})
and, in the case of support-reducing interventions, distribution generalization is therefore achieved by requiring Assumption~\ref{ass:identify_f} to hold.
The intervention $i$ 
with intervention map $\psi^i$
is called 
\begin{align*}
\begin{array}{ll}
&\quad\text{\emph{confounding-preserving}}\phantom{ii}\qquad
	\begin{array}{ll}
	 \text{if there exists a map } \phi^i, \text{ such that }\\
	\psi^i(g, h_2, A^i ,H^i, \ep^i_X ,I^i) = \phi^i(A^i,g(A^i),h_2(H^i, \ep^i_X) ,I^i).
	\end{array}
\end{array}
\end{align*}
Confounding-preserving interventions contain, e.g., \textit{shift interventions} on $X$,
which linearly shift the original assignment by $I^i$, that is,
$\psi^i(g, h_2, A^i,H^i, \ep^i_X ,I^i) = g(A^i) + h_2(H^i, \ep_X^i) +
I^i$. 
The name 
`confounding-preserving'
stems from the fact that the confounding
variables $H$ only enter the intervened structural assignment of $X$
via the term $h_2(H^i, \ep^i_X)$,
which is the same as in the original model.
(This property fails to hold true for the interventions in Proposition~\ref{prop:impossibility_interpolation}.)
If $\II$ consists only of confounding-preserving
interventions, the causal function is generally not a
 minimax solution. 
 However, we will see that, under Assumption~\ref{ass:identify_f}, these types 
of interventions lead to identifiability of the intervention distributions $\P_{M(i)}$, $i \in \II$, and therefore
ensure 
generalization via Proposition~\ref{prop:suff_general}.

Some interventions are 
both
confounding-removing and confounding-preserving, but
not every confounding-removing
intervention is 
confounding-preserving. For example, the
intervention 
$\psi^i(g, h_2, A^i ,H^i, \ep^i_X ,I^i)=\ep^i_X$
is confounding-removing but,
in general,
not confounding-preserving.
Similarly, not all confounding-preserving interventions
are confounding-removing. %
We call a set of interventions $\II$
\textit{well-behaved} either if it consists only of
confounding-preserving interventions or if it contains at least one
confounding-removing intervention.

\subsection{Generalization to interventions on $X$} \label{sec:gen_int_on_X}
We now formally prove in which sense the two types of interventions defined above 
allow for distribution generalization. We will see that this question is closely linked to the
relation between the support of $\P_M$ and the support of the
intervention distributions.  Below, we therefore distinguish between
support-reducing and support-extending interventions on~$X$.

\subsubsection{Support-reducing interventions} \label{sec:supp_reducing_onX}
For support-reducing interventions, Assumption~\ref{ass:identify_f} is sufficient 
for distribution generalization even in nonlinear settings, under a large class of interventions.
\begin{prop}[Generalization to support-reducing interventions on $X$]
    \label{prop:genX_intra}
    Let $\II$ be a well-behaved set of 
    interventions on $X$, and
    assume that $\supp_{\II}(X)\subseteq\supp(X)$. 
    Then, under Assumption~\ref{ass:identify_f}, 
	$(\P_{M}, \cM)$ admits generalization to the interventions $\II$.
	If one of the interventions is confounding-removing, then the causal function is a minimax solution.
\end{prop}
In the case of support-extending interventions, further assumptions are required to ensure distribution generalization.

\subsubsection{Support-extending interventions}\label{sec:support_extending_onX}

If the interventions in $\II$ extend the support of $X$, i.e.,
$\supp_\II(X) \not \subset \supp(X)$, Assumption~\ref{ass:identify_f}
is not sufficient for ensuring distribution generalization. 
This is because there may exist a model $\tilde{M} \in \MM$ which agrees with $M$ on the observational distribution, 
but whose corresponding causal function $\tilde{f}$ differs from $f$ outside of the support of $X$.
In that case, a
support-extending intervention on~$X$ may result in different
dependencies between $X$ and $Y$ in the two models, and therefore
potentially induce a different set of minimax solutions. 
The following assumption on the model class $\FF$
ensures that any $f\in \cF$ is uniquely determined by its values
on $\supp(X)$.
\begin{ass}[Extrapolation of $\FF$] \label{ass:gen_f} For all
  $\tilde{f}, \bar{f} \in \FF$ with $\tilde{f}(x) = \bar{f}(x)$ for all
  $x \in \supp(X)$, it holds that $\tilde{f} \equiv \bar{f}$.
\end{ass}
We will see that this assumption is sufficient
	(Proposition~\ref{prop:genX_extra}) for generalization to well-behaved interventions on $X$. Furthermore, it is also necessary
	(Proposition~\ref{prop:impossibility_extrapolation}) if $\cF$ is sufficiently flexible.
The following proposition  can be seen as an extension of Proposition~\ref{prop:genX_intra}. 
\begin{prop}[Generalization to support-extending interventions on $X$]
    \label{prop:genX_extra}
    Let $\II$ be a well-behaved set of 
    interventions on $X$.
    Then, under Assumptions~\ref{ass:identify_f}~and~\ref{ass:gen_f}, 
	$(\P_{M}, \cM)$ admits generalization to $\II$.
	If one of the interventions is confounding-removing, then the causal function is a minimax solution. %
\end{prop}
Because the interventions 
may change the marginal distribution of~$X$, 
the preceding proposition 
includes examples, 
in which 
distribution generalization is possible
even if 
some of the  considered
joint (test) distributions
are arbitrarily far 
from the training distribution, 
in terms of any reasonable divergence measure over
distributions, such as Wasserstein distance or $f$-divergence.

Proposition~\ref{prop:genX_extra} relies on Assumption~\ref{ass:gen_f}. Even though this assumption
is restrictive, it is satisfied by several reasonable function classes, which therefore allow for generalization
to any set of well-behaved interventions.  
Below, we give two examples
of such 
function classes.

\paragraph{Sufficient conditions for generalization} 

Assumption~\ref{ass:gen_f} states that every function in $\FF$ is globally identified
by its values on $\supp(X)$. This is, for example, satisfied if 
$\mathcal{F}$ is a linear space of functions with domain $\mathcal{D} \subset \R^d$
which are linearly independent on $\supp(X)$. More precisely, 
$\mathcal{F}$ is linearly closed, i.e.,
\begin{align}
f_1, f_2 \in \mathcal{F}, c \in \R, \implies
f_1 + f_2 \in \mathcal{F}, cf_1 \in \mathcal{F},
 \label{eq:linear}
\end{align}
and $\mathcal{F}$ is linearly independent on $\supp(X)$, i.e.,
\begin{align}
 f_1(x) = 0 \quad \forall x \in \supp(X) \;\implies \;f_1(x) = 0 \quad \forall x \in \mathcal{D}.
\label{eq:linearind}
\end{align}
Examples of such classes include (i) globally linear parametric
function classes, i.e., $\FF$ is of the form
\begin{equation*}
  \mathcal{F}^1 \coloneqq\{\fs:\mathcal{D}\rightarrow\R \given \exists\gamma \in \R^k \text{ s.t. } 
  \forall
  x\in\mathcal{D} \st
  \fs(x)=\gamma^\top \nu (x) \},
\end{equation*}
where $\nu = (\nu_1, \dots, \nu_k)$ consists of real-valued, linearly
independent functions satisfying that $\E_M[\nu(X) \nu(X)^\top]$ is
strictly positive definite, and (ii) the class of differentiable
functions that extend linearly outside of $\supp(X)$, that is, $\FF$
is of the form
\begin{equation*}
\FF^2 := \left\lbrace \fs:\mathcal{D}\rightarrow\R \, \bigg\vert \,
\begin{tabular}{@{}l@{}}
$\fs \in C^1 \text{ and } \forall x\in\mathcal{D} \setminus  \supp(X):$
$\fs(x)=\fs(x_b)+\nabla\fs(x_b)(x-x_b)$
\end{tabular}
\right\rbrace
\end{equation*}
where $x_b\coloneqq\argmin_{z\in\supp(X)}\norm{x-z}$ 
and
$\supp(X)$ is assumed to be closed
with non-empty interior.
Clearly, both of the above function classes are linearly closed. To
see that $\FF^1$ satisfies \eqref{eq:linearind}, let $\gamma \in \R^k$
be s.t.\ $\gamma^\top \nu (x) = 0$ for all $x \in \supp(X)$. Then, it
follows that
$0 = \E_M[(\gamma^\top \nu(X))^2] = \gamma^\top \E_M[\nu(X)
\nu(X)^\top] \gamma$ and hence that $\gamma = 0$. To see that $\FF^2$
satisfies \eqref{eq:linearind}, let $\fs \in \FF^2$ and assume that
$\fs(x) = 0$ for all $x \in \supp(X)$.  Then,  $\fs(x) = 0$ for
  all $x \in \mathcal{D}$ and thus  $\FF^2$ uniquely defines the
function on the entire domain $\mathcal{D}$.

By Proposition~\ref{prop:genX_extra}, generalization with respect to
these model classes is possible for any 
well-behaved
set of 
interventions. 
In practice, it may often be more realistic to impose bounds on the
higher order derivatives of the functions in $\FF$.  We now prove that
this still allows for 
what we will call
approximate distribution generalization, 
see Propositions~\ref{prop:extrapolation_bounded_deriv_cr} and~\ref{prop:extrapolation_bounded_deriv}.

\paragraph{Sufficient conditions for approximate generalization} 
For differentiable functions, exact generalization cannot always be achieved. 
Bounding the first derivative, however, allows us to achieve
approximate generalization. We therefore consider 
the following
function class
\begin{equation}\label{eq:boundedder}
\FF^2 := \{  \fs:\mathcal{D} \rightarrow \R \given 
\fs \in C^1 \text{ with } \norm{\nabla\fs}_{\infty} \leq K \}
\end{equation}
  for some fixed $K<\infty$, where $\nabla\fs$ denotes the 
  gradient
  and
  $\mathcal{D}\subseteq\R^d$. We then have the following result.

\begin{prop}[Approx. generalization with bdd. derivatives (confounding-removing)]
  \label{prop:extrapolation_bounded_deriv_cr}
  Let $\mathcal{F}$ be as defined in \eqref{eq:boundedder}.
  Let $\II$ be a set of 
  interventions on $X$ containing at least one confounding-removing
  intervention, and assume that Assumption~\ref{ass:identify_f} holds
  true.
  (In this case, the causal function $f$ is a minimax solution.)
  Then, for all $f^*$
  with $f^*=f$ on $\supp(X)$ and all
  $\tilde{M}\in\mathcal{M}$ with $\P_{\tilde{M}}=\P_{M}$, it holds that
  \begin{align*}\label{eq:gen_cond}
   & \abs[\Big]{\sup_{i\in\II}\E_{\tilde{M}(i)}\big[(Y-f^*(X))^2\big]
    - \inf_{\fs\in\FF}\,
      \sup_{i\in\II}\E_{\tilde{M}(i)}\big[(Y-\fs(X))^2\big]}
      \leq
    4\delta^2K^2+4\delta K \sqrt{\var_M(\xi_Y)},
  \end{align*}
  where
  $\delta :=
  \sup_{x\in\supp_{\cI}^{M}(X)}\inf_{z\in\supp^{M}(X)}\norm{x-z}$. 
  If $\II$ consists only of confounding-removing interventions, 
  the same statement holds when replacing
  the bound
  by
   $4\delta^2K^2$.
\end{prop}
Proposition~\ref{prop:extrapolation_bounded_deriv_cr} states that the
deviation of the worst-case generalization 
error from the best possible value 
is bounded by a term that grows with the
square of $\delta$. 
Intuitively, this means that under the function
class defined in~\eqref{eq:boundedder}, approximate generalization is
reasonable only for interventions that are close to the support of
$X$. 
We now prove a
similar result 
for
cases in which the minimax
solution is not necessarily the causal function. The following
proposition bounds the worst-case generalization error for
arbitrary confounding-preserving interventions. 
Here, the bound 
additionally accounts for the approximation to the minimax solution.
\begin{prop}[Approx. generalization with bdd. derivatives
  (confounding-preserving)]
  \label{prop:extrapolation_bounded_deriv}
  Let $\mathcal{F}$ be as defined in \eqref{eq:boundedder}. Let $\II$
  be a set of confounding-preserving interventions on $X$, and assume
  that Assumption~\ref{ass:identify_f} is satisfied. 
  Let $\epsilon > 0$ and let $f^{*}\in\mathcal{F}$ be such that, 
  \begin{align*}
   \left\vert\sup_{i\in\II}\E_{M(i)}\big[(Y-f^*(X))^2\big]\right.
   \left. - \inf_{\fs\in\FF}\,
      \sup_{i\in\II}\E_{M(i)}\big[(Y-\fs(X))^2\big]\right\vert \leq \epsilon.
  \end{align*}
  Then, for all $\tilde{M}\in\mathcal{M}$ with
  $\P_{\tilde{M}}=\P_{M}$, it holds that
  \begin{align*}
    &\abs[\Big]{\sup_{i\in\II}\E_{\tilde{M}(i)}\big[(Y-f^*(X))^2\big]
      - \inf_{\fs\in\FF}\,
      \sup_{i\in\II}\E_{\tilde{M}(i)}\big[(Y-\fs(X))^2\big]} \\
        &\quad 
        \leq \ep + 12 \delta^2 K^2 + 32 \delta K \sqrt{\var_M(\xi_Y)} +
            4 \sqrt{2} \delta K \sqrt{\ep}
  \end{align*}
  where
  $\delta :=
  \sup_{x\in\supp_{\cI}^{M}(X)}\inf_{z\in\supp^{M}(X)}\norm{x-z}$. 
\end{prop}
We 
can take $f^*$ to be the minimax
solution if it exists. In that case, the terms involving $\ep$ 
disappear from the bound, which 
then becomes more similar to the one in
Proposition~\ref{prop:extrapolation_bounded_deriv_cr}.

\paragraph{Impossibility of generalization without constraints on $\FF$}
If we do not constrain the function class $\FF$, generalization is
impossible. 
Even if
we consider the set
of all continuous functions $\FF$, 
we cannot generalize to
interventions outside the support of $X$.
This statement holds
even if
Assumption~\ref{ass:identify_f} is satisfied.

\begin{prop}[Impossibility of extrapolation]
  \label{prop:impossibility_extrapolation}
  Assume that
  $\FF = \{\fs: \R^d \to \R \mid \fs\text{ is continuous}\}$.  Let $\II$
  be a well-behaved set of  support-extending  interventions on $X$,
  such that $\supp_{\II}(X) \setminus \supp(X)$ has non-empty
  interior.  Then, $(\P_{M}, \cM)$ does not admit generalization to $\II$, 
  even if Assumption~\ref{ass:identify_f} is
  satisfied.  In particular, for any function $\bar{f} \in\FF$ and any
  $c > 0$, there exists a model $\tilde{M} \in \MM$, with
  $\P_{\tilde{M}} = \P_{M}$, such that
  \begin{equation*}
  \abs[\Big]{\sup_{i\in\II}\E_{\tilde{M}(i)}\big[(Y-\bar{f}(X))^2\big]
  - \inf_{\fs \in \FF}\,
  \sup_{i\in\II} \E_{\tilde{M}(i)}\big[(Y-\fs(X))^2\big]} \geq c.
  \end{equation*}
\end{prop}
The above impossibility result is visualized in Figure~\ref{fig:impossibility} (left).

\subsection{Generalization to interventions on $A$} \label{sec:int_onA}

We will see that,
for interventions on $A$, 
parts of the analysis simplify.
Since $A$ influences the system only via 
the covariates $X$, any such intervention
may, in terms of its effect on $(X,Y)$, be 
equivalently expressed as an intervention on $X$
in which
the structural assignment
of $X$ is altered in a way that depends on 
the functional relationship $g$ between $X$ and $A$. 
We can therefore employ several of the results from Section~\ref{sec:gen_int_on_X} by 
imposing an additional assumption on the identifiability of $g$.

\begin{ass}[Identifiability of $g$] \label{ass:identify_g}
For all $\tilde{M} = (\tilde{f}, \tilde{g}, \tilde{h}_1, \tilde{h}_2, \tilde{Q}) \in \MM$ 
with $\P_{\tilde{M}} = \P_M$, it holds that $\tilde{g}(a) = g(a)$ for all $a \in \supp(A) \cup \supp_\II(A)$. 
\end{ass}
Since $g(A)$ is a conditional mean for $X$ given $A$, the values of
$g$ are identified from $\P_M$ for $\P_M$-almost all $a$. If
$\supp_\II(A) \subset \supp(A)$, Assumption~\ref{ass:identify_g} 
therefore holds if, for example, 
$\GG$ contains continuous functions only. 
The
pointwise identifiability of $g$ is necessary, for example, if some of the test
distributions are induced by 
hard interventions on $A$, which
set $A$ to some fixed value $a \in \R^r$. In the case where the
interventions $\II$ extend the support of $A$, we additionally require
the function class $\GG$ to extrapolate from $\supp(A)$ to
$\supp(A) \cup \supp_\II(A)$; this is similar to the
conditions on $\FF$ which we made in
Section~\ref{sec:support_extending_onX}
and requires further restrictions on $\GG$.
 Under
Assumption~\ref{ass:identify_g}, we obtain a result corresponding to
Propositions~\ref{prop:genX_intra}~and~\ref{prop:genX_extra}.
\begin{prop}[Generalization to interventions on $A$]
  \label{prop:genA}
  Let $\II$ be a set of interventions on $A$ and assume
  Assumption~\ref{ass:identify_g} is satisfied. Then, $(\P_{M}, \cM)$
  admits generalization to $\II$ if either $\supp_{\II}(X)\subseteq \supp(X)$
  and Assumption~\ref{ass:identify_f} is satisfied or if both
  Assumptions~\ref{ass:identify_f}~and~\ref{ass:gen_f} are satisfied.
\end{prop}
As becomes clear from the proof of this proposition, in general, the causal function does not need to be a minimax solution.
Further,
Assumption~\ref{ass:identify_f} is not necessary for
generalization. 
In the case where $\FF$, $\GG$, $\HH_1$ and $\HH_2$ consist of linear functions,
anchor regression \cite{rothenhausler2018anchor} and K-class estimators \cite{jakobsen2020distributional}
consider certain sets of interventions on $A$
which render minimax solutions identifiable (and estimate them consistently)
even if 
Assumption~\ref{ass:identify_f} does not hold. 
Similarly, if for a
categorical A, we have $\supp_\II(A) \subseteq \supp(A)$, it is possible to drop Assumption~\ref{ass:identify_f}.

\subsubsection{Impossibility of generalization without constraining $\mathcal{G}$}
\begin{figure}
\centering
\includegraphics[width=\linewidth]{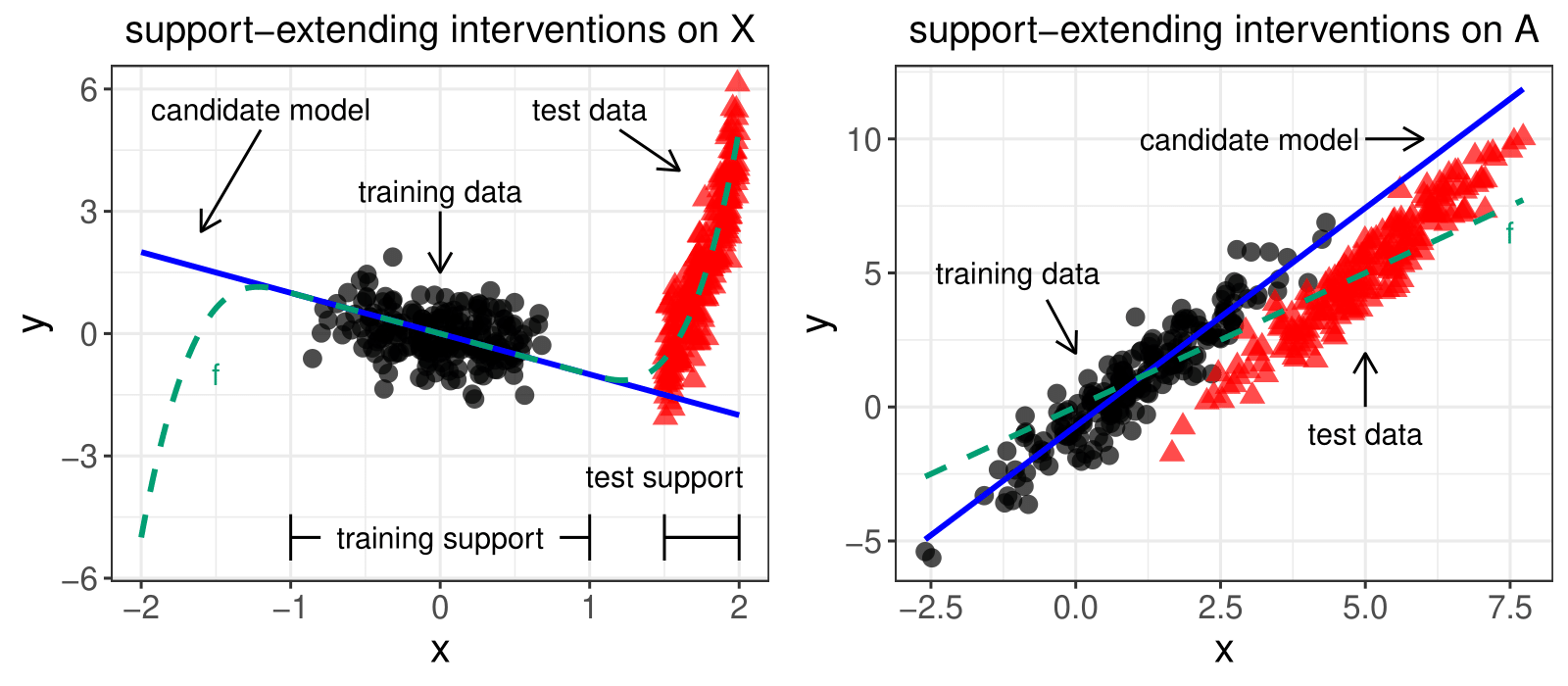} 
\caption{
Plots illustrating the straight-forward idea behind 
the impossibility results in Proposition~\ref{prop:impossibility_extrapolation} (left)
and Proposition~\ref{prop:impossibility_intA} (right).
Both plots visualize the case of univariate variables.
Under well-behaved interventions on $X$ (left; here using confounding-removing interventions)
which extend the support of $X$, generalization 
is impossible without further restrictions on the function class $\FF$. 
This holds true even if Assumption~\ref{ass:identify_f} is
satisfied. Indeed, although the candidate model (blue line) coincides
with the causal model (green dashed curve) on the support of $X$, it
may perform arbitrarily bad on test data generated under
support-extending interventions. Under interventions on $A$ 
(right)
generalization is impossible even under strong assumptions on the
function class $\FF$ (here, $\FF$ is the class of all linear
functions). Any support-extending intervention on $A$ shifts the
marginal distribution of $X$ by an amount which depends on the
(unknown) function $g$, resulting in a distribution of $(X,Y)$ which,
in general,
cannot be identified from the observational distribution. Without
further restrictions on the function class $\GG$, any candidate model
apart from the causal model may result in arbitrarily large worst-case
risk.
}
\label{fig:impossibility}
\end{figure}

Without restrictions on the model class
$\GG$, generalization to interventions on $A$ is impossible.
This holds true even under strong assumptions on the true
causal function (such as $f$ is known to be linear).  Below, we give a
formal impossibility result for hard interventions on $A$, which set $A$ to some fixed value,
and
where $\GG$ is the set of all
continuous functions.
\begin{prop}[Impossibility of generalization to interventions on $A$] \label{prop:impossibility_intA}
Assume that $\FF = \{\fs: \R^d \to \R \given \fs \text{ is linear} \}$ and
$\GG = \{\gs : \R^r \to \R^d \given \gs \text{ is continuous} \}$.
Let $\mathcal{A} \subset \R^r$ be bounded, and let $\II$ denote the
set of all hard interventions which set $A$ to some fixed value from
$\mathcal{A}$.  Assume that $\mathcal{A} \setminus \supp(A)$ has
nonempty interior. Assume further that $\E_M[\xi_X \xi_Y] \neq 0$
(this excludes the case of no hidden confounding).  Then, $(\P_M, \MM)$
does not admit generalization to $\II$. 
In addition,
any
function other than $f$ may perform arbitrarily bad under the
interventions in $\II$.  That is, for any $\bar{f} \neq f$ and $c > 0$, 
there exists a model $\tilde{M} \in \MM$ with
$\P_{\tilde{M}} = \P_M$ such that
\begin{equation*}
  \abs[\Big]{\sup_{i\in\II}\E_{\tilde{M}(i)}\big[(Y-\bar{f}(X))^2\big] - \inf_{\fs \in \FF}\,
  \sup_{i\in\II} \E_{\tilde{M}(i)}\big[(Y-\fs(X))^2\big]} \geq c.
  \end{equation*}
\end{prop}
The above impossibility result is visualized in Figure~\ref{fig:impossibility} (right).
This proposition 
is part of the argument 
showing that anchor regression
  \citep{rothenhausler2018anchor} can be extended to nonlinear
  settings  only under strong assumptions;
  the setting of a linear class $\mathcal{G}$ and a potentially nonlinear class $\mathcal{F}$ is covered in Section~\ref{sec:support_extending_onX}, 
by rewriting interventions on $A$ as interventions on $X$.

An impossibility result similar to the 
proposition above can be shown if
  $A$ is categorical. As long as not all categories have been observed
  during training it is possible that the intervention which sets $A$
  to a previously unseen category can result in a support-extending
  distribution shift on $X$. Using
  Proposition~\ref{prop:impossibility_extrapolation}, it therefore
  follows that generalization can become impossible. 
Since a categorical $A$ can encode 
settings of multi-task learning 
and domain generalization (see Section~\ref{sec:intro_focus}),
this result then 
complements well-known impossibility results for these problems, even under the covariate shift assumption \citep[e.g.,][]{David10}.

\section{Learning generalizing models from data} \label{sec:learning}
So far, our focus has been on the possibility to generalize, that is,
we have investigated under which conditions it is possible to identify
generalizing models from the observational distribution. In practice,
generalizing models need to be estimated from finitely many data.
This task is challenging for several reasons. First, analytical
solutions to the minimax problem \eqref{eq:minimax_problem} are only
known in few cases.  Even if generalization is possible, the
inferential target thus often remains a complicated object, given as a
well-defined but unknown function of the observational
distribution. Second, we have seen that the ability to generalize
depends strongly on whether the interventions extend the support of
$X$, see Propositions~\ref{prop:genX_extra} and
\ref{prop:impossibility_extrapolation}.  In a setting with a finite
amount of data, the empirical support of the data lies within some
bounded region, and suitable constraints on the function class $\FF$
are necessary when aiming to achieve empirical generalization outside
this region, even if $X$ comes from a distribution with full
support. 
As we show in our simulations in  Section~\ref{sec:experiments} (see figures), constraining the function class can  also improve the prediction performance at the boundary of the  support.

In Section~\ref{sec:existingmethods}, we survey existing methods for learning
generalizing models. Often, these methods assume either a globally linear model class $\FF$
or are completely non-parametric and therefore do not generalize outside the empirical 
support of the data. Motivated by this observation, we introduce in Section~\ref{sec:nile}
a novel estimator, which exploits an instrumental variable setup and a particular extrapolation 
assumption to learn a globally generalizing model.

\subsection{Existing methods}
\label{sec:existingmethods}
As discussed in Section~\ref{sec:intro}, a wide range of
methods have been proposed to guard against various types of
distributional changes. Here, we review methods that fit into
the causal framework in the sense that the distributions 
that in the minimax formulation the supremum is taken over %
are induced by interventions.
\begin{table*}
  \centering
  \small
  \renewcommand{\arraystretch}{1.35}
  \begin{tabular}{>{\centering\arraybackslash}p{1.75cm}>{\centering\arraybackslash}p{3.25cm}>{\centering\arraybackslash}p{1.5cm}>{\centering\arraybackslash}p{1.75cm}|>{\centering\arraybackslash}p{3.25cm}}
    \toprule
    model class & interventions & $\supp_{\II}(X)$ & assumptions & algorithm  \\ \toprule
    $\mathcal{F}$ linear & on $X$ or $A$\newline of which at least one is
                           confounding-removing & -- &
                                                       Ass.~\ref{ass:identify_f}
                                                                      & linear
                                                                        IV\newline
                                                                        (e.g., two-stage
                                                                        least
                                                                        squares,
                                                                        K-class or
                                                                        PULSE
                                                                        \cite{Theil1958,jakobsen2020distributional})\\\midrule
    $\FF, \GG$ linear & on $A$ & bounded  strength
                                                   & -- & anchor regression
                                                          \cite{rothenhausler2018anchor}, \newline see also \cite{Theil1958}\\\midrule
    $\mathcal{F}$ smooth & on $X$ or $A$\newline  of which at least one
                           is confounding-removing & support-reducing &
                                                              Ass.~\ref{ass:identify_f}
                                                                      &
                                                                        nonlinear
                                                                        IV\newline %
                                                                        (e.g.,
                                                                        NPREGIV
                                                                        \cite{NPREGIV-CRAN}, 
                                                                        	Deep IV \citep{hartford2017deep},
																			Sieve IV \citep{newey2003instrumental, chen2018optimal},
																			Kernel IV \citep{singh2019kernel})\\\midrule
    $\mathcal{F}$ smooth \newline and linearly extrapolates & on
                                                     $X$ or $A$\newline of which at least
                                                     one is confounding-removing & -- &
                                                                         Ass.~\ref{ass:identify_f}
                                                                      & \textbf{NILE}\newline
                                                                        (Section~\ref{sec:nile})\\\bottomrule
  \end{tabular}
  \caption{List of algorithms to learn the generalizing function from
    data,  the considered model class,
    types of interventions, support under interventions,
and additional
    model assumptions. Sufficient conditions for Assumption~\ref{ass:identify_f}
    are given, for example, in the IV literature by generalized rank
    conditions, see
    Appendix~\ref{sec:IVconditions}.
  }
  \label{tab:learnability}
\end{table*}

For well-behaved interventions on $X$ which contain at least one 
confounding-removing intervention, estimating minimax solutions
reduces to the well-studied problem of estimating causal relationships.
One class of algorithms for this task 
is given by linear instrumental variable
(IV) approaches. They assume that $\mathcal{F}$ is linear and require
identifiability of the causal function
(Assumption~\ref{ass:identify_f}) via a rank condition on the
observational distribution, see
Appendix~\ref{sec:IVconditions}. 
Their target of inference is to
estimate the causal function, which by
Proposition~\ref{prop:minimax_equal_causal} will coincide with the
minimax solution if the set $\II$ consists of well-behaved
interventions with at least one of them being confounding-removing. 
A basic estimator for
linear IV models is the two-stage least squares (TSLS) estimator, 
which minimizes the norm of the prediction residuals
projected onto the subspace spanned by the observed instruments (TSLS objective).
TSLS
estimators are consistent but do not come  with strong finite sample guarantees; e.g., they do
not have finite moments in a just-identified setup  
\cite[e.g.,][]{mariano2001simultaneous}.
K-class estimators
\cite{Theil1958} have been proposed to overcome some of these issues.
They minimize a linear combination of the residual sum of squares (OLS objective) and
the TSLS objective. 
K-class estimators can be seen as utilizing a bias-variance trade-off. 
For fixed and non-trivial relative weights, 
they have,
in a Gaussian setting,  finite moments up to a certain order that depends on the sample-size and the number of predictors used.
If the weights are such that 
the OLS objective is ignored asymptotically, 
they consistently estimate the causal parameter \citep[e.g.,][]{mariano2001simultaneous}.
More recently, PULSE has been proposed
\citep{jakobsen2020distributional}, a 
data-driven procedure for choosing the
relative weights 
such that the prediction residuals `just' pass a test
for simultaneous uncorrelatedness with the instruments.

In cases where the minimax solution does not coincide with the
causal function, only few algorithms exist. 
Anchor regression \citep{rothenhausler2018anchor} is a procedure that
can be used when $\mathcal{F}$ and $\mathcal{G}$ are linear and $h_1$
is additive in the noise component. It finds
the minimax solution if the set $\II$ consists of all 
interventions on $A$ up to a fixed intervention strength,
and is applicable even if
Assumption~\ref{ass:identify_f} is not necessarily satisfied.

In a linear setting, where the
regression coefficients differ between different environments, it is
also possible to minimize the worst-case risk among the observed
environments \citep{meinshausen2015maximin}.  In its current
formulation, this approach does not quite fit into the above
framework, as it does not allow for changing distributions of
the covariates.
A summary of the
mentioned methods and their assumptions
is given in Table~\ref{tab:learnability}.

If $\mathcal{F}$ is a nonlinear or non-parametric class of functions, the task of finding 
minimax solutions becomes more difficult. In cases where the causal function is among 
such solutions, this problem has been studied in the econometrics community. 
For example, \cite{newey2013nonparametric,newey2003instrumental} treat the 
identifiability and estimation of causal functions in non-parametric function classes.
Several non-parametric IV procedures exists, e.g., NPREGIV \cite{NPREGIV-CRAN}
	contains modified implementations of \cite{horowitz2011applied} and \cite{darolles2011nonparametric}, 
	which we will refer to as NPREGIV-1 and NPREGIV-2, respectively. 
	Other procedures include 
	Deep IV \citep{hartford2017deep},
	Sieve IV \citep{newey2003instrumental, chen2018optimal} and
	Kernel IV \citep{singh2019kernel}. 
Identifiability and estimation of the causal function using nonlinear IV methods in parametric function classes is discussed in Appendix~\ref{sec:IVconditions}.
Unlike in the linear case, most of the
methods do not aim to extrapolate 
and only recover the causal function inside the support of $X$,
that is, they cannot be used to predict interventions
outside of 
this domain.
In the following section, we propose a procedure that is able to
extrapolate when $\mathcal{F}$ consists of functions which
extend linearly outside of the support of $X$. 
In principle, any other extrapolation rule may be employed here, 
as long as all functions from $\FF$ are uniquely determined by their values
on the support of $X$, that is, Assumption~\ref{ass:gen_f} is satisfied.

In our simulations, we see that our method can improve the prediction performance on the boundary of the support and outperforms other methods when comparing the estimation on the support. 

\subsection{NILE} \label{sec:nile}
We have seen in Proposition~\ref{prop:impossibility_extrapolation}
that in order to generalize to interventions which extend the support
of $X$, we require additional assumptions on the function class $\FF$.
In this section, we start from such assumptions and verify both
  theoretically and practically that they allows us to perform
  distribution generalization in the considered setup.  Along the way,
  several choices can be made and usually several options are
  possible.  We will see that our choices yield a method with
  competitive performance, but we do not claim optimality of our
  procedure.  Several of our choices were partially 
  made to keep the theoretical exposition simple and the method
  computationally efficient.  
  We first consider the
  univariate case (i.e., $X$ and $A$ are real-valued) and comment
  later on the possibility to extend the methodology to higher
  dimensions.  
  Unless specific
background knowledge is given, it might be reasonable to assume that
the causal function extends linearly outside a fixed interval $[a,b]$. 
  By
additionally imposing differentiability on $\FF$, any function from
$\FF$ is uniquely defined by its values within $[a,b]$, see also
Section~\ref{sec:support_extending_onX}.  Given an estimate $f$ on $[a,b]$,
the linear extrapolation property then yields a global estimate
on the whole of $\mathbb{R}$.
In
principle, any class of differentiable functions can be used. 
Here, 
we assume that, on the interval $[a,b]$, the causal function $f$
is contained in the linear span of a B-spline basis. 
More formally, let $B = (B_1, ..., B_k)$ be a fixed B-spline basis on $[a,b]$, and define $\eta := (a,b,B)$. 
Our procedure assumes that the true causal function $f$ belongs to the function class 
$\FF_{\eta} := \{f_\eta(\cdot; \theta) \st \theta \in \R^k\}$, 
where for every $x \in \R$ and $\theta \in \R^k$, $f_\eta(x; \theta)$ is given as
\begin{equation} \label{eq:f_theta}
f_{\eta}(x;\theta) :=
\begin{cases}
B(a)^\top  \theta +  B^\prime (a)^\top  \theta (x - a) & \text{ if } x < a \\
B(x)^\top  \theta & \text{ if } x \in [a, b] \\
B(b)^\top  \theta + B^\prime (b)^\top \theta (x - b) & \text{ if } x >b,\\
\end{cases}
\end{equation}
where 
$B^\prime := (B_1^\prime, \dots, B_k^\prime)$
denotes
the component-wise derivative of $B$.
In our algorithm, $\eta = (a, b, B)$ is a hyper-parameter, which can be set manually, 
or be chosen from data.

\subsubsection{Estimation procedure} \label{sec:estimation}
We now introduce our estimation procedure for fixed choices of all
hyper-parameters.
Section~\ref{sec:algorithm} describes how these can
be chosen from data in practice.
Let $(\B{X}, \B{Y}, \B{A}) \in \R^{n \times 3}$ be 
$n$ i.i.d.\ realizations sampled from 
a distribution over
$(X,Y,A)$,
let $\eta = (a,b,B)$ be fixed and assume that $\supp(X)\subset [a,b]$. 
Our algorithm aims to learn the causal function $f_\eta(\cdot ; \theta^0) \in \FF_\eta$,
which is determined by
the linear causal parameter $\theta^0$ of 
a $k$-dimensional vector of covariates $(B_1(X), \dots, B_k(X))$. From standard linear IV theory, 
it is known that at least $k$ instrumental variables are required to identify the $k$ causal parameters, see Appendix~\ref{sec:IVconditions}. 
We therefore artificially generate such instruments by nonlinearly transforming $A$, by using another
B-spline basis $C = (C_1, \dots, C_k)$. 
The parameter $\theta^0$ 
can then be identified from the observational distribution under
appropriate rank conditions, see
Section~\ref{sec:consistency}. 
In that case, the hypothesis
  $H_0(\theta) : \theta= \theta^0$ is equivalent to the hypothesis
  $\tilde H_0(\theta) : \E[C(A)(Y - B(X)^\top \theta)] =
  0$.
 Let
$\B{B} \in \R^{n \times k}$ and $\B{C} \in \R^{n \times k}$ be the
associated design matrices, for each $i \in \{1, \dots, n\}$,
$j \in \{1, \dots, k\}$ given as $\B{B}_{ij} = B_j(X_i)$ and
$\B{C}_{ij} = C_{j}(A_i)$. 
A straightforward choice would be to construct the standard TSLS estimator, i.e.,  
$\hat \theta$ as the minimizer of
$\theta \mapsto \norm{\B{P} (\B{Y} - \B{B} \theta)}_2^2$, where
$\B{P}$ is the projection matrix onto the columns of $\B{C}$,
  see also \cite{hall2005generalized}.  Even though this procedure may result in an
asymptotically consistent estimator, there are several reasons why it
may 
be suboptimal
in a finite
sample setting.  First, the above estimator can have large finite
sample bias, in particular if $k$ is large. Indeed, in the extreme
case where $k = n$, and assuming that all columns in $\B{C}$ are
linearly independent, $\B{P}$ is equal to the identity matrix, and
$\hat \theta$ coincides with the OLS estimator.  Second, since
$\theta$ corresponds to the linear parameter of a spline basis, it
seems reasonable to impose constraints on $\theta$ which enforce
smoothness of the resulting spline function. Both of these points can
be addressed by introducing additional penalties into the estimation
procedure.  Let therefore
$\B{K} \in \R^{k \times k}$ 
and $\B{M} \in \R^{k \times k}$ be the matrices that are, for each $i,j \in \{1, \dots, k\}$, 
defined as $\B{K}_{ij} = \int B^{\prime \prime}_i(x) B^{\prime \prime}_j(x) dx$ and 
$\B{M}_{i j} = \int C^{\prime \prime}_{i}(a) C^{\prime \prime}_{j}(a) da$, 
and let $\gamma, \delta > 0$ be the respective penalties associated with $\B{K}$ and $\B{M}$. 
For $\lambda \geq 0$ and with $\mu := (\gamma, \delta, C)$, we then define the estimator 
\begin{equation} \label{eq:thetahat}
\hat \theta^n_{\lambda, \eta, \mu}
:= \argmin_{\theta \in \R^{k}} 
\norm{\B{Y} - \B{B} \theta }_2^2 + \lambda \norm{\B{P}_\delta(\B{Y} - \B{B} \theta)}_2^2 + \gamma \theta^\top \B{K} \theta, \\
\end{equation}
where $\B{P}_\delta := \B{C} (\B{C}^\top \B{C} + \delta \B{M})^{-1} \B{C}^\top$ is the `hat'-matrix 
for a 
penalized regression onto the columns of $\B{C}$. 
By choice of $\B{K}$, the term $\theta^\top \B{K} \theta$ is equal to the integrated squared curvature of the spline
function parametrized by $\theta$. 
The regularization induced by the second summand in~\eqref{eq:thetahat} 
is similar to the one from K-class estimators in linear settings \citep{Theil1958}.
The function class~\eqref{eq:f_theta}
enforces 
linear extrapolation.
In principle, the above approach extends to situations where $X$ and $A$ are higher-dimensional,
in which case
$B$ and $C$ consist of multivariate functions. 
For example, \cite{fahrmeir2013regression} propose the use of tensor product splines, 
and introduce multivariate smoothness penalties based on pairwise first- or second order
parameter differences of basis functions which are close-by with respect to some suitably chosen metric. 
Similarly to \eqref{eq:thetahat}, such penalties result in a convex optimization problem. 
However, due to the large number of involved variables, the optimization procedure 
becomes computationally burdensome already in small dimensions.

Within the function class $\FF_\eta$, the above defines the global
 estimate $f_\eta(x; \hat \theta^n_{\lambda, \eta, \mu})$, 
for every $x \in \R$, given by
\begin{equation} \label{eq:fhat_theta}
f_\eta(x; \hat \theta^n_{\lambda, \eta, \mu}) :=
\begin{cases}
 B(a)^\top \hat \theta^n_{\lambda, \eta, \mu} +  B^\prime (a)^\top \hat \theta^n_{\lambda, \eta, \mu} (x - a) & \text{ if } x < a \\
 B(x)^\top \hat \theta^n_{\lambda, \eta, \mu} & \text{ if } x \in [a,b] \\
 B(b)^\top \hat \theta^n_{\lambda, \eta, \mu} + B^\prime (b)^\top \hat \theta^n_{\lambda, \eta, \mu} (x - b) & \text{ if } x > b. \\
\end{cases}
\end{equation}
We deliberately distinguish between three different groups of
hyper-parameters $\eta$, $\mu$ and $\lambda$.  The parameter
$\eta = (a,b,B)$ defines the function class to which the causal
function $f$ is assumed to belong. To prove consistency of our
estimator, we require this function class to be correctly specified.
In turn, the parameters $\lambda$ and
  $\mu=(\gamma, \delta, C)$ are algorithmic parameters that do not
  describe the statistical model. Their values only affects the finite
  sample behavior of our algorithm, whereas consistency is ensured as
  long as $C$ satisfies certain rank conditions, see Assumption~\ref{ass:RankCondition}
  in Section~\ref{sec:consistency}.
  In practice,
$\gamma$ and $\delta$ are chosen via a cross-validation procedure, see
Section~\ref{sec:algorithm}. The parameter $\lambda$ determines the
relative contribution of the OLS and TSLS losses to the objective
function.
To choose $\lambda$ from data, we use an idea 
similar to the PULSE \citep{jakobsen2020distributional}.

\subsubsection{Algorithm} \label{sec:algorithm}
Let for now $\eta, \mu$ be fixed. 
In the limit $\lambda \to \infty$, our estimation procedure becomes
equivalent to minimizing the
 TSLS loss $\theta \mapsto \norm{\B{P}_\delta (\B{Y} - \B{B}\theta)}_2^2$, 
which may be interpreted as searching for the 
parameter $\theta$ which complies `best' with the hypothesis $\tilde{H}_0(\theta) : \E[C(A)(Y - B(X)^\top \theta)] = 0$. 
For finitely many data,
following the idea introduced in \citep{jakobsen2020distributional}, 
we propose to choose the value for $\lambda$ such that $\tilde{H}_0(\hat \theta^n_{\lambda, \eta, \mu})$ is 
just accepted (e.g., at a significance level $\alpha = 0.05$). 
That is, 
among all $\lambda \geq 0$ which result in an estimator that is not rejected as a 
candidate for the causal parameter, we chose the one which yields maximal contribution of the OLS 
loss to the objective function. 
More formally, let 
for every $\theta \in \R^k$, $T(\theta) = (T_n(\theta))_{n \in \N}$ be
a statistical test 
  at (asymptotic) level $\alpha$ for $\tilde{H}_0(\theta)$ with rejection 
threshold $q(\alpha)$. That is, $T_n(\theta)$ does not reject
$\tilde{H}_0(\theta)$ if and only if $T_n(\theta) \leq q(\alpha)$. The penalty
$\lambda^\star_n$ is then chosen in the following data-driven way
\begin{align*}
\lambda^\star_n := \inf \{\lambda \geq 0 : T_n(\hat \theta^n_{\lambda, \eta, \mu})\leq q(\alpha)\}.
\end{align*}
In general, $\lambda^\star_n$ is not guaranteed to be finite for an arbitrary test statistic $T_n$. 
Even for a reasonable test statistic it might happen 
that $T_n(\hat \theta^n_{\lambda, \eta, \mu} ) > q(\alpha)$ 
for all $\lambda \geq 0$; see \cite{jakobsen2020distributional} for further details. 
We can remedy the problem by reverting 
to another well-defined and consistent estimator, such as
the TSLS (which minimizes the TSLS loss above) 
if $\lambda^\star_n$ 
is not finite.  
Furthermore, if
$\lambda \mapsto T_n(\hat \theta^n_{\lambda, \eta, \mu})$ is
monotonic, $\lambda^\star_n$ can be computed efficiently by a binary
search procedure.  In our algorithm, the test statistic $T$
and rejection threshold $q$ can be supplied by the user. 
Conditions on $T$ that are sufficient to yield a consistent estimator $f_\eta(\cdot , \hat \theta_{\lambda_n^\star, \mu, \eta})$, 
given that $\mathcal{F}_\eta$ is correctly specified,
are presented in Section~\ref{sec:consistency}.
Two choices of test statistics which are implemented in our code package
can be found in Appendix~\ref{sec:test_statistic}.

For every $\gamma \geq 0$, let $\B{Q}_\gamma = \B{B} (\B{B}^\top \B{B} + \gamma \B{K})^{-1} \B{B}^\top$ be the `hat'-matrix 
for the penalized regression onto $\B{B}$. 
Our method then proceeds as shown in Algorithm~\ref{alg:nile}.
\begin{algorithm} \caption{NILE (``Nonlinear Intervention-robust Linear Extrapolator'')} \label{alg:nile}
\textbf{input}: data $(\B{X}, \B{Y}, \B{A}) \in \R^{n \times 3}$\;
\textbf{options}: $k$,  $T$, $q$, $\alpha$\; %
\Begin{
	 $a \leftarrow \min_i X_i$, $b \leftarrow \max_i X_i$\; 
	 construct cubic B-spline bases $B = (B_1, \dots, B_{k})$ and $C = (C_1, \dots, C_k)$ at equidistant knots, 
	 with boundary knots at respective extreme values of $\B{X}$ and $\B{A}$\;
	 define $\hat \eta \leftarrow (a,b,B)$\;
	 choose $\delta^n_{\text{CV}}>0$ by 10-fold CV to minimize the out-of-sample mean squared error of $\hat{\B{Y}} = \B{P}_{\delta} \B{Y}$\;
	 choose $\gamma^n_{\text{CV}}>0$ by 10-fold CV to minimize the out-of-sample mean squared error of $\hat{\B{Y}} = \B{Q}_{\gamma} \B{Y}$\;
	 define $\mu^n_{\text{CV}} \leftarrow (\delta^n_{\text{CV}}, \gamma^n_{\text{CV}}, C)$\;
	 approx. $\lambda^\star_n = \inf\{\lambda \geq 0 : T_n(\hat \theta^n_{\lambda, \mu^n_{\text{CV}}, \hat \eta}) \leq q(\alpha)\}$ by binary search\;
	 update $\gamma^n_{\text{CV}} \leftarrow (1 + \lambda^\star_n) \cdot \gamma^n_{\text{CV}}$\;
	 compute $\hat \theta^n_{\lambda_n^\star, \mu^n_{\text{CV}}, \hat \eta}$ using \eqref{eq:thetahat}\;
}
\textbf{output}: $\hat{f}^n_{\text{NILE}}  := f_{\hat \eta}(\, \cdot \, ; \hat \theta^n_{\lambda_n^\star, \mu^n_{\text{CV}}, \hat \eta})$ defined by \eqref{eq:fhat_theta}\;
\end{algorithm}
The penalty parameter $\gamma^n_{\text{CV}}$ is chosen to minimize the
out-of-sample mean squared error
  of the prediction model
$\hat{\B{Y}} = \B{Q}_{\gamma} \B{Y}$, which corresponds to the
solution of \eqref{eq:thetahat} for $\lambda = 0$.  
After choosing $\lambda_n^\star$, the objective function in \eqref{eq:thetahat} increases 
by the term $\lambda_n^\star \norm{\B{P}_{\delta_{\text{CV}}^n}(\B{Y} - \B{B} \theta)}_2^2$.
In order for the penalty term $\gamma \theta^\top \B{K} \theta$ to impose the same degree of
smoothness in the altered optimization problem, the penalty parameter $\gamma$ needs to be 
adjusted accordingly. 
The heuristic update in our algorithm is motivated by the simple observation that for all $\delta, \lambda \geq 0$,
$\norm{\B{Y} - \B{B} \theta}_2^2 + \lambda \norm{\B{P}_\delta (\B{Y} - \B{B} \theta)}_2^2 \leq (1 + \lambda) \norm{\B{Y} - \B{B} \theta}_2^2$.

\subsubsection{Asymptotic generalization (consistency)} \label{sec:consistency}
We now prove consistency of our estimator in the case where
the hyper-parameters $(\eta, \mu)$ are fixed (rather than data-driven), and the
  function class $\FF_{\eta}$ is correctly specified.
  Fix any $a<b$
and a basis $B=(B_1, \dots, B_k)$. Let $\eta_0 = (a,b,B)$ and let the
model class be given by
$\MM = \FF_{\eta_0} \times \GG \times \HH_1 \times \HH_2 \times
\mathcal{Q}$, where $\FF_{\eta_0}$ is as described in
Section~\ref{sec:nile}. Assume that the data-generating model
$M = ( f_{\eta_0}(\, \cdot \,; \theta^0),g,h_1,h_2,Q) \in \MM$ 
induces
an observational distribution $\P_M$ such that
$\supp^M(X) \subseteq (a,b)$.
Let further $\II$ be a set of 
interventions on $X$ or $A$,
and let $\alpha \in (0,1)$ be a fixed significance level.

We prove asymptotic generalization (consistency) for an idealized
version of the NILE estimator which utilizes $\eta_0$, rather than the
data-driven values. Choose any $\delta,\gamma \geq 0$ 
and basis
$C=(C_1,...,C_k)$ and let $\mu=(\delta,\gamma,C)$. 
We will make use of
the following assumptions.
\begin{enumerate}[label=(B\arabic*),ref=(B\arabic*)]
		\item 
		For all $\tilde{M}\in \cM$ with $\bP_M = \bP_{\tilde M}$ it holds that
		$\sup_{i\in\cI}\E_{\tilde{M}(i)} [X^2] < \infty$ and \newline
	   $\sup_{i\in\cI} \lambda_{\max}(\E_{\tilde{M}(i)} [B(X)B(X)^\top ])<\infty.$ \label{ass:MaximumEigenValueBounded} 
		\item The matrices $\E_M[ B(X)B(X)^\top ]$, $\E_M[ C(A)C(A)^\top ]$ and
		 $\E_M[ C(A)B(X)^\top  ]$ have full rank.\label{ass:RankCondition}
	\end{enumerate}
\begin{enumerate}[label=(C\arabic*),ref=(C\arabic*)]
		\item  $T(\theta)$ has 
		uniform asymptotic power on any compact set of alternatives. \label{ass:ConsistentTestStatistic}
		\item  $\lambda^\star_n := \inf\{\lambda\geq 0 : T_n(\hat \theta^n _{\lambda,\eta_{0},\mu}) \leq q(\alpha)\}$ is almost surely finite. \label{ass:LambdaStarAlmostSurelyFinite}
		\item 
$\lambda \mapsto T_n(\hat \theta^n _{\lambda,\eta_0,\mu})$ is weakly decreasing and $\theta \mapsto T_n(\theta)$ is continuous.
		 \label{ass:MonotonicityAndContinuityOfTest}
	\end{enumerate}
Assumptions \ref{ass:MaximumEigenValueBounded}--\ref{ass:RankCondition} ensure consistency of the estimator as long as $\lambda^\star_n$ tends to infinity. 
Intuitively, in this case, we can apply arguments similar to those
that prove consistency of the TSLS estimator. 
Assumptions \ref{ass:ConsistentTestStatistic}--\ref{ass:MonotonicityAndContinuityOfTest} ensure
that consistency is achieved when choosing $\lambda^\star_n$ in the data-driven fashion described in Section~\ref{sec:algorithm}. 
In Assumption \ref{ass:MaximumEigenValueBounded}, $\lambda_{\max}$ denotes the largest eigenvalue. 
In words, the assumption states that, 
under each model $\tilde M\in \cM$ with $\bP_M = \bP_{\tilde M}$, 
there exists a finite upper bound on the variance of any linear combination of the basis functions $B(X)$, 
uniformly over all distributions induced by $\II$. 
The first two rank conditions of \ref{ass:RankCondition} 
enable certain limiting arguments to be valid and they guarantee that estimators are asymptotically well-defined. 
The last rank condition of \ref{ass:RankCondition} is the so-called rank condition for identification. 
It guarantees that $\theta^0$ is identified from the observational distribution in the sense 
that the hypothesis $H_0(\theta):\theta=\theta^0$ becomes equivalent with $\tilde{H}_0(\theta) : \E_M[C(A)(Y-B(X)^\top \theta)]=0$. 
	\ref{ass:ConsistentTestStatistic} 
	means that for any compact set $K\subseteq \R^k$ with $\theta^0 \not \in K$ it holds that $\lim_{n\to \infty} P(\inf_{\theta\in K} T_n(\theta) \leq q(\alpha)) =0$.
If the considered test 
has, in addition, a level guarantee, 
such as pointwise asymptotic level,
the 
interpretation of the finite sample estimator 
discussed in Section~\ref{sec:algorithm}
remains valid 
(such level guarantee may potentially yield improved
finite sample performance, too).
	\ref{ass:LambdaStarAlmostSurelyFinite} 
is made to simplify the consistency proof. 
As previously discussed in Section~\ref{sec:algorithm}, if \ref{ass:LambdaStarAlmostSurelyFinite} is not satisfied, 
we can output
another well-defined and consistent estimator on the event $(\lambda^\star_n=\infty)$, ensuring that consistency still holds. 

Under these conditions, we have the following asymptotic generalization guarantee.

\begin{prop}[Asymptotic generalization]\label{thm:consis}
  Let $\cI$ be a set of 
  interventions on $X$ or $A$ of which at
  least one is confounding-removing.  If
  assumptions~\ref{ass:MaximumEigenValueBounded}--\ref{ass:RankCondition}
  and
  \ref{ass:ConsistentTestStatistic}--\ref{ass:MonotonicityAndContinuityOfTest}
  hold true, then, for any $\tilde{M} \in \MM$ with
  $\P_{\tilde{M}} = \P_M$, and any $\epsilon > 0$, it holds that
  \begin{align*}
  \P_{M} &\left( \big\vert \sup_{i\in\II}\E_{\tilde{M}(i)}\big[(Y-f_{\eta_0}(X;\hat{\theta}^n_{\lambda^\star_n,\eta_0,\mu}))^2\big] \right.
   - \left. \inf_{\fs\in\FF_{\eta_0}}\,
      \sup_{i\in\II}\E_{\tilde{M}(i)}\big[(Y-\fs(X))^2\big] \big\vert \leq \epsilon \right) \to 1,
  \end{align*}
  as $n\to \infty$. In the above event, only
  $\hat{\theta}^n_{\lambda^\star_n,\eta_0,\mu}$ is stochastic.
\end{prop}

\subsubsection{Experiments} \label{sec:experiments}
\begin{figure}
\centering
\includegraphics[width=\linewidth]{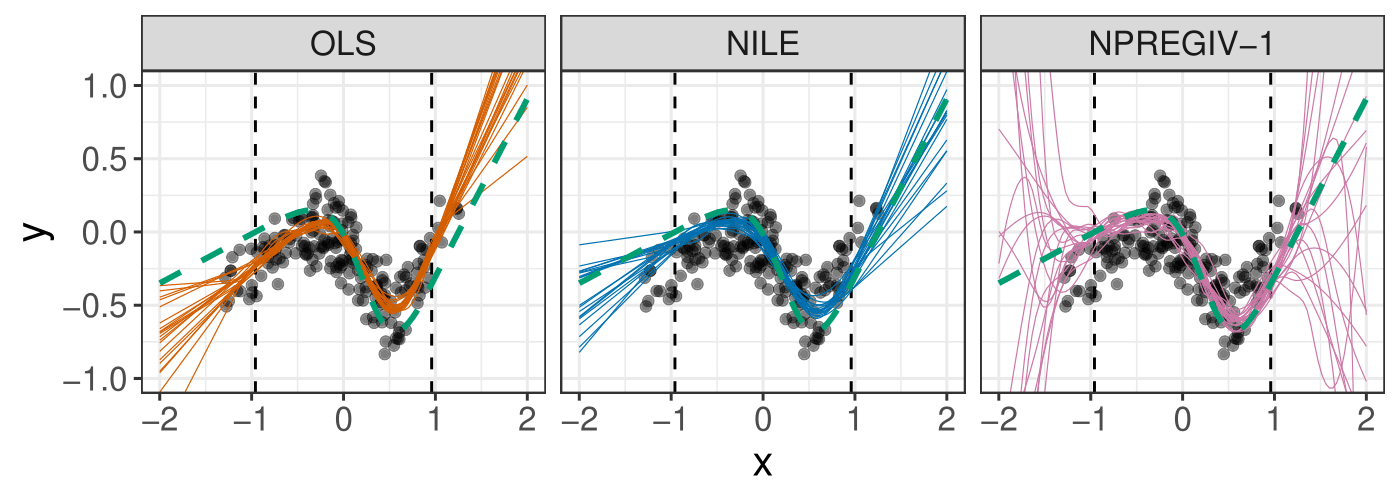}
\caption{A sample dataset from the model \eqref{eq:sim_model} with 
$\alpha_A = \sqrt{1/3}$, $\alpha_H = \sqrt{2/3}$, $\alpha_\epsilon = 0$. The true causal function
is indicated by a green dashed line. For each method, we show 20 estimates of 
this function, each based on an independent sample from \eqref{eq:sim_model}. 
For values within the support of the training data (vertical dashed lines 
mark the inner 90\% quantile range), 
NPREGIV-1 correctly estimates the causal function well. 
As expected, when moving
outside the support of $X$, 
the estimates become unreliable, 
and we gain an increasing advantage by exploiting the linear extrapolation assumed by
the NILE.}
\label{fig:overlay_estimates_and_varying_confounding}
\end{figure}

We now investigate the empirical performance of our proposed
estimator, the NILE, with $k = 50$ spline basis functions.
To choose $\lambda_n^\star$, we use the test statistic
$T_n^2$, 
which tests the slightly stronger hypothesis $\bar{H}_0$, 
see Appendix~\ref{sec:test_statistic}. In all experiments
use the significance level $\alpha = 0.05$. 
We include two other approaches as baseline: (i) the method NPREGIV-1 
(using its default options)
introduced in Section~\ref{sec:existingmethods}, and (ii) a linearly
extrapolating estimator of the ordinary regression of $Y$ on $X$ (which
corresponds to the NILE with $\lambda^\star \equiv 0$).  In all
experiments, we generate data sets of size $n=200$ as independent
replications from %
\begin{align} \label{eq:sim_model}
\begin{split}
A:= \epsilon_A, \quad H:= \epsilon_H, 
\quad X :=  \alpha_A A + \alpha_H H + \alpha_\epsilon \epsilon_X, \quad
Y := f (X) + 0.3  H + 0.2  \epsilon_Y,
\end{split}
\end{align}
where $(\epsilon_A, \epsilon_H, \epsilon_X, \epsilon_Y)$ are jointly 
independent with $\text{Uniform}(-1,1)$ marginals. To make results 
comparable across different parameter settings, we impose the constraint
$\alpha_A^2 + \alpha_H^2  +\alpha_\epsilon^2 = 1$, which ensures that
in all models, $X$ has variance $1/3$. The function $f$ is drawn from 
the linear span of a basis of four natural cubic splines with knots placed equidistantly 
within the $90\%$ inner quantile range of $X$. 
By well-known properties of 
natural splines, any such function extends linearly 
outside the boundary knots. 
Figure~\ref{fig:overlay_estimates_and_varying_confounding} (left) shows an example
data set from \eqref{eq:sim_model}, 
where the causal function is indicated in green. 
We additionally display estimates obtained by each of the considered methods, based
on 20 i.i.d.\ datasets. Due to the confounding variable $H$, the OLS estimator is 
clearly biased. NPREGIV-1 exploits $A$ as an instrumental variable and obtains good results 
within the support of the observed data. Due to its non-parametric nature, however, 
it cannot extrapolate outside this domain. The NILE estimator exploits the linear 
extrapolation assumption on $f$ to produce global estimates. 

\begin{figure}
\centering
\includegraphics[width=\linewidth]{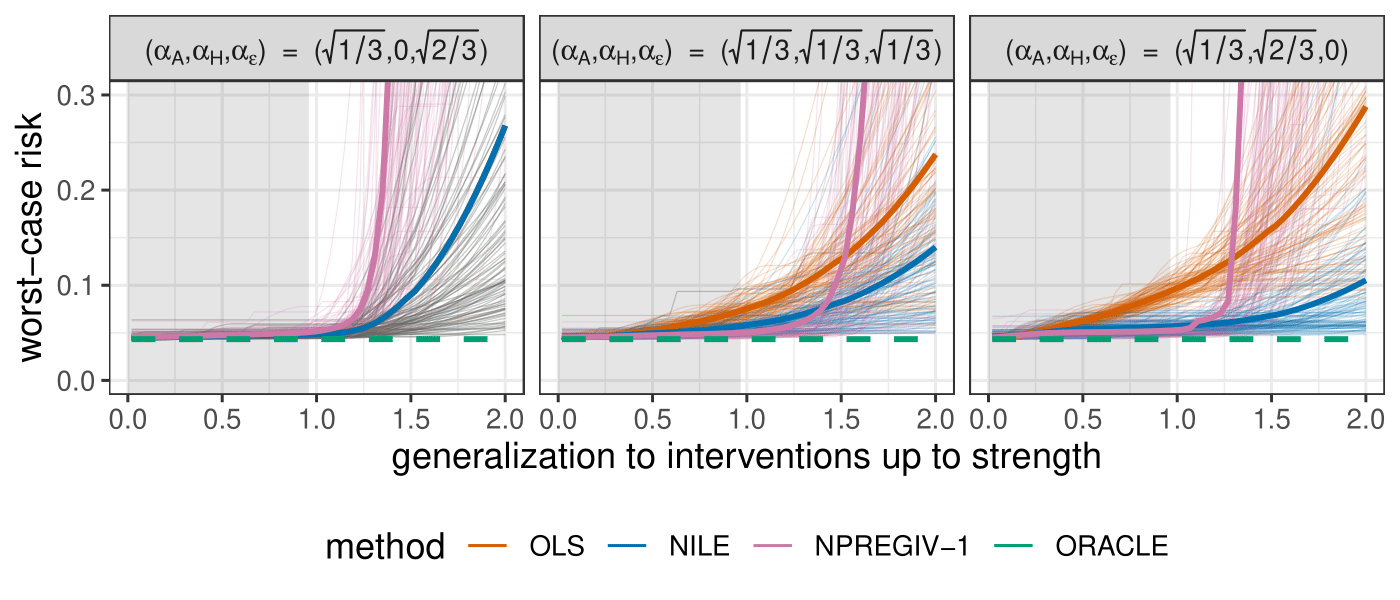}
\caption{
Predictive performance under confounding-removing interventions on $X$
for different confounding- and intervention strengths 
(see alpha values in the grey panel on top).
The right panel corresponds to the same parameter setting as in Figure~\ref{fig:overlay_estimates_and_varying_confounding}. 
The plots in each panel are based on data sets of size $n=200$, 
generated from $N = 100$ different models of the form \eqref{eq:sim_model}. 
For each model, we draw a different function $f$, 
resulting in a different minimax solution (see Appendix~\ref{sec:additional_experiments} for details on the sampling procedure). 
The performances under individual models are shown by thin lines; the average performance \eqref{eq:experiments_risk} 
across all models is indicated by thick lines.
In all considered models, the optimal prediction error (green dashed
line) is equal to $\E[\xi_Y^2]$ 
  (by consistency, for any fixed
  function $f$, NILE's worst-case risk   
  converges pointwise to this value for increasing sample size).
The grey area indicates the inner 90 \% quantile range 
of $X$ in the training distribution; the white area can be seen as an area of generalization.
}
\label{fig:overlay_estimates_and_varying_confounding2}
\end{figure}

We further investigate the empirical worst-case risk %
across several different models of the form \eqref{eq:sim_model}. That is, for a fixed set of parameters
$(\alpha_A, \alpha_H, \alpha_\ep)$, we construct several models $M_1, \dots, M_N$ of the form 
\eqref{eq:sim_model} by randomly sampling causal functions $f_1, \dots, f_N$ 
(see Appendix~\ref{sec:additional_experiments} for further details on the sampling procedure). 
For every $x \in [0,2]$, let $\II_x$ denote the set of hard interventions
which set $X$ to some fixed value in $[-x,x]$. 
We then characterize the performance of each method using the average (across different models) 
worst-case risk (across the interventions in $\II_x$), i.e., for each estimator $\hat{f}$, we consider 
\begin{align} \label{eq:experiments_risk}
\begin{split}
\frac{1}{N} \sum_{j=1}^N \sup_{i \in \II_x} \E_{M_j(i)} \big[ (Y - \hat{f}(X))^2 \big]
= \E[\xi_Y^2] + \frac{1}{N} \sum_{j=1}^N \sup_{\tilde{x} \in [-x,x]}  (f_j(\tilde{x}) - \hat{f}(\tilde{x}))^2,
\end{split}
\end{align}
where $\xi_Y :=  0.3  H + 0.2  \epsilon_Y$ is the noise term for $Y$ (which is fixed across all experiments).
In practice, we evaluate the functions $\hat{f}$, $f_1, \dots, f_N$ on a fine grid on $[-x,x]$ to approximate
the above supremum. 
Figure~\ref{fig:overlay_estimates_and_varying_confounding2}
plots the average worst-case risk versus intervention strength 
for varying degree of confounding ($\alpha_H$).
The optimal worst-case risk $\E[\xi_Y^2]$ is indicated by a green dashed line. 
The results show that the linear extrapolation
property of the NILE estimator is beneficial in particular for strong interventions. 
In the case of no confounding ($\alpha_H = 0$), the minimax solution coincides with the 
regression of $Y$ on $X$, hence even the OLS estimator yields good predictive performance. 
In this case, the hypothesis $\bar{H}_0(\hat \theta^n_{\lambda, \delta^n_{\text{CV}}, \gamma^n_{\text{CV}}})$ 
is accepted already for small values of $\lambda$
(in this experiment, the empirical average of $\lambda^\star_n$ equals 0.015),
and the 
NILE estimator becomes indistinguishable from the OLS. As the confounding strength increases, the OLS 
becomes increasingly biased, and the NILE objective function 
differs more notably
 from the OLS 
(average $\lambda^\star_n$ of 2.412 and 5.136, respectively). The method NPREGIV-1 slightly 
outperforms the NILE inside the support of the observed data, but drops in performance 
for stronger interventions. 
We believe that the increase in extrapolation performance of the NILE for stronger 
confounding (increasing $\alpha_H$) 
might stem from
the fact that, as the $\lambda_n^\star$ increases, also the smoothness
penalty $\gamma$ increases, see Algorithm~\ref{alg:nile}. While this results in slightly worse 
in-sample prediction, it seems beneficial for extrapolation (at least for the particular function class 
that we consider). We do not claim that our algorithm has theoretical guarantees which 
explain this increase in performance.

\begin{figure}
\centering
\includegraphics[width=\linewidth]{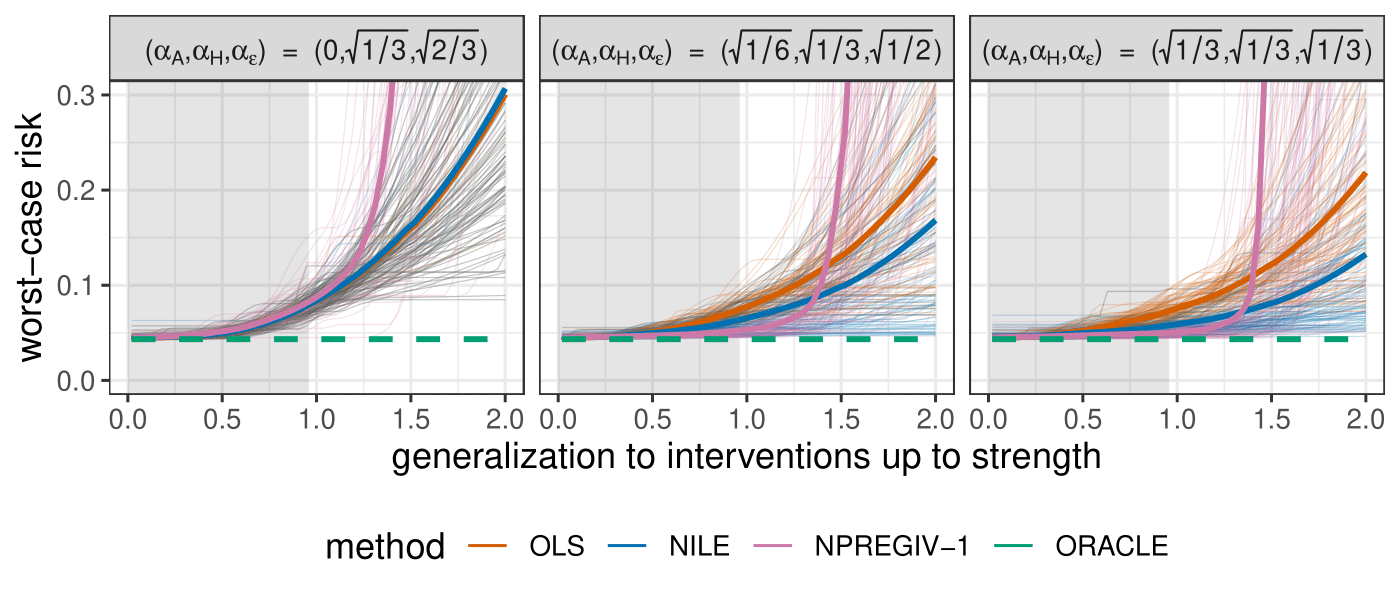}
\caption{Predictive performance for varying instrument strength. 
  If the
  instruments have no influence on $X$ ($\alpha_A = 0$), the second
  term in the objective function \eqref{eq:thetahat} is effectively
  constant in $\theta$, and the NILE therefore coincides with the OLS
  estimator (which uses $\lambda=0$).  This guards the NILE against
  the large variance which most IV estimators suffer from in a weak
  instrument setting.  For increasing influence of $A$, it clearly
  outperforms both alternative methods for large intervention
  strengths.  %
  }
\label{fig:weak_instruments}
\end{figure}

Figure~\ref{fig:weak_instruments} shows the worst-case risk for varying instrument strength ($\alpha_A$).
In the case where all exogenous noise comes from the unobserved variable $\epsilon_X$ (i.e., $\alpha_A$ = 0),
the NILE coincides with the OLS estimator. In such settings, standard IV methods are known to perform poorly, 
although also the NPREGIV-1 method seems 
robust to such scenarios. As the instrument strength
increases, the NILE %
 clearly outperforms OLS and NPREGIV-1 for interventions on $X$ which 
include values outside the training data.

\begin{figure}
\centering
\includegraphics[width=.7\columnwidth]{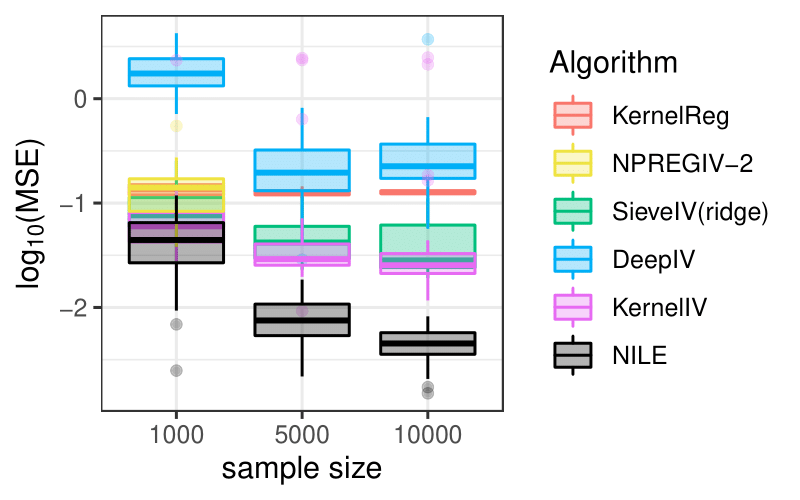}
\caption{
Comparison between the NILE and several alternative procedures for learning 
    a nonlinear causal function, 
    based on the same
    experimental setup as in \cite{singh2019kernel}. 
    The estimated functions are evaluated
    on the support (no generalization).
    NILE outperforms
    the competing methods.
}
\label{fig:kernelIV}
\end{figure}

We further 
compare NILE's ability to estimate the causal function 
on the support of the covariate $X$
in a nonlinear IV setting 
and compare it with the results from
other state-of-the-art procedures for nonlinear IV estimation, 
following the experimental setup by \cite{singh2019kernel}. Here, the authors consider a predictor variable $X \sim \text{Uniform}(0,1)$ which causally 
influences the target variable $Y$ via the structural assignment $Y := f(X) + \xi_Y$, where $f$ is the 
nonlinear causal function $f(x) = \log( \vert 16 x- 8 \vert +1) \cdot \text{sgn}(x-1/2)$, and $\xi_Y$
is an additive error term which is correlated with $X$. 
They compare their proposed procedure Kernel IV %
to the methods NPREGIV-2 (\cite{singh2019kernel} refer to this method as `Smooth IV'), 
Sieve IV and Deep IV (see Section~\ref{sec:existingmethods}). 
As a baseline, they also include a method for standard kernel ridge regression (`Kernel Reg') \citep{saunders1998ridge}, 
which ignores the existence of hidden confounders. 
Each procedure yields a different estimator $\hat{f}$.
Based on $40$ independent simulations, 
the 
estimators are then compared in terms of 
the average squared distance between $f$ and $\hat{f}$ across 1000 equidistant points in the interval $[0,1]$.
We refer to \cite[][Appendix~A.11]{singh2019kernel} %
for a precise description of the experimental setup. 
Figure~\ref{fig:kernelIV} shows the results of the above experiment 
(corresponding to Figure~2 in \citep{singh2019kernel}), where we have also included the NILE.
Our method outperforms all other procedures, in particular for large
sample sizes. 
There is slight
	difference in the way the different algorithms use the available data.
	In order to reduce finite sample bias, \cite{singh2019kernel} use sample splitting, 
	where the first and second step of the two-stage-least-squares
	procedure are performed on disjoint data sets. The NILE, in contrast, uses
	all of the data at once. However, even when running our procedure on only half 
	of the data, we still outperform the other procedures by a distinct margin, see Figure~\ref{fig:nile50_vs_kernelIV}.
We believe that the superior MSE performance of NILE could be due to 
the different approaches of regularization. For example, 
	NILE uses causal regularization similar to that of PULSE, i.e., a data-driven K-class regularization;
	in linear IV settings,
this type of regularization 
often yields a 
smaller MSE than 
standard IV methods such as TSLS
 \cite{jakobsen2020distributional}.

\section{Discussion and future work}
  In many real world
problems, the test distribution may differ from the training
distribution.  This requires statistical methods that come with a
provable guarantee in such a setting.  It is possible to characterize
robustness by considering predictive performance for distributions that
are close to the training distribution in terms of standard
divergences or metrics, such as KL divergences or Wasserstein
distance.  
As an alternative view point, we have introduced a novel
framework that formalizes the task of distribution generalization
when
considering distributions that are induced by a set of interventions.
Based on the concept of modularity,
interventions modify parts of the joint distribution and leave other
parts invariant.  
Thereby, they impose constraints on the changes of the distributions that are qualitatively different from considering balls in the above metrics. 
As such, we see them as a useful
  language to describe realistic changes between training and test
  distributions.

  Our framework is general in that it allows us to
model a wide range of causal models and interventions, which do not
need to be known beforehand. 
We have proved several generalization
guarantees, some of which show robustness for distributions that are
not close to the training distribution by considering almost any of
the standard metrics.  
Here, generalization can be obtained by causal functions, but also by non-causal functions; in general, however, the minimizer changes when the 
intervention class is altered (or misspecified).
  We have further
proved impossibility results that indicate the limits of what is
possible to learn from the training distribution.  In particular, in
nonlinear models, strong assumptions are required for
distribution generalization to a different support of the
covariates.
As such,
methods such as
anchor regression cannot be expected
to work in nonlinear models,
unless strong restrictions are placed on the function class $\GG$.

Our work can be extended into several directions.  
It
may, for example, be worthwhile to investigate the sharpness of the
bounds we provide in Section~\ref{sec:support_extending_onX} and other
extrapolation assumptions on $\mathcal{F}$. 
Our results make use of the form of the squared loss and it remains an open question to which extent they hold for general convex loss functions.
While our results can be applied to situations where causal background
knowledge is available, via a transformation of SCMs, our analysis is
deliberately agnostic about such information.  It would be interesting
to see whether stronger theoretical results can be obtained by
including causal background information.  We showed that the
  type of the interventions play a crucial role in determining whether
  the causal function is a minimax optimal solution. Building on this,
  it would be interesting to find empirical procedures which test
  whether an intervention is confounding-removing,
  confounding-preserving or neither.
Finally, 
it could be worthwhile to investigate whether NILE,
which outperforms existing approaches with 
respect to extrapolation,
can be combined with non-parametric 
methods to further improve in-sample performance.
While our current framework already contains certain settings of multi-task learning and domain generalization, it could be instructive to additionally include the possibility to model unlabeled data in the test task.
Finally, our results concern the infinite sample case, but we believe that they can 
form the basis for a corresponding analysis involving rates
or even finite sample results.

We view our work as a step towards understanding the problem of
distribution generalization.
We hope that considering the concepts of
interventions may help to shed further light 
into the question 
of generalizing 
knowledge that was
acquired during training to a different test distribution.

\section*{Acknowledgments}
We thank Thomas Kneib for helpful discussions and two anonymous reviewers for valuable comments. 
RC and JP were supported by 
a research grant (18968) from VILLUM FONDEN; MEJ and JP were supported by the Carlsberg Foundation. 

\appendix


\section{Transforming causal models}
\label{sec:causal_relations_X}

As illustrated in Remark~\ref{rem:model}, our framework can
  also be applied 
  in situations where training and test distributions are generated from 
  an SCM with a different structure than~\eqref{eq:SCMmodelreduced}.
  Below, we show that
  a general class of SCMs can be transformed into our reduced
  setting. 
  To this end, assume the true underlying causal
  structure is given by the SCM
\begin{align}\label{eq:SCM_full_causal}
  \begin{split}
    A\coloneqq \ep_A\qquad\qquad
    &X\coloneqq w(X, Y) + g(A) + h_2(H, \ep_X)\\
    H\coloneqq \ep_H\qquad\qquad
    &Y\coloneqq f(X) + h_1(H, \ep_Y),
  \end{split}
\end{align}
where, as before, $f,g,w,h_1$ and $h_2$ are measurable functions. 
First, we show how to transform the above SCM into the reduced form \eqref{eq:SCMmodelreduced}
without changing the induced observational distribution. In Appendix~\ref{sec:int_type}, we then 
discuss how to transform interventions in \eqref{eq:SCM_full_causal} to interventions in the reduced 
model.

Throughout this appendix, we assume that~\eqref{eq:SCM_full_causal}
 is uniquely solvable in the sense that
  there exists a unique function $F$ such that
  $(A, H, X, Y)=F(\ep_A,\ep_H,\ep_X,\ep_Y)$ almost surely, see \cite{Bongers2016b}
  for more details.  
  Denote by $F_X$ the coordinates of $F$
  that correspond to the $X$ variable (i.e., the coordinates from
  $r+q+1$ to $r+q+d$). 
  We further assume that
  there exist functions $\tilde{g}$ and $\tilde{h}_2$ such that
  \begin{equation}
    \label{eq:linear_decomposition_inverse}
    F_X(\ep_A,\ep_H,\ep_X,\ep_Y)=\tilde{g}(\ep_A) + \tilde{h}_2((\ep_H, \ep_Y), \ep_X).
  \end{equation}
This decomposition 
  is not always possible, but 
it exists in the following settings, for example:
(i) \emph{There are no $A$ variables.} 
In these
    cases, the additive decomposition
    \eqref{eq:linear_decomposition_inverse} becomes trivial.
  (ii) \emph{There are further constraints on the original SCM.}
The additive
    decomposition \eqref{eq:linear_decomposition_inverse} holds if,
    for example, 
    $w$ is a linear function or $A$ only enters the
    structural assignments of covariates $X$ which have at most $Y$ as
    a descendant.

Using the decomposition in \eqref{eq:linear_decomposition_inverse}, we can define the following reduced SCM
\begin{align}\label{eq:SCM_full_causal_simple}
\begin{split}
  A\coloneqq \ep_A\qquad\qquad
     &X\coloneqq \tilde{g}(A) + \tilde{h}_2(\tilde{H}, \ep_X)\\
  \tilde{H}\coloneqq \ep_{\tilde{H}}\qquad\qquad
     &Y\coloneqq f(X) + h_1(\tilde{H}),
\end{split}
\end{align}
where $\ep_{\tilde{H}}$ has the same distribution as $(\ep_H, \ep_Y)$
in \eqref{eq:SCM_full_causal}. This model fits the framework from 
Section~\ref{sec:modeling_int_ind_distr}, where the noise term in $Y$ is now taken to
be constantly zero. 
Both SCMs~\eqref{eq:SCM_full_causal}
and~\eqref{eq:SCM_full_causal_simple} induce the same observational
distribution and 
the same function $f$ appears in the assignments of $Y$.

If 
one intends to use 
interventions 
in the original SCM
  (i.e., \eqref{eq:SCM_full_causal}) to
model  
  the test distributions, one needs  
  to also transform these
  interventions. 
  We discuss how this can be done in the following
  subsection. 

\subsection{Transforming interventions}\label{sec:int_type}
For SCMs of the form \eqref{eq:SCM_full_causal} (and which satisfy \eqref{eq:linear_decomposition_inverse}), 
any distribution arising from an intervention on a 
subset of covariates from $X$ can be equivalently expressed using an intervention on all of $X$
in the corresponding reduced model 
\eqref{eq:SCM_full_causal_simple}. To see this, let $i$ be such an intervention 
in the original SCM, 
and let $\P^i$ be the induced interventional 
distribution over $(X,Y,A)$. We can then generate the same intervention distribution
in 
\eqref{eq:SCM_full_causal_simple} 
using the intervention $X:= \epsilon_X^i$, where the distribution of
$\epsilon_X^i$ coincides with the marginal of $X$ in $\P^i$.  Note,
however, that this type of transformation may fail for some
  model classes,
  for example,
 this may happen  
if the original SCM contains a hidden variable which is a descendant of
some (intervened) $X$ variables and a cause of $Y$.  
Also, even in situations where the above
transform is possible,
  the interventions can change
their intervention targets, become non-well-behaved or change their
support. In order to apply the developed methodology, one needs to
check whether the transformed interventions are a well-behaved (this
is not necessarily the case, even if the original intervention was
well-behaved) and how the support of all $X$ variables behaves under
that specific intervention.

\textbf{Intervention type}\quad First, we consider which types of
interventions in~\eqref{eq:SCM_full_causal} translate to well-behaved
interventions in~\eqref{eq:SCM_full_causal_simple}. A simple example
is given by interventions on $A$ in the original SCM, which result in the
same interventions on $A$ also in the reduced SCM. Similarly,
performing hard interventions on all components of $X$ in the original SCM
leads to the same intervention in the reduced SCM, which is in
particular both confounding-removing and confounding-preserving. For
interventions on subsets of the $X$, this is not always the case. To
see that, consider the following example 
\begin{center}
  \begin{minipage}{0.40\columnwidth}
    \centering
    {
      \begin{align*}
        \begin{split}
          A&\coloneqq \ep_A\\
          X_1 &\coloneqq \ep_1\\
          X_2 &\coloneqq Y + \epsilon_2\\
          Y &\coloneqq X_1 + \epsilon_Y
        \end{split}
      \end{align*}}
\end{minipage}%
\begin{minipage}{0.18\columnwidth}
  {\Large
    \begin{equation*}
      \xrightarrow{\text{transform}} \, \,
    \end{equation*}}
  \qquad
  $ $
\end{minipage}%
\begin{minipage}{0.40\columnwidth}
  \centering
  {
      \begin{align*}
        \begin{split}
          A &\coloneqq \ep_A\\
          H &\coloneqq \ep_Y \\
          X &\coloneqq (\ep_1,  H + \ep_1 + \ep_2)\\
          Y &\coloneqq X_1 + H,
        \end{split}
      \end{align*}}
  \end{minipage}%
\end{center}
with
$\ep_A, \epsilon_1, \epsilon_2, \ep_Y$
i.i.d.\ noise innovations.
Here, the left hand
  side represents the original SCM and the right hand side
corresponds to the reduced SCM fitting in our framework. Consider now, in
the original SCM, the intervention $X_1 \coloneqq i$, for some
$i\in\R$. In the reduced SCM, this intervention corresponds to the
intervention $X = (X_1, X_2) \coloneqq (i, H + i + \ep_2)$, which is
neither confounding-preserving nor confounding-removing.\footnote{
  This may not come as a surprise since, without the help of an
  instrument, it is impossible to distinguish whether a covariate is
  an ancestor or a descendant of $Y$.  } On the other hand, any
intervention on $X_2$ or $A$ in the original SCM model corresponds to the
same intervention in the reduced SCM. We can generalize these
observations to the following statements 
\begin{itemize}
\item \emph{Interventions on $A$:} If we intervene on $A$ in the original
  SCM \eqref{eq:SCM_full_causal} (i.e., by replacing the structural
  assignment of $A$ with $\psi^i(I^i, \ep_A^i)$), then this translates
  to the same intervention on $A$ in the reduced
  SCM \eqref{eq:SCM_full_causal_simple}.
\item \emph{Shift intervention on
    $X_j$ which are not ancestors of
    $Y$:} If we perform a shift intervention on
  $X_j$ in the original SCM \eqref{eq:SCM_full_causal} (assuming no
  confounding $H$) and $X_j$ is not an ancestor of
  $Y$, then this corresponds to a confounding-preserving intervention
  in the reduced SCM \eqref{eq:SCM_full_causal_simple}.
\item \emph{Hard interventions on all $X$:} If we intervene on all $X$
  in the original SCM \eqref{eq:SCM_full_causal} by replacing the
  structural assignment of $X$ with an independent random variable
  $I\in\R^d$, then this translates to the same intervention in the
  reduced SCM \eqref{eq:SCM_full_causal_simple} which is
  confounding-removing.
\item \emph{No $X$ is a descendant of $Y$ and there is no unobserved
    confounding $H$:} If we intervene on $X$ in the original SCM
  \eqref{eq:SCM_full_causal} (i.e., by replacing the structural
  assignment of $X$ with $\psi^i(g, A^i, \ep^i_X ,I^i)$), then this
  translates to a potentially different but confounding-removing
  intervention in the reduced SCM
  \eqref{eq:SCM_full_causal_simple}. This is because the reduced SCM
  \eqref{eq:SCM_full_causal_simple} does not include unobserved
  variables $H$ in this case.
\item \emph{Hard interventions on a variable $X_j$ which has at most $Y$
    as a descendant:} 
  If we intervene on $X_j$ in the original SCM \eqref{eq:SCM_full_causal}
  by replacing the structural assignment of $X_j$ with an independent
  random variable $I$, then this intervention translates to a
  potentially different but confounding-preserving intervention.
\end{itemize}
Other settings may yield well-behaved interventions, too, but may
require more assumptions on the original SCM model
\eqref{eq:SCM_full_causal} or further restrictions on the intervention
classes.

\textbf{Intervention support}\quad 
A support-reducing intervention in the original SCM can translate to a
support-extending intervention in the reduced SCM. Consider the
following example
\begin{center}
  \begin{minipage}{0.40\columnwidth}
    \centering
    {
      \begin{align*}
        \begin{split}
          X_1 &\coloneqq \ep_1 \\
          X_2 &\coloneqq X_1 + \mathbf{1}\{X_1 = 0.5\}\\
          Y &\coloneqq X_2 + \epsilon_Y
        \end{split}
      \end{align*}}
\end{minipage}%
\begin{minipage}{0.18\columnwidth}
  {\Large
    \begin{equation*}
      \xrightarrow{\text{transform}}  \, \,
    \end{equation*}}
  \qquad 
  $ $
\end{minipage}%
\begin{minipage}{0.40\columnwidth}
  \centering
  {
      \begin{align*}
        \begin{split}
          X &\coloneqq (\ep_1, \ep_1 + \mathbf{1}\{\ep_1 = 0.5\})\\
          Y &\coloneqq X_2 + \ep_Y,
        \end{split}
      \end{align*}}
  \end{minipage}%
\end{center}
with
$\ep_1, \ep_Y\overset{i.i.d.}{\sim} \mathcal{U}(0,1)$.  As before,
the left hand side
represents the
original SCM, whereas 
the right hand
  side
  corresponds to the reduced
SCM converted to fit our framework.  Under the observational
distribution, the support of $X_1$ and $X_2$ is equal to the open
interval $(0, 1)$.  Consider now the support-reducing intervention
$X_1:= 0.5$ in original SCM.
Within our framework, such an intervention would correspond to the
intervention $X = (X_1, X_2) := (0.5, 1.5)$, which is
support-extending. This example is rather special in that the SCM
consists of a function that changes on a null set of the observational
distribution. With appropriate assumptions to exclude similar
degenerate cases, it is possible to show that support-reducing
interventions in \eqref{eq:SCM_full_causal} correspond to
support-reducing interventions within our
framework~\eqref{eq:SCM_full_causal_simple}.

\section{Sufficient conditions for Assumption~1 in IV settings}
\label{sec:IVconditions}

Assumption~\ref{ass:identify_f} states that $f$ is identified on the
support of $X$ from the observational distribution of $(Y,X,A)$. 
Whether this assumption is satisfied
depends on the structure of
$\cF$ but also on the other function
  classes $\cG,\cH_1,\cH_2$ and $\mathcal{Q}$ that
make up the model class $\cM$ from which we assume that the
distribution of $(Y,X,A)$ is generated.

Identifiability of the causal
function in the presence of 
instrumental variables
is a well-studied problem in econometrics
literature. Most prominent is the literature on identification in
linear SCMs
\cite[e.g.,][]{fisher1966identification,greene2003econometric}.
However, identification 
has also been studied for various other
parametric function classes. 
We say that $\mathcal{F}$ is a parametric
function class if it can be parametrized by some finite dimensional
parameter set $\Theta \subseteq \R^p$. 
We here consider classes of the form
\begin{align*}
	\mathcal{F} := \{ f(\cdot ,\theta):\R^d \to \R\,\vert\, \Theta \ni \theta \mapsto f(x,\theta) \in C^2, \forall  x\in \R^d \}.
\end{align*}
Consistent estimation of the 
parameter
$\theta_0$ using instrumental variables in such function classes has
been studied extensively in the econometric literature
\cite[e.g.,][]{amemiya1974nonlinear,jorgenson1974efficient,kelejian1971two}. %
These works also contain rigorous results on how instrumental variable
estimators of $\theta_0$ are constructed and under which conditions
consistency (and thus identifiability) holds. Here, we give an
argument on why the presence of the exogenous variables $A$ yields identifiability under certain
regularity conditions. Assume that $\E[h_1(H, \epsilon_Y)|A]=0$, which implies
that the true causal function $f(\cdot,\theta_0)$ satisfies the
population orthogonality condition
\begin{align} \label{Eq:PopOrthCondNonLinearIV}
	\E[l(A)^\top (Y-f(X,\theta_0))] = \E\big[l(A)^\top \E[h_1(H, \epsilon_Y)|A]\big]= 0,
\end{align}
for some measurable mapping $l:\R^q\to \R^g$,
 for some $ g \in \mathbb{N}_{>0}$.
  Clearly, $\theta_0$ is identified from the
  observational distribution if the map
  $\theta \mapsto \E[l(A)^\top (Y-f(X,\theta))]$ is zero if and only
  if $\theta=\theta_0$.   Furthermore, since
$\theta\mapsto f(x,\theta)$ is differentiable for all $x\in \R^d$,
 the
mean value theorem yields that, for any $\theta\in \Theta$ and
$x\in \R^d$,
 there exists an intermediate point
$\tilde{\theta}(x,\theta,\theta_0)$ on the line segment between
$\theta$ and $\theta_0$ such that
\begin{equation*}
f(x,\theta) - f(x,\theta_0) = D_\theta f(x,\tilde{\theta}(x,\theta,\theta_0))(\theta-\theta_0),
\end{equation*}
where, 
for each
$x\in \R^d$,
$D_\theta f(x,\theta)\in\R^{1\times p}$ is the derivative of
$\theta\mapsto f(x,\theta)$ evaluated in $\theta$.  Composing the above expression with the random vector
$X$, multiplying with $l(A)$ and taking expectations yields that
\begin{align*}
\E[l(A)(Y-f(X,\theta_0))] - \E[l(A)(Y-f(X,\theta))] 
=  \E[l(A)D_\theta f(X,\tilde{\theta}(X,\theta,\theta_0))](\theta_0-\theta).
\end{align*}
Hence, if
$ \E[l(A)D_\theta f(X,\tilde{\theta}(X,\theta,\theta_0))]\in
\R^{g\times p}$ is of rank $p$ for all $\theta\in\Theta$
(which implies $g \geq p$), 
then
$\theta_0$ is identifiable as it is the only parameter that satisfies
the population orthogonality condition of
\eqref{Eq:PopOrthCondNonLinearIV}.  As $\theta_0$ uniquely determines
the entire function, we get identifiability of
$f\equiv f(\cdot,\theta_0)$, not only on the support of $X$ but the
entire domain $\R^d$, i.e., both
  Assumptions~\ref{ass:identify_f} and~\ref{ass:gen_f} are satisfied.
In the case that $\theta \mapsto f(x,\theta)$ is linear, i.e.\
$f(x,\theta) = f(x)^T \theta$ for all $x\in \R^d$, the above rank
condition reduces to $\E[l(A)f(X)^T]\in \R^{g\times p}$ having rank $p$
(again, implying that $g \geq p$). Furthermore, when
$(x,\theta)\mapsto f(x,\theta)$ is bilinear, a reparametrization of
the parameter space ensures that $f(x,\theta)= x^T \theta$ for
$\theta\in\Theta \subseteq \R^d$. In this case, the rank condition can
be reduced to the well-known rank condition for identification in a
linear SCM, namely that
$\E[AX^T] \in \R^{q\times p}$ is of rank $p$.

Finally, identifiability and methods of consistent estimation of the causal function have also been studied for non-parametric function classes. The conditions for identification are rather technical, however, and we refer the reader to  \cite{newey2013nonparametric,newey2003instrumental} for further details.

\section{Choice of test statistic} 
\label{sec:test_statistic}
By considering the variables $B(X) = (B_1(X), \dots, B_k(X))$ and $C(A) = (C_1(A), \dots, C_k(A))$
as vectors of covariates and instruments, respectively, our setting in Section~\ref{sec:nile} reduces to the classical
(just-identified) linear IV setting. 
We could therefore 
use a test statistics similar to the one propsed by the PULSE \cite{jakobsen2020distributional}. 
With a notation that is slightly adapted to our setting, this estimator tests $\tilde{H}_0(\theta)$
using the test statistic
\begin{align*}
T^1_n(\theta) = c(n) \frac{\norm{\B{P} (\B{Y} - \B{B}\theta)}_2^2}{\norm{\B{Y} - \B{B}\theta}_2^2},
\end{align*}
where $\B{P}$ is the projection onto the columns of $\B{C}$, and $c(n)$ is some function with $c(n) \sim n$ as $n\to\infty$. 
Under the null hypothesis, $T^1_n$ converges in distribution to the $\chi^2_{k}$ distribution, and diverges to infinity in 
probability under the general alternative. Using this test statistic, $\tilde{H}_0(\theta)$ is rejected if and only if $T^1_n(\theta)> q(\alpha)$, 
where $q(\alpha)$ is the $(1-\alpha)$-quantile of the $\chi^2_{k}$ distribution. The acceptance region of this test statistic 
is asymptotically equivalent with the confidence region of the Anderson-Rubin test \cite{anderson1949estimation} for the 
causal parameter $\theta^0$. Using the above test results in a consistent estimator for $\theta^0$ 
\citep[][Theorem~3.12]{jakobsen2020distributional}; the proof exploits the particular form of $T^1_n$ without 
explicitly imposing that assumptions~\ref{ass:ConsistentTestStatistic} and \ref{ass:LambdaStarAlmostSurelyFinite} hold.

If the number $k$ of basis functions is large, however, numerical experiments suggest that the above 
test has low power in finite sample settings. As default, our algorithm  therefore 
uses a different test based on a penalized regression approach. 
This test has been proposed in \cite{chen2014note}
for inference in nonparametric regression models. 
We now introduce this procedure
with a notation that is adapted to our setting. For every $\theta \in \R^k$, let 
$R_\theta = Y - B(X)^\top \theta$ be the residual associated with $\theta$. 
We then test the slightly stronger hypothesis
\begin{equation*}
\bar{H}_0(\theta): \exists \,  \sigma_\theta^2>0 \text{ s.t. }  \E[R_\theta \given A] \eqas 0  \text{ and } \text{Var}[R_\theta \given A] = \sigma_\theta^2
\end{equation*}
against the alternative that $\E[R_\theta \given A] = m(A)$ for some
smooth function $m$.  To see that the above hypothesis implies
$\tilde{H}_0(\theta)$ (and therefore $H_0(\theta)$, see
  Section~\ref{sec:estimation}),
  let $\theta \in \R^k$ be such that
$\bar{H}_0(\theta)$ holds true.  Then,
\begin{align*}
\E[C(A)(Y - B(X)^\top \theta)] 
= \E[C(A) R_\theta] = \E[\E[C(A) R_\theta \given A]] 
= \E[C(A) \E[R_\theta \given A]] = 0,
\end{align*}
showing that also $\tilde{H}_0(\theta)$ holds true. 
Thus, if $\tilde{H}_0(\theta)$ is false, then also $\bar H_0(\theta)$ is false.
As a test statistic $T^2_n(\theta)$ for $\bar{H}_0(\theta)$, we use (up to a normalization) 
the squared
norm of a penalized regression estimate of $m$, evaluated at the data $\B{A}$, i.e., 
 the TSLS loss $\norm{\B{P}_\delta (\B{Y} - \B{B}\theta)}_2^2$.
In the fixed design case, where $\B{A}$ is non-random, it has been shown that,
under $\bar{H}_0(\theta)$ and certain additional regularity conditions, it holds that 
\begin{align*}
\frac{\norm{\B{P}_\delta (\B{Y} - \B{B}\theta)}_2^2 - \sigma_\theta^2 c_n}{\sigma_\theta^2 d_n} \stackrel{\text{d}}{\longrightarrow} \mathcal{N}(0,1),
\end{align*}
where $c_n$ and $d_n$ are known functions of $\B{C}$, $\B{M}$ and $\delta$ \citep[][Theorem~1]{chen2014note}.
The authors further state that the above convergence is unaffected by exchanging $\sigma_\theta^2$
with a consistent estimator $\hat{\sigma}_\theta^2$, which motivates our use of the test statistic
\begin{equation*}
T^2_n(\theta) := \frac{\norm{\B{P}_\delta (\B{Y} - \B{B}\theta)}_2^2 - \hat \sigma_{\theta,n}^2 c_n}{\hat \sigma_{\theta,n}^2 d_n},
\end{equation*}
where $\hat \sigma_{\theta,n}^2 := \frac{1}{n-1} \sum_{i=1}^n \norm{(\B{I}_n - \B{P}_\delta)(\B{Y} - \B{B} \theta)}_2^2$. 
As a
rejection threshold $q(\alpha)$ we use the $1-\alpha$ quantile of a standard normal distribution. 
For results on the asymptotic power of the test defined by $T^2$, we refer to Section~2.3 in \cite{chen2014note}. 

In our software package, both of the above tests are available options.

\section{Addition to experiments} 
\label{sec:additional_experiments}

\subsection{Sampling of the causal function} \label{sec:exp_sampling}
\begin{figure}
\centering
\includegraphics[width=\linewidth]{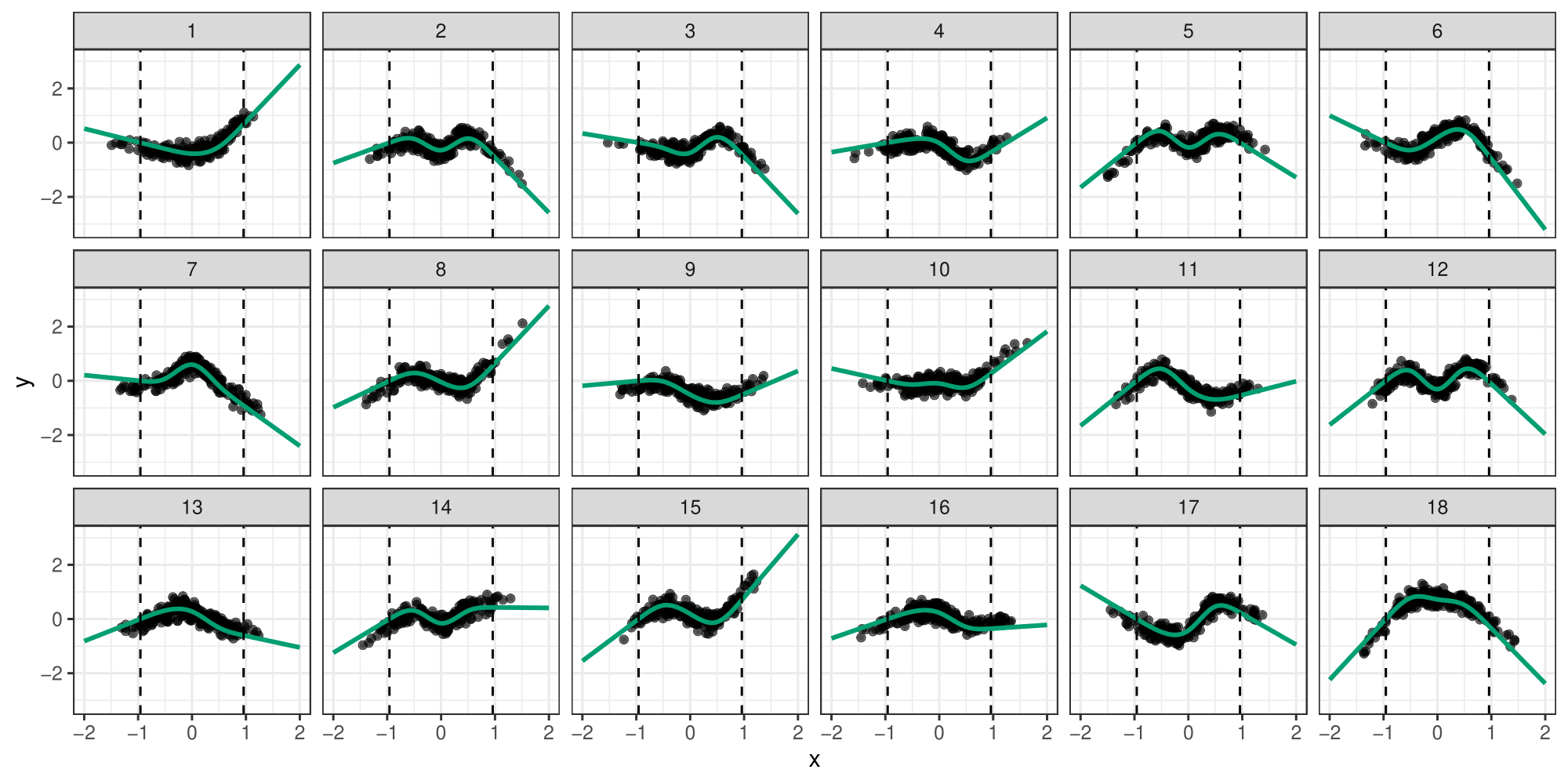}
\caption{
The plots show independent realizations of the causal function that
    is used in all our experiments. 
    These are sampled from a 
    linear space of natural cubic splines, as described in
    Appendix~\ref{sec:exp_sampling}. To ensure a fair comparison
    with the alternative method, NPREGIV, the true causal 
    function is chosen from a model class different from 
    the one assumed by the NILE.}
\label{fig:f_samples}
\end{figure}

To ensure linear extrapolation of the causal function, we have chosen a function
class consisting of natural cubic splines, which, by construction, extrapolate linearly
outside the boundary knots. We now describe in detail how we sample 
functions from this class for the experiments in Section~\ref{sec:experiments}.
Let $q_{\min}$ and $q_{\max}$ be the respective $5\%$- and $95\%$ quantiles of $X$, 
and let $B_1, \dots, B_4$ be a basis of natural cubic splines corresponding to 5 knots 
placed equidistantly between $q_{\min}$ and $q_{\max}$. We then sample coefficients 
$\beta_i \iid \text{Uniform}(-1,1)$, $i = 1, \dots, 4$, and construct $f$ as 
$f = \sum_{i=1}^4 \beta_i B_i$. For illustration, we have included 18 
realizations in Figure~\ref{fig:f_samples}. 

\subsection{Violations of the linear extrapolation assumption} \label{sec:exp_violation_lin_extrap}
\begin{figure}[t]
\centering
\includegraphics[width=\linewidth]{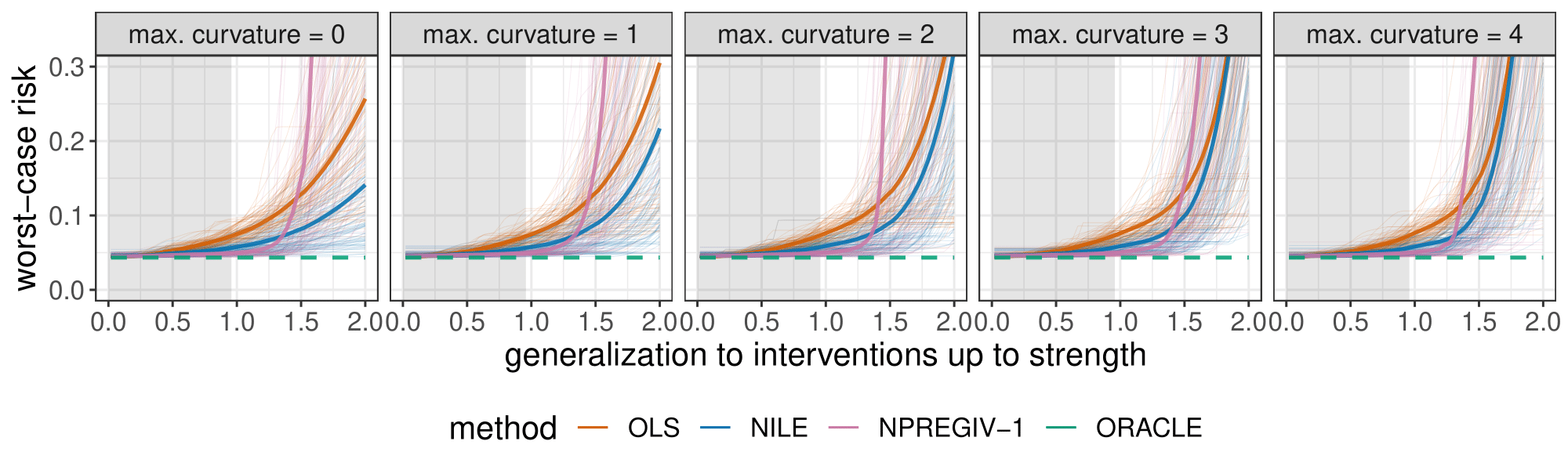}
\caption{Worst-case risk for increasingly strong violations of the linear extrapolation assumption. 
The grey area marks the inner 90 \% quantile range of $X$ in the training distribution.
As the 
curvature of $f$
outside the domain of the observed data increases, 
it becomes 
difficult 
to predict the interventional behavior of $Y$ for strong interventions. However, even in situations
where the linear extrapolation assumption is strongly violated,  it remains beneficial to
extrapolate linearly. %
}
\label{fig:violation_lin_extrap}
\end{figure}

We have assumed that the true causal function extrapolates linearly 
outside the 90\% quantile range of $X$. We now investigate the performance
of our method for violations of this assumption. To do so, we again sample from 
the model \eqref{eq:sim_model}, with $\alpha_A = \alpha_H = \alpha_\epsilon = 1/\sqrt{3}$. 
For each data set, the causal function is sampled as follows. Let $q_{\min}$ and $q_{\max}$
be the $5\%$- and $95\%$ quantiles of $X$. We first generate a function $\tilde{f}$
that linearly extrapolates outside $[q_{\min}, q_{\max}]$ as described in 
Section~\ref{sec:exp_sampling}. For a given threshold $\kappa$, we then draw 
$k_1, k_2 \iid \text{Uniform}(-\kappa, \kappa)$ and construct $f$ for every $x \in \R$ by
\begin{equation*}
f(x) = \tilde{f}(x) + \frac{1}{2} k_1 ((x-q_{\min})_{-})^2 + \frac{1}{2} k_2 ((x-q_{\max})_{+})^2,
\end{equation*}
such that the curvature of $f$ on $(-\infty, q_{\min}]$ and $[q_{\max}, \infty)$ is $k_1$ and $k_2$, respectively. 
Figure~\ref{fig:violation_lin_extrap} shows results for $\kappa = 0,1,2,3,4$. As the curvature increases, 
the ability to generalize decreases. 

\subsection{Running NILE on half of the available data}
\begin{figure}[t]
\centering
\includegraphics[width=.7\columnwidth]{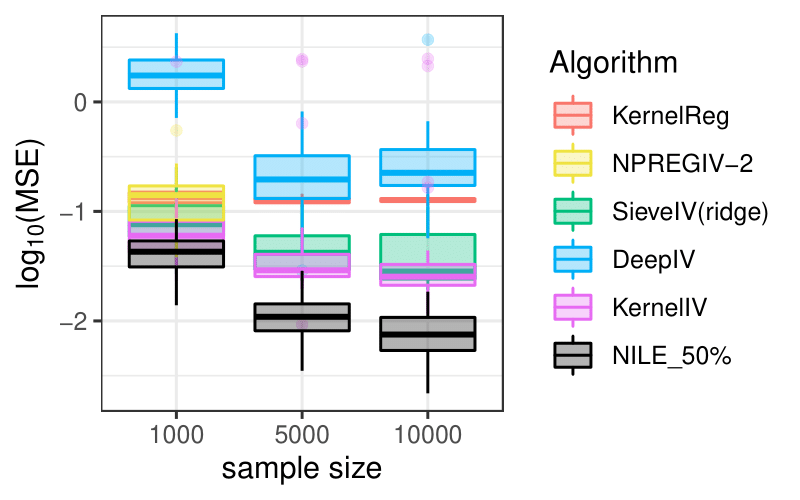}
\caption{Same results as shown in Figure~\ref{fig:kernelIV}, except that here, NILE is run only on 
half of the available data.}
\label{fig:nile50_vs_kernelIV}
\end{figure}
In Section~\ref{sec:experiments}, we compared the NILE to several alternative procedures for 
estimating a non-linear causal function. As mentioned, these procedure use a sample-splitting
strategy, where the two steps of the the two-stage-least-squares procedure are run on 
disjoint data sets. The NILE, on the other hand, uses all of the available data for the model fitting. 
Figure~\ref{fig:nile50_vs_kernelIV} shows that, even when using only half of the available data, 
the NILE still outperforms the other methods considerably. 

\section{Proofs}\label{app:proofs}

\subsection{Proof of Proposition~\ref{prop:minimax_equal_causal}}\label{sec:prop:minimax_equal_causal}

\begin{proof}
  Assume that $\cI$ is a set of interventions on $X$ with at least one
  confounding-removing intervention. Let $i \in \II$ and
  $\fs \in \FF$, then we have the following expansion
  \begin{align} \label{eq:decomp_equation}
    \E_{M(i)}[(Y - \fs(X))^2] = 
    \E_{M(i)}[(f(X) - \fs(X))^2]
    +\E_{M(i)}[\xi_Y^2]  
    +2\E_{M(i)}[\xi_Y(f(X)-\fs(X))], 
  \end{align}
  where $\xi_Y=h_1(H, \epsilon_Y)$. For any intervention $i\in \cI$
  the causal function $f$ always yields an identical loss. In particular,
  it holds that
  \begin{align} \label{eq:SupLossCausalFunction}
    \sup_{i\in\II} \E_{M(i)}[(Y - f(X))^2] =\sup_{i\in\II} \E_{M(i)}[\xi_Y^2] = \E_{M}[\xi_Y^2],
  \end{align}
  where we used that the distribution of $\xi_Y$ is not affected by an
  intervention on $X$. The loss of the causal function can
  never be better than the minimax loss, that is,
  \begin{align} \label{eq:lowerbdd_prop1}
    \inf_{\fs\in\FF}\sup_{i\in\II}\E_{M(i)}[(Y - \fs(X))^2]
 \leq\sup_{i\in\II}\E_{M(i)}[(Y - f(X))^2]   =\E_{M}[\xi_Y^2].
  \end{align}
  In other words, the minimax solution (if it exists) is always better
  than or equal to the causal function. We will now show that when
  $\cI$ contains at least one confounding-removing intervention, then
  the minimax loss is dominated by any such intervention.

  Fix $i_0\in \cI$ to be a confounding-removing intervention and let
  $(X,Y,H,A)$ be generated by the SCM $M(i_0)$.  Recall that there
  exists a map $\psi^{i_0}$ such that
  $X:= \psi^{i_0}(g, h_2, A, H, \ep_X ,I^{i_0})$ 
 and that
  $X\independent H$ as $i_0$ is a confounding-removing intervention.
  Furthermore, since 
  the vectors $A$, $H$, $\ep_X$, $\ep_Y$ and
  $I^{i_0}$ are mutually independent, we have that
  $(X,H)\independent \ep_Y$ which together with $X\independent H$
  implies $X, H$ and $\ep_Y$ are mutually independent, and hence
  $X \independent h_1(H,\ep_Y)$. Using this independence we get that
  $\E_{M(i_0)}[\xi_Y(f(X)-\fs(X))]=\E_{M}[\xi_Y]\E_{M(i_0)}[(f(X)-\fs(X))]$. Hence,
  \eqref{eq:decomp_equation} for the intervention $i_0$ together with
  the modeling assumption $\E_{M}[\xi_Y]=0$ implies 
  that
  for all
  $\fs \in \FF$, 
  \begin{align*}
\E_{M}[\xi_Y^2] &\leq \E_{M(i_0)}[(f(X) - \fs(X))^2]+\E_{M}[\xi_Y^2] =  \E_{M(i_0)}[(Y - \fs(X))^2] .
  \end{align*}
  This proves that the smallest loss at a confounding-removing
  intervention is achieved by the causal function. 
  Denoting the non-empty subset of confounding-removing interventions by $\cI_{\text{cr}}\subseteq \cI$, this implies
    \begin{align} 
    	\E_{M}[\xi_Y^2] &=  \inf_{\fs\in\FF} \E_{M(i_0)}[(Y - \fs(X))^2] \notag
    	\leq \inf_{\fs\in\FF} \sup_{i\in \cI_{\text{cr}}} \E_{M(i)}[(Y - \fs(X))^2] \notag \\
    	&\leq  \inf_{\fs\in\FF} \sup_{i\in\II}\E_{M(i)}[(Y - \fs(X))^2].   \label{eq:upperbdd_prop1}
  \end{align}
  Combining \eqref{eq:lowerbdd_prop1} and \eqref{eq:upperbdd_prop1} it
  immediately follows that
  \begin{align*}
    \inf_{\fs\in\FF} \sup_{i\in\II}\E_{M(i)}[(Y - \fs(X))^2] =  \sup_{i\in\II} \E_{M(i)}[(Y - f(X))^2],
  \end{align*}
  and hence
  \begin{equation*}
    f\in\argmin_{\fs \in \FF} \sup_{i \in \II} \E_{M(i)}[(Y-\fs(X))^2],
  \end{equation*}
  which completes the proof of Proposition~\ref{prop:minimax_equal_causal}.
\end{proof}

\subsection{Proof of Proposition~\ref{prop:shift_interventions}}

\begin{proof}
  Let $\FF$ be the class of all linear functions and let $\II$ denote
  the set of interventions on $X$ that satisfy
  \begin{equation*}
    \sup_{i\in \cI} \lambda_{\min}\big(\E_{M(i)}\big[XX^\top\big]\big) =\infty.
  \end{equation*}
  We claim that the causal function $f(x)=b^\top x$ is the unique
  minimax solution of \eqref{eq:minimax_problem}. We prove the result
  by contradiction. 
  Let $\bar{f}\in\mathcal{F}$ (with $\bar{f}(x)=\bar{b}^{\top}x$) be such that 
    \begin{equation*}
    \sup_{i\in\II}\E_{M(i)}[(Y-\bar{b}^\top X)^2]\leq\sup_{i\in\II}\E_{M(i)}[(Y-b^\top X)^2],
  \end{equation*}
  and assume that $\norm{\bar{b}-b}_2>0$.
  For a fixed $i\in \cI$, we get the following bound
  \begin{align*}
    \E_{M(i)}[(b^\top X-\bar{b}^\top X)^2]
    &=(b-\bar{b})^\top \E_{M(i)}[XX^{\top}](b-\bar{b})
    \geq \lambda_{\min}(\E_{M(i)}[XX^{\top}]) \|b-\bar{b} \|_2^2.
  \end{align*}
  Since we assumed that the minimal eigenvalue is unbounded, this
  means that we can choose $i\in\II$ such that
  $\E_{M(i)}[(b^\top X-\bar{b}^\top X)^2]$ can be arbitrarily large. However,
  applying Proposition~\ref{prop:difference_to_causal_function}, this
  leads to a contradiction since
  $\sup_{i \in \II} \E_{M(i)}[(b^\top X-\bar{b}^\top X)^2]\leq 4\operatorname{Var}_{M}(\xi_Y)$
  cannot be satisfied. Therefore, it must holds that $\bar{b}=b$, which
  moreover implies that $f$ is indeed a solution to the minimax
  problem
  $\argmin_{\fs \in\mathcal{F}}\sup_{i\in\II}\E_{M(i)}[(Y-\fs(X))^2]$,
  as it achieves the lowest possible objective value. This
  completes the proof of Proposition~\ref{prop:shift_interventions}.
\end{proof}

\subsection{Proof of
  Proposition~\ref{prop:difference_to_causal_function}}\label{sec:prop:difference_to_causal_function}

\begin{proof}
  Let $\II$ be a set of interventions on $X$ or $A$ and let
  $\fs\in\FF$ with
  \begin{equation}
    \label{eq:better_than_causal_cond}
    \sup_{i\in\II}\E_{M(i)}[(Y-\fs(X))^2]\leq\sup_{i\in\II}\E_{M(i)}[(Y-f(X))^2].
  \end{equation}
  For any $i\in\II$, the Cauchy-Schwartz inequality implies that
  \begin{align*}
    & \, \E_{M(i)}[(Y-\fs(X))^2]
    = \E_{M(i)}[(f(X)+\xi_Y-\fs(X))^2]\\
    &\,=   \E_{M(i)}[(f(X)-\fs(X))^2]+\E_{M(i)}[\xi_Y^2]
     +2\E_{M(i)}[\xi_Y(f(X)-\fs(X))]\\
    & \,\geq \E_{M(i)}[(f(X)-\fs(X))^2]+\E_{M}[\xi_Y^2] -2\left(\E_{M(i)}[(f(X)-\fs(X))^2]\E_{M}[\xi_Y^2]\right)^{\frac{1}{2}}.
  \end{align*}
  A similar computation shows that the causal function $f$ satisfies
  \begin{equation*}
    \E_{M(i)}[(Y-f(X))^2]=\E_{M}[\xi_Y^2].
  \end{equation*}
  So by condition \eqref{eq:better_than_causal_cond} this 
  implies for any $i\in\II$ that
  \begin{align*}
    \E_{M}[\xi_Y^2] \geq &\,  \E_{M(i)}[(f(X)-\fs(X))^2]+\E_{M}[\xi_Y^2] 
      -2\left(\E_{M(i)}[(f(X)-\fs(X))^2]\E_{M}[\xi_Y^2]\right)^{\frac{1}{2}},
  \end{align*}
  which is equivalent to
  \begin{align*}
  &\E_{M(i)}[(f(X)-\fs(X))^2] \leq 2\sqrt{\E_{M(i)}[(f(X)-\fs(X))^2]\E_{M}[\xi_Y^2]},
  \end{align*}
i.e. $\E_{M(i)}[(f(X)-\fs(X))^2]\leq 4\E_{M}[\xi_Y^2]$.
  As this inequality holds for all $i\in\II$, we can take the supremum
  over all $i\in\II$, which completes the proof of
  Proposition~\ref{prop:difference_to_causal_function}.  
\end{proof}

\subsection{Proof of Proposition~\ref{prop:misspecification_minimax}}
  \begin{proof}
  As argued before, we have that for all $i \in \II_1$, 
    \begin{equation*}
      \E_{M(i)}\big[(Y-f(X))^2\big]=\E_{M(i)}\big[\xi_Y^2\big]=\E_{M}\big[\xi_Y^2\big].
    \end{equation*}
Let now
    $f_1^*\in\mathcal{F}$ be a minimax solution w.r.t.\ $\II_1$. 
    Then, using that the causal function $f$ lies in $\FF$, it holds that
    \begin{align*}
      \sup_{i\in\II_1}\E_{M(i)}\big[(Y-f_1^*(X))^2\big] &\leq \sup_{i\in\II_1}\E_{M(i)}\big[(Y-f(X))^2\big] = \E_{M}\big[\xi_Y^2\big].
    \end{align*}
    Moreover, if $\II_2\subseteq\II_1$, then it must also hold that
    \begin{align*}
      \sup_{i\in\II_2}\E_{M(i)}\big[(Y-f_1^*(X))^2\big] &\leq \E_{M}\big[\xi_Y^2\big] =\sup_{i\in\II_2}\E_{M(i)}\big[(Y-f(X))^2\big].
    \end{align*}
    To prove the second part, we give a one-dimensional example. Let $\mathcal{F}$ be
    linear (i.e., $f(x)=b x$) and let $\II_1$ consist of shift
    interventions on $X$ of the form
    \begin{equation*}
      X^i\coloneqq g(A^i) + h_2(H^i, \epsilon_X^i)+ c,
    \end{equation*}
    with 
    $c\in [0, K]$. 
    Then, the minimax solution $f^*_1$
    (where $f^*_1(x)=b^*_1 x$) with respect to $\II_1$ is not equal to the
    causal function $f$ as long as $\cov(X, \xi_Y)$ is 
    strictly positive.
     This can be seen by explicitly computing the OLS
    estimator for a fixed shift $c$ and observing that the worst-case
 risk is attained at $c=K$. %
    Now let $\cI_2$ be a set of
    interventions of the same form as $\cI_1$ but including shifts with $c>K$
    such that
    $\cI_2 \not \subseteq \cI_1$. Since $\cF$ consists of linear
    functions, we know that the loss
    $\E_{M(i)}\big[(Y-f_1^*(X))^2\big]$ can become arbitrarily large,
    since
        \begin{align*}
      & \, \E_{M(i)}\big[(Y-f_1^*(X))^2\big]\\
      & \,=(b-b^*_1)^2\E_{M(i)}[X^2]+\E_{M}[\xi_Y^2]+2(b-b^*_1)\E_{M(i)}[\xi_Y X]\\
      & \,=(b-b^*_1)^2(c^2+\E_{M}[X^2]+2c\E_{M}[X])+\E_{M}[\xi_Y^2]
       +2(b-b^*_1)(\E_{M}[\xi_Y X]+\E_{M}[\xi_Y]c),
    \end{align*}
    and $(b-b^*)^2>0$. In contrast, the loss for the causal function
    is always $\E_{M}[\xi_Y^2]$, so the worst-case risk of $f^*_1$
    becomes arbitrarily worse than that of $f$. This completes the
    proof of Proposition~\ref{prop:misspecification_minimax}.
  \end{proof}

\subsection{Proof of Proposition~\ref{prop:suff_general}}

\begin{proof}
Let $\epsilon > 0$. By definition of the infimum, we can find $f^* \in \FF$ such that
  \begin{align*}
    &\left| \sup_{i\in\II}\E_{M(i)}\big[(Y-f^*(X))^2\big]
     - \inf_{\fs\in\FF}\,
      \sup_{i\in\II}\E_{M(i)}\big[(Y-\fs(X))^2\big] \right| \leq \epsilon.
  \end{align*}
  Let now $\tilde{M} \in \MM$ be s.t.\ $\P_{\tilde{M}} = \P_M$. By assumption, the left-hand side 
  of the above inequality is unaffected by substituting $M$ for $\tilde{M}$, and the 
  result thus follows. 
\end{proof}

\subsection{Proof of Proposition~\ref{prop:impossibility_interpolation}}

\begin{proof}
  We first show that the causal parameter $\beta$ is not a minimax solution.
  Let $u := \sup \II < \infty$, since $\II$ is bounded,
  and
  take $b = \beta + 1/(\sigma u)$. 
  By an explicit computation we get that 
 \begin{align*}
 \inf_{\bes \in \R}\,
 \sup_{i\in\II} \E_{M(i)}\big[(Y-\bes X)^2\big]   \leq &\, \sup_{i\in\II} \E_{M(i)}\big[(Y-b X)^2\big]
 =  \sup_{i\in\II} \E_{M(i)}\big[(\ep_Y + \tfrac{1}{\sigma}H - 
 \tfrac{1}{\sigma u}iH)^2\big]\\
 = &\, \sup_{i\in\II} \left[1 + \left(1 - \tfrac{i}{u}\right)^2\right] 
 <  2
 = \sup_{i\in\II}\E_{M(i)}\big[(Y-\beta X)^2\big],
  \end{align*}
  where the last inequality holds because $0 < 1 + (1 - i/u)^2 < 2$ for all
  $i \in \II$, and since $\II\subset \R_{>0}$ is compact with upper bound $u$. Hence,
  \begin{align*}
  \sup_{i\in\II}\E_{M(i)}\big[(Y-\beta X)^2\big]
  - \inf_{\bes \in \R}\,
  \sup_{i\in\II} \E_{M(i)}\big[(Y-\bes X)^2\big] > 0,
  \end{align*}
proving that the causal parameter is not a minimax solution for model $M$
w.r.t. $(\cF, \cI)$. Recall that in order to prove that $(\bP_{M},\cM)$ does not generalize with respect to $\cI$ we have to show that there exists an $\ep >0$ such that for all $b\in \R$ it holds that
\begin{align*}
&\sup_{\tilde M: \bP_{\tilde{M}}= \bP_M}\big|   \sup_{i\in\II}\E_{\tilde{M}(i)}\big[(Y-b X)^2\big]
 - \inf_{\bes \in \R}\,
\sup_{i\in\II} \E_{\tilde{M}(i)}\big[(Y-\bes X)^2\big]\big| \geq \ep.
\end{align*}
Thus, it remains to show that for all $b \not = \beta$ there exists a model $\tilde M \in \cM$ with $\bP_M = \bP_{\tilde M}$ such that the generalization loss is bounded below uniformly by a positive constant. We will show the stronger statement that 
  for any $b \neq \beta$, there exists a model $\tilde{M}$ with
  $\P_{\tilde{M}} = \P_M$, such that under $\tilde{M}$, 
  $b$ results in arbitrarily large generalization error.
  Let $c > 0$ and $i_0 \in \II$. Define 
  \begin{align*}
  \tilde\sigma \coloneqq \frac{\sign{((\beta - b)i_0)}\sqrt{1 + c} - 1}{
  (\beta - b)i_0} > 0,
  \end{align*}
  and let 
  $\tilde M \coloneqq M(\gamma, \beta, \tilde{\sigma}, Q)$. By construction of the model class $\MM$, 
  it holds that $\P_{\tilde{M}} = \P_{M}$.
  Furthermore, by an explicit computation we get that 
  \begin{align}\label{eq:proof_imp_a1}
  \begin{split}
  \sup_{i\in\II} \E_{\tilde{M}(i)}\big[(Y-bX)^2\big]
  \geq & \, \E_{\tilde{M}(i_0)}\big[(Y-bX)^2\big]
  = \E_{\tilde{M}(i_0)}\big[((\beta-b)i_0H+\ep_Y+
  \tfrac{1}{\tilde\sigma}H)^2\big]\\
  =& \, \E_{\tilde{M}(i_0)}\big[([(\beta-b)i_0\tilde{\sigma}+1] \ep_H+\ep_Y )^2\big]  
  = [(\beta-b)i_0\tilde{\sigma}+1]^2 + 1\\
  = & \, ((\beta - b)i_0\tilde\sigma)^2 + 2(\beta - b)i_0\tilde\sigma +
  2\\
  =& \, (\sign{((\beta-b)i_0)}\sqrt{1+c}-1)^2 
  + 2\sign{((\beta-b)i_0)}\sqrt{1+c}\\
  =& \,  c + 2.
  \end{split}
  \end{align}
  Finally, by definition of the infimum, it holds that
  \begin{align}\label{eq:proof_imp_b1}
    \begin{split}
      \inf_{\bes \in \R}\,\sup_{i\in\II} \E_{\tilde{M}(i)}\big[(Y-\bes
      X)^2\big] \leq \sup_{i\in\II}\E_{\tilde{M}(i)}\big[(Y-\beta
      X)^2\big] = 2.
    \end{split}
  \end{align}
  Combining~\eqref{eq:proof_imp_a1} and~\eqref{eq:proof_imp_b1} 
  yields that the generalization error is bounded below by $c$. That is,
  \begin{align*}
  &\big|   \sup_{i\in\II}\E_{\tilde{M}(i)}\big[(Y-b X)^2\big] 
  - \inf_{\bes \in \R}\,
  \sup_{i\in\II} \E_{\tilde{M}(i)}\big[(Y-\bes X)^2\big]\big| \geq c.
  \end{align*}
  The above results make no assumptions on $\gamma$, and hold true, 
  in particular, if $\gamma \neq 0$ (in which case Assumption~\ref{ass:identify_f} 
  is satisfied, see Appendix~\ref{sec:IVconditions}).
  This completes the proof of Proposition~\ref{prop:impossibility_interpolation}.
  
\end{proof}

\subsection{Proof of Proposition~\ref{prop:genX_intra}}

\begin{proof}
  Let $\II$ be a well-behaved set of interventions on $X$. We
  consider two cases; (A) all interventions in $\II$ are
  confounding-preserving and (B) there is at least one intervention in
  $\II$ that is confounding-removing.

  \textbf{Case (A):} In this case, we prove the result in two steps:
  (i) We show that $(A, \xi_X, \xi_Y)$ is identified from the
  observational distribution $\P_M$. (ii) We show that 
  this 
  implies that the intervention distributions $(X^i, Y^i)$, $i \in \II$, are also
  identified from the observational distribution, and conclude by using
  Proposition~\ref{prop:suff_general}.   
    Some of the details will be slightly technical
  because we allow for a large class of
  distributions (e.g., there is no assumption on the existence of densities).
  
  We begin with step (i). In this case, $\II$ is a set of confounding-preserving
  interventions on $X$, and we have that
  $\supp_{\II}(X)\subseteq\supp(X)$. Fix
  $\tilde{M} =(\tilde{f},\tilde{g},\tilde{h}_1,\tilde{h}_2,\tilde{Q})
  \in \MM$ such that $\P_{\tilde{M}} = \P_M$ and let
  $(\tilde{X}, \tilde{Y},\tilde{H},\tilde{A})$ be generated by the SCM
  of $\tilde{M}$. We have that
  $(X,Y,A) \eqd (\tilde{X}, \tilde{Y},\tilde{A})$ and by
  Assumption~\ref{ass:identify_f}, we have that $f \equiv \tilde{f}$
  on $\supp(X)$, hence $f(X) \eqas \tilde{f}(X)$. Further, fix any
  $B\in \cB(\R^p)$ (i.e., in the Borel sigma-algebra on $\R^p$) and note
  that
  \begin{align*}
    \bE_M[\mathbbm{1}_{B}(A)X|A]
    &=\bE_M[\mathbbm{1}_{B}(A)g(A)
    +\mathbbm{1}_{B}(A)h_2(H,\ep_X)|A] \\
    &= \bE_M[\mathbbm{1}_{B}(A)g(A)|A]
    + \mathbbm{1}_{B}(A)\bE[h_2(H,\ep_X)]
    = \mathbbm{1}_{B}(A)g(A),
  \end{align*}
  almost surely. Here, we have used our modeling assumption $\E[h_2(H, \epsilon_X)] = 0$. 
  Hence, by similar arguments for
  $\bE_{\tilde{M}}(\mathbbm{1}_{B}(\tilde{A})\tilde{X}|\tilde{A})$ and
  the fact that
  $(X,Y,A) \eqd (\tilde{X}, \tilde{Y},\tilde{A})$ we have
  that
  \begin{align*}
    \mathbbm{1}_{B}(A)g(A) &\eqas\bE_M (\mathbbm{1}_{B}(A)X|A) 
    \eqd \bE_{\tilde{M}}(\mathbbm{1}_{B}(\tilde{A})\tilde{X}|\tilde{A}) \eqas \mathbbm{1}_{B}(\tilde{A})\tilde{g}(\tilde{A}).
  \end{align*}
  We conclude that
  $\mathbbm{1}_{B}(A)g(A)\eqd
  \mathbbm{1}_{B}(\tilde{A})\tilde{g}(\tilde{A})$ for any
  $B\in \cB(\R^p)$.
   Let $\bP$ and
  $\tilde{\bP}$ denote the respective background probability measures on which
  the random elements $(X,Y,H,A)$ and
  $(\tilde{X},\tilde{Y},\tilde{H},\tilde{A})$ are defined. 
  Fix any $F\in \sigma(A)$ (i.e., in the sigma-algebra
  generated by $A$) and note that there exists a $B\in \cB(\R^p)$ such
  that $F=\{A\in B\}$. Since $A \eqd \tilde{A}$, we have that,
  \begin{align*}
    \int_F g(A) \, \mathrm{d} \bP &= \int \mathbbm{1}_{B}(A) g(A) \, \mathrm{d} \bP = \int \mathbbm{1}_{B}(\tilde{A})\tilde{g}(\tilde{A})  \, \mathrm{d} \tilde{\bP}  = \int  \mathbbm{1}_{B}(A)\tilde{g}(A)  \, \mathrm{d}\bP = \int_F \tilde{g}(A)  \, \mathrm{d}\bP.
  \end{align*}
  Both $g(A)$ and $\tilde{g}(A)$ are $\sigma(A)$-measurable and they
  agree integral-wise over every set $F\in \sigma(A)$, so we must have
  that $g(A) \eqas \tilde{g}(A)$. With
  $\eta(a,b,c)= (a,c-\tilde{f}(b),b-\tilde{g}(a))$ we have that
  \begin{align*}
    (A,\xi_Y,\xi_X)   &\eqas (A,Y-\tilde{f}(X),X-\tilde{g}(A)) 
                    =\eta(A,X,Y) 
                    \eqd\eta(\tilde{A},\tilde{X},\tilde{Y}) 
                    = (\tilde{A},\tilde{\xi}_Y,\tilde{\xi}_X),
  \end{align*}
  so $(A,\xi_Y,\xi_X)
  \eqd(\tilde{A},\tilde{\xi}_Y,\tilde{\xi}_X)$. This
  completes step (i).

  Next, we proceed with step (ii). Take an arbitrary intervention
  $i \in \II$ and let $\phi^i, I^i, \tilde{I}^i$ with
  $I^i \eqd\tilde{I}^i$,
  $I^i\independent (\ep_X^i,\ep_Y^i,\ep_H^i,\ep_A^i) \sim Q$ and
  $\tilde{I}^i \independent
  (\tilde{\ep}^i_X,\tilde{\ep}^i_Y,\tilde{\ep}^i_H,\tilde{\ep}^i_A)
  \sim \tilde{Q}$ be such that the structural assignments for $X^i$ and
  $\tilde{X}^i$ in $M(i)$ and $\tilde{M}(i)$, respectively,  are given as
  \begin{align*}
    X^i &:= \phi^i(A^i,g(A^i), h_2(H^i, \epsilon_X^i), I^i), \quad \tilde{X}^i := \phi^i(\tilde{A}^i,\tilde{g}(\tilde{A}^i), \tilde{h}_2(\tilde{H}^i, \tilde{\epsilon}_X^i), \tilde{I}^i).
  \end{align*}
  Define $\xi_X^i := h_2(H^i, \epsilon_X^i)$,
  $\xi_Y^i := h_1(H^i, \epsilon_Y^i)$,
  $\tilde{\xi}_X^i := \tilde{h}_2(\tilde{H}^i, \tilde{\epsilon}_X^i)$
  and
  $\tilde{\xi}_Y^i := \tilde{h}_1(\tilde{H}^i,
  \tilde{\epsilon}_Y^i)$.
  Then, it holds that
  \begin{equation*}
    (A^i, \xi_X^i, \xi_Y^i) \eqd (A,\xi_X, \xi_Y)  \eqd (\tilde{A},  \tilde{\xi}_X, \tilde{\xi}_Y)  \eqd (\tilde{A}^i, \tilde{\xi}_X^i, \tilde{\xi}_Y^i),
  \end{equation*}
  where we used step (i), that $(A^i, \xi_X^i, \xi_Y^i)$ and
  $(A,\xi_X, \xi_Y)$ are generated by identical functions of the noise
  innovations and that $ (\ep_X,\ep_Y,\ep_H,\ep_A) $ and
  $(\ep_X^i,\ep_Y^i,\ep_H^i,\ep_A^i)$ have identical distributions.
  Adding a random variable with the same distribution, that is
  mutually independent with all other variables, on both sides does
  not change the distribution of the bundle, hence
  \begin{align*}
    (A^i, \xi_X^i, \xi_Y^i,I^i)  \eqd (\tilde{A}^i, \tilde{\xi}_X^i, \tilde{\xi}_Y^i, \tilde{I}^i).
  \end{align*}
  Define
  $\kappa(a,b,c,d) :=
  (\phi^i(a,\tilde{g}(a),b,d),\tilde{f}(\phi^i(a,\tilde{g}(a),b,d))+c)$. As
  shown in step (i) above, we have that $g(A^i)\eqas \tilde{g}(A^i)$. Furthermore,
  since $\supp(X^i) \subseteq\supp(X)$ we have that
  $f(X^i) \eqas \tilde{f}(X^i)$, and hence
  \begin{align*}
    (X^i, Y^i) \eqas &\, (X^i,\tilde{f}(X^i)+\xi_Y^i) \\
    =& \,  (\phi^i(A^i,g(A^i), \xi_X^i , I^i) ,  \, \, \tilde{f}(\phi^i(A^i,g(A^i), \xi_X^i, I^i))+\xi_Y^i) \\
    \eqas  &\,(\phi^i(A^i,\tilde{g}(A^i), \xi_X^i , I^i) , \, \, \tilde{f}(\phi^i(A^i,\tilde{g}(A^i), \xi_X^i, I^i))+\xi_Y^i) \\
   =& \, \kappa(A^i, \xi_X^i, \xi_Y^i,I^i)
                \eqd \kappa(\tilde{A}^i, \tilde{\xi}_X^i, \tilde{\xi}_Y^i, \tilde{I}^i)
               = (\tilde{X}^i, \tilde{Y}^i).
  \end{align*}
  Thus, $\bP_{M(i)}^{(X,Y)} = \bP_{\tilde{M}(i)}^{(X,Y)}$, which
  completes step (ii).  Since $i \in \II$ was arbitrary, the result now
  follows from Proposition~\ref{prop:suff_general}.

  \textbf{Case (B):} Assume that the set of interventions $\II$
  contains at least one confounding-removing intervention. Let
  $\tilde{M} =(\tilde{f},\tilde{g},\tilde{h}_1,\tilde{h}_2,\tilde{Q})
  \in \MM$ be such that $\P_{\tilde{M}} = \P_M$. Then, by
  Proposition~\ref{prop:minimax_equal_causal}, it follows that the
  causal function $\tilde{f}$ is a minimax solution w.r.t.\
  $(\tilde{M}, \II)$.  By Assumption~\ref{ass:identify_f}, we further
  have that $\tilde{f}$ and $f$ coincide on
  $\supp(X) \supseteq \supp_{\II}(X)$. Hence, it follows that
  \begin{align*}
  \inf_{\fs \in \FF} \sup_{i \in \II} \E_{\tilde{M}(i)}[(Y - \fs(X))^2]&= \sup_{i \in \II} \E_{\tilde{M}(i)}[(Y - \tilde{f}(X))^2] = \sup_{i \in \II} \E_{\tilde{M}(i)}[(Y - f(X))^2],
  \end{align*}
  showing that also $f$ is a minimax solution w.r.t.\
  $(\tilde{M}, \II)$. This completes the proof of
  Proposition~\ref{prop:genX_intra}.  
\end{proof}

\subsection{Proof of Proposition~\ref{prop:genX_extra}}

\begin{proof}
Let $\tilde{M} \in \MM$ be such that $\P_{\tilde{M}} = \P_M$. By Assumptions~\ref{ass:identify_f}~and~\ref{ass:gen_f}, 
it holds that $f \equiv \tilde{f}$. The proof now proceeds analogously to that of Proposition~\ref{prop:genX_intra}. 
\end{proof}

\subsection{Proof of Proposition~\ref{prop:extrapolation_bounded_deriv_cr}}\label{sec:prop:extrapolation_bounded_deriv_cr}

\begin{proof} 
  By Assumption~\ref{ass:identify_f}, $f$ is identified on
  $\supp^{M}(X)$ by the observational distribution $\bP_{M}$. Let
  $\cI$ be a set of interventions containing at least one
  confounding-removing intervention.
  For any
  $\tilde{M}=(\tilde{f},\tilde{g},\tilde{h}_1,\tilde{h}_2,\tilde{Q})\in
  \cM$, Proposition~\ref{prop:minimax_equal_causal} yields that the
  causal function is a minimax solution. That is,
  \begin{align} \notag
    \inf_{\fs\in\mathcal{F}}\sup_{i\in\cI}\E_{\tilde{M}(i)}\big[(Y-\fs(X))^2\big]
    &= \sup_{i\in\cI}\E_{\tilde{M}(i)}\big[(Y-\tilde{f}(X))^2\big] \\
    &= \sup_{i\in \cI }\E_{\tilde{M}(i)}[\xi_Y^2] =\E_{\tilde{M}}[\xi_Y^2], \label{eq:propboundedderiv_causalfunctionsolvesminimax}
  \end{align}
  where we used that any intervention $i\in \cI$ does not affect the
  distribution of $\xi_Y=\tilde{h}_2(H,\ep_Y)$. Now, assume that
  $\tilde{M}=(\tilde{f},\tilde{g},\tilde{h}_1,\tilde{h}_2,\tilde{Q})\in
  \cM$ satisfies $\bP_{\tilde{M}} = \bP_M$. Since $(\P_M,\cM)$
  satisfies Assumption~\ref{ass:identify_f}, we have that
  $f \equiv \tilde{f}$ on $\supp^M(X)=\supp^{\tilde{M}}(X)$. Let $f^*$
  be any function in $\FF$ such that $f^*=f$ on $\supp^M(X)$.  We
  first show that
  $\norm{\tilde{f} - f^*}_{\cI,\infty} \leq 2\delta K$, where
  $\|f\|_{\cI,\infty} := \sup_{x\in\supp_{\cI}^M(X)}\|f(x)\|$.  By the
  mean value theorem, for all $\fs \in\FF$ it holds that
  $\abs{\fs(x) - \fs(y)} \leq K\norm{x - y}$, for all
  $x, y\in \mathcal D$.  For any $x \in \supp^M_\cI(X)$ and
  $y \in \supp^M(X)$ we have
  \begin{align*}
    \abs[\big]{\tilde{f}(x) - f^*(x)} 
    &=  \abs[\big]{\tilde{f}(x) - \tilde{f}(y) + f^*(y) - f^*(x)}\\
    &\leq  \abs[\big]{\tilde{f}(x) - \tilde{f}(y)} + \abs[\big]{f^*(y) - f^*(x)}\\
    &\leq 2 K \norm{x - y},
  \end{align*}
  where we used the fact that $\tilde{f}(y)= f(y) =f^*(y)$, for all
  $y\in \supp^M(X)$.  In particular, it holds that
  \begin{align}\label{eq:unif_norm}
    \begin{split}
      \norm{\tilde{f} - f^*}_{\cI,\infty} 
      &=\sup_{x\in \supp^M_\cI(X)} \abs[\big]{\tilde{f}(x) - f^*(x)}\\
      &\leq  2K\sup_{x\in \supp^M_\cI(X)} \inf_{y\in\supp^M(X)}\norm{x - y}\\
      &= 2\delta K.
    \end{split}
  \end{align}
  For any $i\in\II$ we have that
  \begin{align} \notag 
    \E_{\tilde{M}(i)}\big[(Y- f^*(X))^2\big] 
    = & \, \E_{\tilde{M}(i)}\big[(\tilde{f}(X)+\xi_Y-f^*(X))^2\big]\nonumber\\
    =&\, \E_{\tilde{M}}\big[\xi_Y^2\big] +
      \E_{\tilde{M}(i)}\big[(\tilde{f}(X)-f^*(X))^2\big]\nonumber\\
      &\ + 2\E_{\tilde{M}(i)}\big[\xi_Y(\tilde{f}(X)-f^*(X))\big].\label{eq:starting_pt_split}
  \end{align}
  Next, we can use Cauchy-Schwarz,
  \eqref{eq:propboundedderiv_causalfunctionsolvesminimax} and
  \eqref{eq:unif_norm} in \eqref{eq:starting_pt_split} to get that
  \begin{align} \nonumber
    \bigg|\sup_{i\in\II}&\, \E_{\tilde{M}(i)}\big[(Y-f^*(X))^2\big]
     -\inf_{\fs\in\FF}\sup_{i\in\II}\E_{\tilde{M}(i)}\big[(Y-\fs(X))^2\big]\nonumber \bigg|\\
    =& \,  \sup_{i\in\II} \E_{\tilde{M}(i)}\big[(Y-f^*(X))^2\big]-\E_{\tilde{M}}[\xi_Y^2]\nonumber \\ \nonumber
     = & \,  \sup_{i\in\II} \big(
      \E_{\tilde{M}(i)}\big[(\tilde{f}(X)-f^*(X))^2\big] +2\E_{\tilde{M}(i)}\big[\xi_Y(\tilde{f}(X)-f^*(X))\big] \big) \nonumber\\
    \leq &\,  4\delta^2K^2+4\delta K \sqrt{\var_M(\xi_Y)},\label{eq:ineq_prt}
  \end{align}
  proving the first statement.  Finally, if $\II$ consists only of
  confounding-removing interventions, then the bound in
  \eqref{eq:ineq_prt} can be improved by using that $\E[\xi_Y]=0$
  together with $H\independent X$. In that case, we get that
  $\E_{\tilde{M}(i)}\big[\xi_Y(\tilde{f}(X)-f(X))\big]=0$ and hence
  the bound becomes $4\delta^2 K^2$. This completes the proof of
  Proposition~\ref{prop:extrapolation_bounded_deriv_cr}.
\end{proof}

\subsection{Proof of Proposition~\ref{prop:extrapolation_bounded_deriv}}\label{sec:prop:extrapolation_bounded_deriv}
\begin{proof}
    By Assumption~\ref{ass:identify_f},
	$f$ is identified on $\supp^{M}(X)$ by the observational
	distribution $\bP_{M}$. Let $\cI$ be a set of confounding-preserving
	interventions.  For a fixed $\epsilon>0$, let $f^*\in\mathcal{F}$ be
	a function satisfying
	\begin{align}
	&\big|\sup_{i\in\cI}\E_{M(i)}\big[(Y-f^*(X))^2)\big]
		 -\inf_{\fs\in\mathcal{F}}\sup_{i\in\cI}\E_{M(i)}\big[(Y-\fs(X))^2)\big] \big| \leq  \epsilon. \label{eq:fstarDiffEpsilon}
	\end{align}
	Fix any secondary model
        $\tilde{M}=(\tilde{f},\tilde{g},\tilde{h}_1,\tilde{h}_2,\tilde{Q})\in
        \cM$ with $\bP_{\tilde{M}} = \bP_M$.  The general idea is to
        derive an upper bound for
        $\sup_{i\in\II}\E_{\tilde{M}(i)}[(Y-f^*(X))^2]$ and a lower
        bound for
        $\inf_{\fs\in\FF}\,
        \sup_{i\in\II}\E_{\tilde{M}(i)}[(Y-\fs(X))^2]$ which will
        allow us to bound the absolute difference of interest.
	
	Since $(\P_M,\cM)$ satisfies
	Assumption~\ref{ass:identify_f}, we have that
	$f \equiv \tilde{f}$ on
	$\supp^M(X)=\supp^{\tilde{M}}(X)$. We first show that
	$\norm{\tilde{f} - f}_{\cI,\infty} \leq 2\delta K$, where $\|f\|_{\cI,\infty} := \sup_{x\in\supp_{\cI}^M(X)}\|f(x)\|$.  By the mean
	value theorem, for all $\fs \in\FF$ it holds that
	$\abs{\fs(x) - \fs(y)} \leq K\norm{x - y}$, for all
	$x, y\in \mathcal D$.  For any $x \in \supp^M_\cI(X)$ and
	$y \in \supp^M(X)$ we have
	\begin{align*}
	\abs[\big]{\tilde{f}(x) - f(x)} 
	&=  \abs[\big]{\tilde{f}(x) - \tilde{f}(y) + f(y) - f(x)}\\
	&\leq  \abs[\big]{\tilde{f}(x) - \tilde{f}(y)} + \abs[\big]{f(y) - f(x)}\\
	&\leq 2 K \norm{x - y},
	\end{align*}
	where we used the fact that $\tilde{f}(y)=f(y)$, for all
	$y\in \supp_M(X)$.  In particular, it holds that
	\begin{align}\label{eq:unif_norm2}
	\begin{split}
	\norm{\tilde{f} - f}_{\cI,\infty} 
	&= \sup_{x\in \supp^M_\cI(X)} \abs[\big]{\tilde{f}(x) - f(x)}\\
	&\leq  2K\sup_{x\in \supp^M_\cI(X)} \inf_{y\in\supp^M(X)}\norm{x - y}\\
	&= 2\delta K.
	\end{split}
	\end{align}
	Let now $i\in\II$ be fixed. The term $\xi_Y = h_1(H, \epsilon_Y)$ is
	not affected by the intervention $i$. Furthermore,
	$\P^{(X,\xi_Y)}_{M(i)}=\P^{(X,\xi_Y)}_{\tilde{M}(i)}$ since $i$ is
	confounding-preserving (this can be seen by a slight modification to the
	arguments from case (A) in the proof of
	Proposition~\ref{prop:genX_intra}). Thus, for any $\fs\in \cF$ we
	have that
	\begin{align}
	&\E_{\tilde{M}(i)}\big[(Y-\fs(X))^2\big] \nonumber\\
	&=\E_{\tilde{M}(i)}\big[(\tilde{f}(X)+\xi_Y-\fs(X)+f(X)-f(X))^2\big]\nonumber\\
	&=\E_{\tilde{M}(i)}\big[\xi_Y^2\big] +
	\E_{\tilde{M}(i)}\big[(f(X)-\fs(X))^2\big]
	+
	\E_{\tilde{M}(i)}\big[(\tilde{f}(X)-f(X))^2\big]\nonumber\\
	&\qquad\qquad + 2\E_{\tilde{M}(i)}\big[\xi_Y(f(X)-\fs(X))\big]\nonumber\\
	&\qquad\qquad +
	2\E_{\tilde{M}(i)}\big[(\tilde{f}(X)-f(X))(f(X)-\fs(X))\big]\nonumber\\
	&\qquad\qquad +
	2\E_{\tilde{M}(i)}\big[\xi_Y(\tilde{f}(X)-f(X))\big]\nonumber\\
	&=\E_{M(i)}\big[\xi_Y^2\big] +
	\E_{M(i)}\big[(f(X)-\fs(X))^2\big]+
	\E_{M(i)}\big[(\tilde{f}(X)-f(X))^2\big]\nonumber\\
	&\qquad\qquad +2\E_{M(i)}\big[\xi_Y(f(X)-\fs(X))\big]\nonumber\\
	&\qquad\qquad +
	2\E_{M(i)}\big[(\tilde{f}(X)-f(X))(f(X)-\fs(X))\big]\nonumber\\
	&\qquad\qquad + 2\E_{M(i)}\big[\xi_Y(\tilde{f}(X)-f(X))\big]\nonumber\\
	&=\E_{M(i)}\big[(Y-\fs(X))^2\big] + L_1^i(\tilde{f}) +
	L_2^i(\tilde{f},\fs) + L_3^i(\tilde{f}),\label{eq:decomposition_intoLs}
	\end{align}
	where, we have made the following definitions
	\begin{align*}
	L_1^i(\tilde{f})&\coloneqq \E_{M(i)}\big[(\tilde{f}(X)-f(X))^2\big],\\
	L_2^i(\tilde{f},\fs)&\coloneqq
	2\E_{M(i)}\big[(\tilde{f}(X)-f(X))(f(X)-\fs(X))\big],\\
	L_3^i(\tilde{f})&\coloneqq 2\E_{M(i)}\big[\xi_Y(\tilde{f}(X)-f(X))\big].
	\end{align*}
	Using \eqref{eq:unif_norm2} it follows that
	\begin{equation}
	\label{eq:estimate_L1}
	0\leq L_1^i(\tilde{f}) \leq 4\delta^2 K^2,
	\end{equation}
	and by the Cauchy-Schwarz inequality it follows that
	\begin{align} 
	\abs[\big]{L_3^i(\tilde{f})} &\leq 2 \sqrt{\var_M(\xi_Y)4\delta^2K^2} =4\delta K \sqrt{\var_M(\xi_Y)}.	\label{eq:estimate_L3}
	\end{align}
	Let now $\fs\in\FF$ be 
any function
	such that
	\begin{equation} \label{eq:ConditionBetterThanTildeCausal}
	\sup_{i\in\II}\E_{\tilde M(i)}\big[(Y-\fs(X))^2)\big]\leq\sup_{i\in\II}\E_{\tilde M(i)}\big[(Y-\tilde{f}(X))^2)\big],
	\end{equation}
	then by \eqref{eq:unif_norm2}, the Cauchy-Schwarz inequality and
	Proposition~\ref{prop:difference_to_causal_function}, it holds for all $i\in \cI$ that
	\begin{align} \nonumber
	L_2^i(\tilde{f},\fs) =& \, 2\E_{M(i)}\big[(\tilde{f}(X)-f(X))(f(X)-\fs(X))\big]\\ 
	= & \, 2\E_{\tilde{M}(i)}\big[(\tilde{f}(X)-f(X))(f(X)-\fs(X))\big]\nonumber\\
	= & \,  -2\E_{\tilde{M}(i)}\big[(\tilde{f}(X)-f(X))^2\big]
	+2\E_{\tilde{M}(i)}\big[(\tilde{f}(X)-f(X))(\tilde{f}(X)-\fs(X))\big]\nonumber\\
	\geq &\, -8\delta^2K^2 -
	2\sqrt{4\delta^2K^2}\sqrt{4\var_M(\xi_Y)}\nonumber\\
	= & \, -8\delta^2K^2- 8\delta K \sqrt{\var_M(\xi_Y)}, \label{eq:estimate_L2_BetterThanTildeCausal}
	\end{align}
	where, in the third equality, we have added and subtracted the
        term
        $2\E_{\tilde{M}(i)}\big[(\tilde{f}(X)-f(X))\tilde{f}(X)\big]$.
        Now let
        $\cS := \{\fs \in \cF :
        \sup_{i\in\II}\E_{\tilde{M}(i)}\big[(Y-\fs(X))^2\big] \leq
        \sup_{i\in\II}\E_{\tilde{M}(i)}\big[(Y-\tilde{f}(X))^2\big]
        \}$ be the set of all functions satisfying
        \eqref{eq:ConditionBetterThanTildeCausal}. Due to
        \eqref{eq:decomposition_intoLs}, \eqref{eq:estimate_L1},
        \eqref{eq:estimate_L3} and
        \eqref{eq:estimate_L2_BetterThanTildeCausal} we have the
        following lower bound of interest
	\begin{align} \notag
	&\inf_{\fs\in\FF}\,	\sup_{i\in\II}\E_{\tilde{M}(i)}\big[(Y-\fs(X))^2\big] \\\notag
	=& \inf_{\fs\in\cS}\,	\sup_{i\in\II}\E_{\tilde{M}(i)}\big[(Y-\fs(X))^2\big] \\\notag
	=& \inf_{\fs\in \cS}\,
	\sup_{i\in\II}\big\{ \E_{M(i)}\big[(Y-\fs(X))^2\big] + L_1^i(\tilde{f}) +
	L_2^i(\tilde{f},\fs) + L_3^i(\tilde{f}) \big\} \\\notag
	\geq& \inf_{\fs\in \cS}\,
	\sup_{i\in\II} \E_{M(i)}\big[(Y-\fs(X))^2\big]  -8\delta^2K^2- 8\delta K \sqrt{\var_M(\xi_Y)} - 4\delta K \sqrt{\var_M(\xi_Y)} \\ \label{eq:lowerboundoninf}
	\geq & \inf_{\fs\in \cF}\,
	\sup_{i\in\II} \E_{M(i)}\big[(Y-\fs(X))^2\big]  
	 - 8 \delta^2 K^2 - 12 \delta K \sqrt{\var_M(\xi_Y)}.
	\end{align}
	Next, we construct the aforementioned upper bound of interest. To that end, note that 
	\begin{align} \notag
	&\sup_{i\in\II} \E_{\tilde{M}(i)}\big[(Y-f^*(X))^2\big] \\ \label{eq:QuantityForUpperBound}
	&\quad = \sup_{i\in\II} \left\{ \E_{M(i)}\big[(Y-f^*(X))^2\big]
	+ L_1^i(\tilde{f}) +
	L_2^i(\tilde{f},f^*) + L_3^i(\tilde{f}) \right\},
	\end{align}	
	by \eqref{eq:decomposition_intoLs}. We have already
        established upper bounds for $L_1^i(\tilde f)$ and
        $L_3^i(\tilde{f})$ in \eqref{eq:estimate_L1} and
        \eqref{eq:estimate_L3}, respectively. In order to control
        $L_2^i(\tilde{f},f^*)$ we introduce an auxiliary function. Let
        $\bar{f}^*\in \cF$ satisfy
	\begin{equation} \label{eq:BarfStarBetterThanCausal}
	\sup_{i\in\II}\E_{M(i)}\big[(Y-\bar{f}^*(X))^2)\big]\leq\sup_{i\in\II}\E_{M(i)}\big[(Y-f(X))^2)\big],
	\end{equation}
	and
	\begin{align}
	&\bigg|\sup_{i\in\II}\E_{M(i)}\big[(Y-\bar{f}^*(X))^2\big] - \inf_{\fs\in\FF}\,
		\sup_{i\in\II}\E_{M(i)}\big[(Y-\fs(X))^2\big]\bigg|\leq \epsilon. \label{eq:epsilonineqforBarfStar}
	\end{align}
	Choosing such a $\bar f^*\in \cF$ is always possible. If $f$
        is an $\ep$-minimax solution, i.e., it satisfies
        \eqref{eq:epsilonineqforBarfStar}, then choose $\bar
        f^*=f$. Otherwise, if $f$ is not a $\ep$-minimax solution,
        then choose any $\bar f^*\in \cF$ that is an $\ep$-minimax
        solution (which is always possible). In this case we have that
	\begin{align*}
	&\sup_{i\in\II}\E_{M(i)}\big[(Y-\bar{f}^*(X))^2\big] - \inf_{\fs\in\FF}\,
	\sup_{i\in\II}\E_{M(i)}\big[(Y-\fs(X))^2\big] \leq \ep ,
	\end{align*}
	and
	\begin{align*}
	&\sup_{i\in\II}\E_{M(i)}\big[(Y-f(X))^2\big]
	  - \inf_{\fs\in\FF}\,
	\sup_{i\in\II}\E_{M(i)}\big[(Y-\fs(X))^2\big] \geq \ep ,
	\end{align*}
	which implies that \eqref{eq:BarfStarBetterThanCausal} is
        satisfied. We can now construct an upper bound on
        $L_2^i(\tilde{f},f^*)$ in terms of $L_2^i(\tilde{f},\bar f^*)$
        by noting that for all $i\in\cI$
	\begin{align} \notag
	\abs[\big]{L_2^i(\tilde{f},f^*)}
	= & \,  2\big|\E_{M(i)}\big[(\tilde{f}(X)-f(X))(f(X)-f^*(X))\big] \big| \notag \\
	\notag
	\leq & \,  2\big|\E_{M(i)}\big[(\tilde{f}(X)-f(X))(f(X)-\bar{f}^*(X))\big]\big| \\ \notag
	& \,  +2\E_{M(i)}\big|(\tilde{f}(X)-f(X))(\bar{f}^*(X)-f^*(X))\big| \\ \notag
	= & \, \abs[\big]{L_2^i(\tilde{f},\bar f^*)} +2\E_{M(i)}\big|(\tilde{f}(X)-f(X))(\bar{f}^*(X)-f^*(X))\big| \\  \notag
	\leq& \, 2\sqrt{\E_{M(i)}\left[(\tilde{f}(X)-f(X))^2\right]\E_{M(i)}\left[(\bar{f}^*(X)-f^*(X))^2 \right]} +  \abs[\big]{L_2^i(\tilde{f},\bar f^*)} \\ \label{eq:UpperboundOfL2TildefStarf}
	\leq & \,  \abs[\big]{L_2^i(\tilde{f},\bar f^*)} +4\delta K\sqrt{\E_{M(i)}\left[(\bar{f}^*(X)-f^*(X))^2 \right]},
	\end{align}
	where we used the triangle inequality, Cauchy-Schwarz inequality and  \eqref{eq:unif_norm2}. Furthermore, 
	\eqref{eq:unif_norm2} and \eqref{eq:BarfStarBetterThanCausal} together with Proposition~\ref{prop:difference_to_causal_function} yield the following bound
	\begin{align}
	|L_2^i(\tilde{f},\bar f^*)|
	= &\, 2\big|\E_{M(i)}\big[(\tilde{f}(X)-f(X))(f(X)-\bar f^*(X))\big] \big|\nonumber\\
	=&\, 2 \sqrt{\E_{M(i)}\big[(\tilde{f}(X)-f(X))^2\big] \E_{M(i)}\big[(f(X)-\bar f^*(X))^2\big]}\nonumber\\
	\leq &\,  2\sqrt{4\delta^2K^2} \sqrt{4\var_M(\xi_Y)}\nonumber \\
	=  &\, 8\delta K \sqrt{\var_M(\xi_Y)}, \label{eq:estimate_L2}
	\end{align}
	for any $i\in\cI$. Thus, it suffices to construct an upper
        bound on the second term in the final expression in
        \eqref{eq:UpperboundOfL2TildefStarf}. Direct computation leads
        to
	\begin{align*}
	\E_{M(i)}\big[(Y-f^*(X))^2\big]   = &\, \E_{M(i)}\big[(Y-\bar{f}^*(X))^2\big] \\
	&\,  +\E_{M(i)}\big[(\bar{f}^*(X)-f^*(X))^2\big] \\
	&\,   +2\E_{M(i)}\big[(Y-\bar{f}^*(X))(\bar{f}^*(X)-f^*(X))\big].
	\end{align*}
	Rearranging the terms and applying the triangle inequality and Cauchy-Schwarz results in
	\begin{align*}
	\E_{M(i)}&\, \big[(\bar{f}^*(X)-f^*(X))^2\big] \\
	 =&\,\E_{M(i)}\big[(Y-f^*(X))^2\big]-\E_{M(i)}\big[(Y-\bar{f}^*(X))^2\big]\\
	& \,  -2\E_{M(i)}\big[(Y-\bar{f}^*(X))(\bar{f}^*(X)-f^*(X))\big]\\
	\leq &\,\big|\E_{M(i)}\big[(Y-f^*(X))^2\big]- \inf_{\fs\in\FF}\,
	\sup_{i\in\II}\E_{M(i)}\big[(Y-\fs(X))^2\big]\big| \\
	& \, + \big| \inf_{\fs\in\FF}\,
	\sup_{i\in\II}\E_{M(i)}\big[(Y-\fs(X))^2\big] -\E_{M(i)}\big[(Y-\bar{f}^*(X))^2\big] \big| \\
	&  \, + 2\E_{M(i)}\big|(Y-\bar{f}^*(X))(\bar{f}^*(X)-f^*(X))\big|\\
	\leq&\, 2\epsilon
	 +2\sqrt{\E_{M(i)}\big[(Y-\bar{f}^*(X))^2\big]}\sqrt{\E_{M(i)}\big[(\bar{f}^*(X)-f^*(X))^2\big]}\\
	\leq &\, 2\epsilon+2\sqrt{\var_{M}(\xi_Y)}\sqrt{\E_{M(i)}\big[(\bar{f}^*(X)-f^*(X))^2\big]},
	\end{align*}
	for any $i\in \cI$. Here, we used that both $f^*$ and $\bar f^*$ are $\ep$-minimax solutions with respect to $M$ and that $\bar f^*$ satisfies \eqref{eq:BarfStarBetterThanCausal} which implies that
	\begin{align*}
	&\, \E_{M(i)}\big[(Y-\bar{f}^*(X))^2)\big]
	\leq\sup_{i\in\II}\E_{M(i)}\big[(Y-f(X))^2)\big] =  \sup_{i\in\II} \E_{M(i)}\big[\xi_Y^2\big] = \var_{M}(\xi_Y),
	\end{align*}
	for any $i\in \cI$, as $\xi_Y$ is unaffected by an intervention on $X$.
	Thus, $\E_{M(i)}\big[(\bar{f}^*(X)-f^*(X))^2\big]$ must satisfy $\ell(\E_{M(i)}\big[(\bar{f}^*(X)-f^*(X))^2\big])\leq 0$, where $\ell:[0,\infty)\to \R$ is given by $\ell(z)=z-2\ep-2\sqrt{\var_{M}(\xi_Y)} \sqrt{z}$.
	The linear term of $\ell$ grows faster than the square
	root term, so the largest allowed value of $\E_{M(i)}\big[(\bar{f}^*(X)-f^*(X))^2\big]$  coincides with the largest root of $\ell(z)$. The largest root is given by
	\begin{align*}
	C^2:=2\ep +2\var_{M}(\xi_Y) + 2\sqrt{\var_{M}(\xi_Y)^2+2\ep\var_{M}(\xi_Y)},
	\end{align*}
	where $(\cdot)^2$ refers to the square of $C$. Hence, for any
        $i\in \cI$ it holds that
	\begin{equation} \label{eq:LastFactorUpperBound}
	\E_{M(i)}\big[(\bar{f}^*(X)-f^*(X))^2\big]\leq C^2.
	\end{equation}
	Hence by \eqref{eq:UpperboundOfL2TildefStarf}, \eqref{eq:estimate_L2} and  \eqref{eq:LastFactorUpperBound} we have that the following upper bound is valid for any $i\in \cI$.
	\begin{align} 
	\abs[\big]{L_2^i(\tilde{f},f^*)} &\leq  8\delta K \sqrt{\var_M(\xi_Y)} + 4\delta K C \label{eq:estimate_L2_fstar}.
	\end{align}
	Thus, using \eqref{eq:QuantityForUpperBound} with
	\eqref{eq:estimate_L1}, \eqref{eq:estimate_L3} and
	\eqref{eq:estimate_L2_fstar}, we get the following upper bound
	\begin{align} \notag
	&\sup_{i\in\II} \E_{\tilde{M}(i)}\big[(Y-f^*(X))^2\big] \\
	&\quad  \leq \sup_{i\in\II}\E_{M(i)}\big[(Y-f^*(X))^2\big] + 4\delta^2K^2+4\delta K C+ 12\delta K \sqrt{\var_M(\xi_Y)}.\label{eq:upperboundsup}
	\end{align}	
	Finally, by combining the bounds \eqref{eq:lowerboundoninf}
        and \eqref{eq:upperboundsup} together with
        \eqref{eq:fstarDiffEpsilon} we get that
	\begin{align}
	\bigg|\sup_{i\in\II}&\,\E_{\tilde{M}(i)} \big[(Y-f^*(X))^2\big]
	 	-
		\inf_{\fs\in\FF}\,
		\sup_{i\in\II}\E_{\tilde{M}(i)}\big[(Y-\fs(X))^2\big] \bigg|\nonumber\\ \nonumber	
	\leq &\, \sup_{i\in\II}\E_{M(i)}\big[(Y-f^*(X))^2\big] - 
	\inf_{\fs\in\FF}\,
	\sup_{i\in\II}\E_{M(i)}\big[(Y-\fs(X))^2\big]\nonumber\\ \nonumber
	&\, +  4\delta^2K^2+4\delta K C+ 12\delta K \sqrt{\var_M(\xi_Y)}\\ \nonumber
	& \,  + 8 \delta^2 K^2 + 12 \delta K \sqrt{\var_M(\xi_Y)} \\
	  \leq &\,  \ep + 12 \delta^2 K^2 + 24 \delta K \sqrt{\var_M(\xi_Y)} + 4 \delta K C. \label{eq:CorrectUpperBound}
	\end{align}
	Using that all terms are positive, we get that
		\begin{align*}
	C &= \sqrt{\var_{M}(\xi_Y)} + \sqrt{\var_{M}(\xi_Y) + 2 \ep}
	\leq 2\sqrt{\var_{M}(\xi_Y)} + \sqrt{2 \ep}
	\end{align*}
	Hence, \eqref{eq:CorrectUpperBound} is bounded above by
	\begin{align*}
          \ep + 12 \delta^2 K^2 + 32 \delta K \sqrt{\var_M(\xi_Y)} +
            4 \sqrt{2} \delta K \sqrt{\ep}.
	\end{align*}
         This
        completes the proof of
        Proposition~\ref{prop:extrapolation_bounded_deriv}.
\end{proof}

\subsection{Proof of Proposition~\ref{prop:impossibility_extrapolation}}

\begin{proof}
  Let $\bar{f}\in \FF$ and $c > 0$. By assumption, 
  $\II$ is a well-behaved set of
  support-extending interventions on $X$.
  Since $\supp_{\II}^{M}(X) \setminus \supp^{M}(X)$ has non-empty
  interior, there exists an intervention $i_0 \in \II$ and $\ep>0$
  such that $\P_{M(i_0)}(X\in B) \geq \epsilon$, 
  for some open subset
  $B \subsetneq \bar{B}$,   such that $\mathrm{dist}(B,\R^d \setminus \bar{B})>0$, where $\bar{B}:=\supp_{\II}^{M}(X) \setminus \supp^{M}(X)$.
  Let $\tilde{f}$ be any continuous 
  function satisfying that, for all $x \in B\cup(\R^d\setminus\bar B)$,
  \begin{align*}
    \tilde{f}(x) =
    \begin{cases}
      \bar{f}(x) + \gamma, \quad & x \in B\\
      f(x), \quad & x \in \R^d \setminus \bar{B},
    \end{cases}
  \end{align*}
  where
  $\gamma := \epsilon^{-1/2} \left\{(2 \E_{\tilde{M}}[\xi_{Y}^2] +
    c)^{1/2} + (\E_{\tilde{M}}[\xi_{Y}^2])^{1/2}\right\}$.
  
  Consider a secondary model
  $\tilde{M} = (\tilde{f}, g, h_1, h_2, Q)\in\cM$. Then, by
  Assumption~\ref{ass:identify_f}, it holds that
  $\P_{M} = \P_{\tilde{M}}$.  Since $\II$ only consists of
  interventions on $X$, it holds that
  $\P_{M(i_0)}(X\in B) = \P_{\tilde{M}(i_0)}(X\in B)$ (this
  holds since all components of $\tilde{M}$ and $M$ are equal, except for the
  function $f$, which is not allowed to enter in the intervention on
  $X$). Therefore,
  \begin{align}\label{eq:proof_imp_a}
  \E_{\tilde{M}(i_0)}\big[(Y-\bar{f}(X))^2\big]
  &\geq   \E_{\tilde{M}(i_0)}\big[(Y-\bar{f}(X))^2 \mathbbm{1}_{B}(X)\big] \nonumber\\
  &= \E_{\tilde{M}(i_0)}\big[(\gamma + \xi_Y)^2 \mathbbm{1}_{B}(X)\big] \nonumber\\
  &\geq   \gamma^2\epsilon + 2 \gamma \E_{\tilde{M}(i_0)}\big[\xi_Y
  \mathbbm{1}_{B}(X) \big]\nonumber\\
  &\geq  \gamma^2\epsilon - 2 \gamma \left(\E_{\tilde{M}}\big[\xi_Y^2\big]
  \epsilon\right)^{1/2} \nonumber\\
  &=  c + \E_{\tilde{M}}[\xi_{Y}^2],
  \end{align}
where the third inequality follows from Cauchy–Schwarz.
  Further, by the definition of the infimum it holds that
  \begin{align}\label{eq:proof_imp_b}
  \begin{split}
  &\inf_{\fs \in \FF}\,\sup_{i\in\II} \E_{\tilde{M}(i)}\big[(Y-\fs(X))^2\big]
  \leq  \sup_{i\in\II}\E_{\tilde{M}(i)}\big[(Y-\tilde{f}(X))^2\big] 
  = \E_{\tilde{M}}[\xi_{Y}^2].
  \end{split}
  \end{align}
Therefore, combining~\eqref{eq:proof_imp_a} and~\eqref{eq:proof_imp_b}, the 
claim follows.
\end{proof}

\subsection{Proof of Proposition~\ref{prop:genA}}

\begin{proof}
  We prove the result by showing that under
  Assumption~\ref{ass:identify_g} it is possible to express
  interventions on $A$ as confounding-preserving interventions on $X$
  and applying
  Propositions~\ref{prop:genX_intra}~and~\ref{prop:genX_extra}.  To
  avoid confusion, we will throughout this proof denote the true model
  by $M^0 = (f^0, g^0, h_1^0, h_2^0, Q^0)$.   Fix an intervention
  $i\in\II$. Since it is an intervention on $A$, there exist $\psi^i$
  and $I^i$ such that for any $M = (f,g,h_1, h_2, Q) \in \MM$, the
  intervened SCM $M(i)$ is of the form
  \begin{align*}
    A^i &:= \psi^i(I^i, \ep_A^i),  \quad H^i := \ep_H^i,
     \quad X^i := g(A^i) + h_2(H^i, \ep_X^i),  \quad Y^i  := f(X^i) + h_1(H^i,\ep_Y^i),
  \end{align*}
  where $(\ep^i_X, \ep^i_Y, \ep^i_A, \ep^i_H) \sim Q$.  We now define
  a confounding-preserving intervention $j$ on $X$, such that, for all
  models $\tilde{M}$ with $\P_{\tilde{M}} = \P_M$, the distribution of
  $(X,Y)$ under $\tilde{M}(j)$ coincides with that under
  $\tilde{M}(i)$. To that end, define the intervention function
  \begin{equation*}
    \bar{\psi}^j(h_2, A^j, H^j, \ep^j_X ,I^j) \coloneqq g^0(\psi^i(I^j, A^j)) +
    h_2(H^j, \ep_X^j),
  \end{equation*}
  where $g^0$ is the fixed function corresponding to model $M$, and
  therefore not an argument of $\bar{\psi}^j$.    Let now $j$ be
  the intervention on $X$ satisfying that, for all
  $M = (f,g,h_1, h_2, Q) \in \MM$, the intervened model $M(j)$ is
  given as
  \begin{align*}
    &A^{j} := \ep_A^{j}, \quad  H^{j} := \ep_H^{j},
     \quad X^{j} := \bar{\psi}^j(h_2, A^{j}, H^{j}, \ep^{j}_X ,I^{j}), \quad Y^{j}:= f(X^{j}) + h_1(H^{j},\ep_Y^{j}),
  \end{align*}
  where $(\ep^j_X, \ep^j_Y, \ep^j_A, \ep^j_H) \sim Q$ and where $I^j$
  is chosen such that $I^j\eqd I^i$.  By definition, $j$ is
  a confounding-preserving intervention. Let now
  $\tilde{M} = (\tilde{f}, \tilde{g}, \tilde{h}_1, \tilde{h}_2,
  \tilde{Q})$ be such that $\P_{\tilde{M}} = \P_M$, and let
  $(\tilde{X}^i, \tilde{Y}^i)$ and $(\tilde{X}^j, \tilde{Y}^j)$ be
  generated under $\tilde{M}(i)$ and $\tilde{M}(j)$, respectively. By
  Assumption~\ref{ass:identify_g}, it holds for all
  $a \in \supp(A) \cup \supp_{\II}(A)$ that $\tilde{g}(a) = g^0(a)$.
  Hence, we get that
  \begin{align*}
    (\tilde{X}^i, \tilde{Y}^i)
    \eqd &
      (\tilde{g}(\psi^i(I^i, \tilde{\ep}_A^i)) + \tilde{h}_2(\tilde{\ep}^i_H, \tilde{\ep}_X^i),
      \tilde{f}(\tilde{g}(\psi^i(I^i, \tilde{\ep}_A^i)) + \tilde{h}_2(\tilde{\ep}^i_H, \tilde{\ep}_X^i)) + \tilde{h}_1(\tilde{\ep}_H^i, \tilde{\ep}^i_Y)) \\
    = &
      (g^0(\psi^i(I^i, \tilde{\ep}_A^i)) + \tilde{h}_2(\tilde{\ep}^i_H, \tilde{\ep}_X^i),
      \tilde{f}(g^0(\psi^i(I^i, \tilde{\ep}_A^i)) + \tilde{h}_2(\tilde{\ep}^i_H, \tilde{\ep}_X^i)) + \tilde{h}_1(\tilde{\ep}_H^i, \tilde{\ep}^i_Y)) \\
    \eqd &
      (g^0(\psi^i(I^j, \tilde{\ep}_A^j)) + \tilde{h}_2(\tilde{\ep}^j_H, \tilde{\ep}_X^j),
      \tilde{f}(g^0(\psi^i(I^j, \tilde{\ep}_A^j)) + \tilde{h}_2(\tilde{\ep}^j_H, \tilde{\ep}_X^j)) + \tilde{h}_1(\tilde{\ep}_H^j, \tilde{\ep}^j_Y)) \\
    \eqd &
      (\bar{\psi}^j(\tilde{h}_2, \tilde{\ep}_A^j, \tilde{\ep}_H^j, \tilde{\ep}^j_X ,I^j), 
      \tilde{f}(\bar{\psi}^j(\tilde{h}_2, \tilde{\ep}_A^j, \tilde{\ep}_H^j, \tilde{\ep}^j_X ,I^j)) + \tilde{h}_1(\tilde{\ep}_H^j, \tilde{\ep}^j_Y)) \\
    \eqd &
      (\tilde{X}^j, \tilde{Y}^j),
  \end{align*}
  as desired. Since $i \in \II$ was arbitrary, we have now shown that
  there exists a mapping $\pi$ from $\II$ into a set $\mathcal{J}$ of
  confounding-preserving (and hence a well-behaved set) of
  interventions on $X$, such that for all $\tilde{M}$ with
  $\P_{\tilde{M}} = \P_M$,
  $\P^{(X,Y)}_{\tilde{M}(i)} = \P_{\tilde{M}(\pi(i))}^{(X,Y)}$. Hence,
  we can rewrite Equation~\eqref{eq:def_generalization} in
  Definition~\ref{defi:general} in terms of the set $\mathcal{J}$. The
  result now follows from
  Propositions~\ref{prop:genX_intra}~and~\ref{prop:genX_extra}.
\end{proof}

\subsection{Proof of Proposition~\ref{prop:impossibility_intA}}
  
\begin{proof}
  Let $b \in \R^d$ be such that $f(x) = b^\top x$ for all
  $x \in \R^d$. We start by characterizing the error
  $\E_{\tilde{M}(i)}\big[(Y-\fs(X))^2\big]$. %
  Let us consider models of the form
  $\tilde{M} = (f, \tilde{g}, h_1, h_2, Q) \in \MM$ for some function
  $\tilde{g} \in \GG$ with $\tilde{g}(a) = g(a)$ for all
  $a \in \supp_M(A)$. Clearly, any such model satisfies that
  $\P_{\tilde{M}} = \P_M$. For every $a \in \mathcal{A}$, let
  $i_a \in \II$ denote the corresponding hard intervention on $A$.
For every $a \in \mathcal{A}$ and $\bes \in \R^d$, we then have
\begin{equation} \label{eq:linear_mse}
\begin{aligned}
&\ \E_{\tilde{M}(i_a)}\big[(Y - \bes^\top X)^2\big]  \\
= &\ \E_{\tilde{M}(i_a)}\big[(b^\top X + \xi_Y - \bes^\top X)^2\big] \\
																	= &\ (b - \bes)^\top \E_{\tilde{M}(i_a)}[X X^\top] (b - \bes) 
																	+ 2(b - \bes)^\top \E_{\tilde{M}(i_a)}[X \xi_Y] + \E_{\tilde{M}(i_a)}\big[\xi_Y^2] \\
= &\ (b - \bes)^\top \underbrace{(\tilde{g}(a) \tilde{g}(a)^\top + \E_{M}[\xi_X \xi_X^\top])}_{=: K_{\tilde{M}}(a)} (b - \bes) + 2(b - \bes)^\top \E_{M}[\xi_X \xi_Y] + \E_{M}\big[\xi_Y^2],
\end{aligned}
\end{equation}
where we have used that, under $i_a$, the distribution of 
$(\xi_X, \xi_Y)$
is unaffected.
We now show that, for any $\tilde{M}$ with the above form, the causal
function $f$ does not minimize the worst-case 
risk
across interventions
in $\II$.  The idea is to show that the worst-case risk~\eqref{eq:linear_mse}
strictly decreases at
$\bes = b$ in the direction
$u := \E_{M}[\xi_X \xi_Y] / \norm{\E_{M}[\xi_X \xi_Y]}_2$.  For every
$a \in \mathcal{A}$ and $s \in \R$, define
\begin{align*}
\ell_{\tilde{M},a}(s) :&= \E_{\tilde{M}(i_a)}\big[(Y-(b + s u)^\top X)^2\big]
= u^\top K_{\tilde{M}}(a) u \cdot s^2  - 2 u^\top \E_{M}[\xi_X \xi_Y] \cdot s + \E_{M}\big[\xi_Y^2].
\end{align*}
For every $a$,
$\ell_{\tilde{M},a}^\prime (0) = -2 \norm{\E_{M}[\xi_X \xi_Y]}_2 < 0$,
showing that $\ell_{\tilde{M},a}$ is strictly decreasing at $s=0$ 
(with a derivative that is bounded away from 0 across all $a \in \mathcal{A}$).
By boundedness of $\mathcal{A}$ and by the continuity of $a \mapsto \ell_{\tilde{M},a}^{\prime \prime } (0)  = 2 u^\top K_{\tilde{M}}(a) u$, 
it further follows that $\sup_{a \in \mathcal{A}} \card{\ell^{\prime \prime}_{\tilde{M},a} (0)} < \infty$.
Hence, we can find $s_0 > 0$ such that for all $a \in \mathcal{A}$, $\ell_{\tilde{M}, a}(0) > \ell_{\tilde{M}, a}(s_0)$.  
It now follows by continuity of $(a, s) \mapsto \ell_{\tilde{M},a}(s)$ that
\begin{align*}
&\sup_{i \in \II} \E_{\tilde{M}(i)}\big[(Y-b^\top X)^2\big] = \sup_{a \in \mathcal{A}} \ell_{\tilde{M}, a}(0) > \sup_{a \in \mathcal{A}} \ell_{\tilde{M}, a}(s_0) = \sup_{i \in \II} \E_{\tilde{M}(i)}\big[(Y-(b+s_0 u)^\top X)^2\big],
\end{align*}
showing that $b+s_0 u$ attains a lower worst-case risk than $b$. 

We now show that all functions other than $f$ may result in an
arbitrarily large error.  Let $\bar{b} \in \R^d \setminus \{ b \}$ be
given, and let $j \in \{1, \dots, d\}$ be such that
$b_j \neq \bar{b}_j$.  The idea is to construct a function
$\tilde{g} \in \GG$ such that, under the corresponding model
$\tilde{M} = (f, \tilde{g}, h_1, h_2, Q) \in \MM$, some hard
interventions on $A$ result in strong shifts of the $j$th coordinate
of $X$.  Let $a \in \mathcal{A}$. Let $e_j \in \R^d$ denote the $j$th unit vector, and assume
that $\tilde{g}(a) = n e_j$ for some $n \in \N$.  Using
\eqref{eq:linear_mse}, it follows that
\begin{align*}
&\E_{\tilde{M}(i_a)}\big[(Y-\bar{b}^\top X)^2\big] \\
& = n^2 (\bar{b}_j - b_j)^2 + (\bar{b} - b)^\top \E_{M}[\xi_X \xi_X^\top] (\bar{b} - b)  + 
2
(\bar{b} - b)^\top \E_{M}[\xi_X \xi_Y] + \E_{M}\big[\xi_Y^2].
\end{align*}
By letting $n \to \infty$, we see that the above error may become
arbitrarily large. Given any $c > 0$, we can therefore construct
$\tilde{g}$ such that
$\E_{\tilde{M}(i_a)}\big[(Y-\bar{b}^\top X)^2\big] \geq c
+\E_{M}\big[\xi_Y^2]$.  By 
carefully
choosing
$a \in \text{int}(\mathcal{A} \setminus \supp_M(A))$, this can be done
such that $\tilde{g}$ is continuous and $\tilde{g}(a) = g(a)$
  for all $a \in \supp_M(A)$, ensuring that $\bP_{\tilde M}= \bP_M$. It follows that
\begin{align*}
c 	&\leq \E_{\tilde{M}(i_a)}\big[(Y-\beb^\top X)^2\big]  - \E_{M}\big[\xi_Y^2] \\
	&= \E_{\tilde{M}(i_a)}\big[(Y-\beb^\top X)^2\big]  -  \sup_{i \in \II}\E_{\tilde{M}(i)} \big[(Y- b^\top X)^2] \\
	&\leq \E_{\tilde{M}(i_a)}\big[(Y-\beb^\top X)^2\big]  - \inf_{\bes \in \R^d} \sup_{i \in \II}\E_{\tilde{M}(i)} \big[(Y-\bes^\top X)^2]  \\
	&\leq \sup_{i \in \II} \E_{\tilde{M}(i)}\big[(Y-\beb^\top X)^2\big]
	 - \inf_{\bes \in \R^d} \sup_{i \in \II}\E_{\tilde{M}(i)} \big[(Y-\bes^\top X)^2],
\end{align*}
which completes the proof of Proposition~\ref{prop:impossibility_intA}.
\end{proof}

\subsection{Proof of Proposition~\ref{thm:consis}}

\begin{proof}
By assumption, $ \cI$ is a set of 
interventions on $X$ or $A$ of which at least one is confounding-removing. 
Now fix any
$$
\tilde M=(f_{\eta_0}(x;\tilde{\theta}),\tilde g,\tilde h_1,\tilde h_2,\tilde Q)\in \cM,
$$ 
with $\bP_{M}=\bP_{\tilde M}$.
By Proposition~\ref{prop:minimax_equal_causal}, we have that a minimax solution is given by the causal function. That is,
\begin{align*}
&\inf_{\fs\in\FF_{\eta_0}}\,
\sup_{i\in \II}\E_{\tilde{M}(i)}\big[(Y-\fs(X))^2\big]
 =\sup_{i\in \II}\E_{\tilde{M}(i)}\big[(Y-
f_{\eta_0}(X;\tilde{\theta})
)^2\big] = \E_{M}[\xi^2_Y],
\end{align*}
where we used that $\xi_Y$ is unaffected by an intervention on $X$.
By the support restriction $\supp^M(X) \subseteq (a,b)$ we know that
\begin{align*}
f_{\eta_0}(x;\theta^0)&=B(x)^\top \theta^0, \quad
f_{\eta_0}(x;\tilde \theta)=B(x)^\top \tilde \theta, \quad f_{\eta_0}(x;\hat \theta_{\lambda^\star_n,\eta_0,\mu}^n)=B(x)^\top  \hat \theta_{\lambda^\star_n,\eta_0,\mu}^n,
\end{align*}
for all $x\in \supp^M(X)$. Furthermore, as $Y=B(X)^\top \theta^0+\xi_Y$ $\bP_{M}$-almost surely, we have that
\begin{align}
\E_M\left[ C(A) Y \right] &= \E_M\left[ C(A) B(X)^\top \theta^0\right] + \E_M\left[C(A) \xi_Y \right] = \E_M\left[ C(A) B(X)^\top \right]\theta^0, \label{eq:ExpansionOfECY_0}
\end{align}
where we used the assumptions that $\E\left[ \xi_Y \right] =0$ and $A \independent \xi_Y$ by the exogeneity of $A$. Similarly,
\begin{align*} %
\E_{\tilde M}\left[ C(A) Y \right] = \E_{\tilde M}\left[ C(A) B(X)^\top \right]\tilde\theta.
\end{align*}
As $\bP_M = \bP_{\tilde M}$, we have that $
\E_M[ C(A) Y ] = \E_{\tilde M}[ C(A) Y ]$ and $\E_M[ C(A) B(X)^\top ]= \E_{\tilde{M}}[ C(A) B(X)^\top ]$, hence 
\begin{align*}
&\E_{ M}\left[ C(A) B(X)^\top \right]\tilde\theta = \E_{M}\left[ C(A) B(X)^\top \right]\theta^0
 \iff \tilde \theta = \theta^0,
\end{align*}
by assumption \ref{ass:RankCondition},
which states that $ \E[ C(A) B(X)^\top ]$ is of full rank
(bijective). 
In other words, the causal function parameterized by
$\theta^0$ is identified from the observational distribution. Assumptions~\ref{ass:identify_f}~and~\ref{ass:gen_f} are therefore satisfied.
Furthermore, we also have that
	\begin{align*}
	\sup_{i\in  \II}&\E_{\tilde{M}(i)}\big[(Y-f_{\eta_0}(X;\hat{\theta}^n_{\lambda^\star_n,\eta_0,\mu}))^2\big] \\
	 = &\, \sup_{i\in\cI} \big\{ \E_{\tilde{M}(i)} \big[(f_{\eta_0}(X;\theta^0)-f_{\eta_0}(X;\hat{\theta}^n_{\lambda^\star_n,\eta_0,\mu}))^2\big]
	  +  \E_{\tilde{M}(i)}\big[\xi_Y^2\big] \\
	&\quad   + 2 \E_{\tilde{M}(i)}\big[\xi_Y (f_{\eta_0}(X;\theta^0)-f_{\eta_0}(X;\hat{\theta}^n_{\lambda^\star_n,\eta_0,\mu})) \big]  \big\} \\
	 \leq &\,  \sup_{i\in\cI} \big\{ \E_{\tilde{M}(i)} \big[(f_{\eta_0}(X;\theta^0)-f_{\eta_0}(X;\hat{\theta}^n_{\lambda^\star_n,\eta_0,\mu}))^2\big] 
	   +  \E_{\tilde{M}(i)}\big[\xi_Y^2\big] \\
	&\quad + 2 \sqrt{ \E_{\tilde{M}(i)}\big[\xi_Y^2 \big] \E_{\tilde{M}(i)} \big[(f_{\eta_0}(X;\theta^0)-f_{\eta_0}(X;\hat{\theta}^n_{\lambda^\star_n,\eta_0,\mu}))^2 \big] } \big\} \\
	 \leq &\, \sup_{i\in\cI}  \E_{\tilde{M}(i)} \big[(f_{\eta_0}(X;\theta^0)-f_{\eta_0}(X;\hat{\theta}^n_{\lambda^\star_n,\eta_0,\mu}))^2\big] +  \E_{M}\big[\xi_Y^2\big] \\
    &\quad    + 2 \sqrt{ \E_{M}\big[\xi_Y^2 \big] \sup_{i\in\cI}  \E_{\tilde{M}(i)} \big[(f_{\eta_0}(X;\theta^0)-f_{\eta_0}(X;\hat{\theta}^n_{\lambda^\star_n,\eta_0,\mu}))^2 \big] },
	\end{align*}
	by Cauchy-Schwarz inequality, where we additionally used that
	$ \E_{\tilde M(i)} [ \xi_Y^2 ] = \E_{M} [ \xi_Y^2 ]$ as $\xi_Y$ is
	unaffected by interventions on $X$. Thus,
	\begin{align*}
	\big\vert \sup_{i\in\II}&\E_{\tilde{M}(i)}\big[(Y-f_{\eta_0}(X;\hat{\theta}^n_{\lambda^\star_n,\eta_0,\mu}))^2\big]
	- \inf_{\fs\in\FF_{\eta_0}}\,
	\sup_{i\in\II}\E_{\tilde{M}(i)}\big[(Y-\fs(X))^2\big] \big\vert \\
&\leq  \sup_{i\in\cI}  \E_{\tilde{M}(i)} \big[(f_{\eta_0}(X;\theta^0)-f_{\eta_0}(X;\hat{\theta}^n_{\lambda^\star_n,\eta_0,\mu}))^2\big]  \\
&\qquad  + 2 \sqrt{ \E_{M}\big[\xi_Y^2 \big] \sup_{i\in\cI}  \E_{\tilde{M}(i)} \big[(f_{\eta_0}(X;\theta^0)-f_{\eta_0}(X;\hat{\theta}^n_{\lambda^\star_n,\eta_0,\mu}))^2 \big] } .
	\end{align*}
For the next few derivations let $\hat \theta =\hat{\theta}^n_{\lambda^\star_n,\eta_0,\mu}$ for notational simplicity. Note that, for all $x \in \R$,
\begin{align*}
(f_{\eta_0}(x;\theta^0)-f_{\eta_0}(x;\hat \theta ))^2
\leq &\, (\theta^0 -\hat \theta)^\top B(x)B(x)^\top (\theta^0 -\hat \theta) \\
  &+ (B(a)^\top (\theta^0-\hat \theta ) + B'(a)^\top (\theta^0-\hat \theta )(x-a))^2 \\
  &+ (B(b)^\top (\theta^0-\hat \theta ) + B'(b)^\top (\theta^0-\hat \theta )(x-b))^2.
\end{align*}
The second term has the following upper bound
\begin{align*}
 (B(&a)^\top (\theta^0-\hat \theta ) + B'(a)^\top (\theta^0-\hat \theta )(x-a))^2 \\
   =&\, (\theta^0 -\hat \theta)^\top B(a)B(a)^\top (\theta^0 -\hat \theta) \\
 &  +  (x-a)^2(\theta^0 -\hat \theta)^\top B'(a)B'(a)^\top (\theta^0 -\hat \theta)  \\
 &  + 2(x-a) (\theta^0 -\hat \theta)^\top B'(a)B(a)^\top (\theta^0 -\hat \theta) \\
 \leq &\, \lambda_{\mathrm{m}}(B(a)B(a)^\top) \|\theta^0 - \hat \theta\|_2^2 \\
 &  + (x-a)^2 \lambda_{\mathrm{m}}(B'(a)B'(a)^\top ) \|\theta^0 - \hat \theta\|_2^2 \\
 &  + (x-a) \lambda_{\mathrm{m}}(B'(a)B(a)^\top+B(a)B'(a)^\top  ) \|\theta^0 - \hat \theta\|_2^2,
\end{align*}
where $\lambda_{\mathrm{m}}$ denotes the maximum eigenvalue. An analogous upper bound can be constructed for the third term. 
Thus, by combining these two upper bounds with a similar upper bound for the first term, we arrive at
\begin{align*}
\E&_{\tilde{M}(i)} \big[(f_{\eta_0}(X;\theta^0)-f_{\eta_0}(X;\hat{\theta}))^2\big] \\
\leq &\, \lambda_{\mathrm{m}}(\E_{\tilde{M}(i)} [B(X)B(X)^\top])\|\theta^0 -\hat \theta\|_2^2 \\
& + \lambda_{\mathrm{m}}(B(a)B(a)^\top) \|\theta^0 - \hat \theta\|_2^2 \\
&  + \E_{\tilde{M}(i)} [(X-a)^2] \lambda_{\mathrm{m}}(B'(a)B'(a)^\top ) \|\theta^0 - \hat \theta\|_2^2 \\
&  + \E_{\tilde{M}(i)} [X-a] \lambda_{\mathrm{m}}(B'(a)B(a)^\top +B(a)B'(a)^\top  ) \|\theta^0 - \hat \theta\|_2^2\\
& + \lambda_{\mathrm{m}}(B(b)B(b)^\top) \|\theta^0 - \hat \theta\|_2^2 \\
&  + \E_{\tilde{M}(i)} [(X-b)^2] \lambda_{\mathrm{m}}(B'(b)B'(b)^\top ) \|\theta^0 - \hat \theta\|_2^2 \\
&  + \E_{\tilde{M}(i)} [X-b] \lambda_{\mathrm{m}}(B'(b)B(b)^\top  +B(b)B'(b)^\top   ) \|\theta^0 - \hat \theta\|_2^2.
\end{align*}
Assumption \ref{ass:MaximumEigenValueBounded}
imposes that $\sup_{i\in\cI}\E_{\tilde{M}(i)} [X^2]$ 
and $\sup_{i\in\cI}\lambda_{\mathrm{m}}(\E_{\tilde{M}(i)} [B(X)B(X)^\top])$ are finite. Hence, the supremum of each of 
the above terms is finite. That is, there exists a constant $c>0$ such that 

	\begin{align*}
	&\left\vert \sup_{i\in\II}\E_{\tilde{M}(i)}\big[(Y-f_{\eta_0}(X;\hat{\theta}^n_{\lambda^\star_n,\eta_0,\mu}))^2\big] 
	 - \inf_{\fs\in\FF_{\eta_0}}\,
	\sup_{i\in\II}\E_{\tilde{M}(i)}\big[(Y-\fs(X))^2\big] \right\vert \\
&\quad \leq  c \|\theta^0 - \hat{\theta}^n_{\lambda^\star_n,\eta_0,\mu} \|_2^2 + 2 \sqrt{ \E_{M}\big[\xi_Y^2 \big] c }\|\theta^0 - \hat{\theta}^n_{\lambda^\star_n,\eta_0,\mu} \|_2 .
	\end{align*}
It therefore suffices to show that
\begin{align*}
\hat{\theta}^n_{\lambda^\star_n,\eta_0,\mu} \underset{n\to\infty}{\stackrel{P}{\longrightarrow}} \theta^0,
\end{align*}
with respect to the distribution induced by $M$. To simplify
  notation, we henceforth drop the $M$ subscript in the expectations
  and probabilities.
Note that by the rank conditions in \ref{ass:RankCondition}, and the law of large numbers, we may assume that the corresponding sample product moments satisfy the same conditions. That is, for the 
purpose of the following arguments, it suffices that the sample product moment only satisfies these rank 
conditions asymptotically with probability one. 

Let $B:= B(X)$, $C:= C(A)$, let
$\fB$ and $\fC$ be row-wise stacked i.i.d. copies of $B(X)^\top$ and
$C(A)^\top $, and recall the definition
$\fP_\delta := \fC \left( \fC^\top \fC + \delta \fM \right)^{-1}
\fC^\top$. By convexity of the objective function we can find a closed form expression for our estimator of $\theta^0$ by solving the corresponding normal equations. The closed form expression is given by
\begin{align*}
\hat \theta^n_{\lambda, \eta, \mu}
: &= \argmin_{\theta \in \R^{k}} 
\norm{\B{Y} - \B{B} \theta }_2^2 + \lambda \norm{\B{P}_\delta(\B{Y} - \B{B} \theta)}_2^2 + \gamma \theta^\top \B{K} \theta \\
&=  \left(  \frac{\fB^\top  \fB}{n}   + \lambda^\star_n  \frac{\fB^\top \fP_\delta  \fP_\delta   \fB}{n} + \frac{\gamma \fK}{n} \right)^{-1} 
 \left( \frac{\fB^\top \fY }{n} + \lambda^\star_n \frac{\fB^\top \fP_\delta  \fP_\delta   \fY}{n} \right),
\end{align*}
where we used that $\lambda^\star_n \in [0,\infty)$ almost surely by \ref{ass:LambdaStarAlmostSurelyFinite}. 
Consequently
(using standard convergence arguments and that $n^{-1} \gamma \fK $ and  $n^{-1} \delta \fM$ converges to zero in 
probability),
if $\lambda^\star_n$ diverges to infinity in probability as $n$ tends to infinity, then
\begin{align*}
\hat{\theta}^n_{\lambda^\star_n,\eta_0,\mu} %
&\stackrel{P}{\to} \left( \E\left[ BC^\top   \right]\E\left[ CC^\top   \right]^{-1}\E\left[ CB^\top   \right] \right)^{-1}
 \E \left[ BC^\top \right] \E\left[ CC^\top   \right]^{-1} \E\left[ C Y   \right]  \\
&= \theta^0.
\end{align*}
Here, we also used that the terms multiplied by $\lambda^\star_n$ are the only asymptotically relevant terms. 
These are the standard arguments that the K-class estimator (with minor penalized regression modifications) is consistent as long as the parameter $\lambda^\star_n$ converges to infinity, or, equivalently, $\kappa_n^\star= \lambda^\star_n/(1+\lambda^\star_n)$ converges to one in probability.

We now consider two cases: \textit{(i)} $\E[B\xi_Y]\not =0$ and \textit{(ii)} $\E[B\xi_Y]=0$, corresponding to the case with unmeasured confounding and without, respectively. For \textit{(i)} we show that $\lambda^\star_n$ converges to infinity in probability and for \textit{(ii)} we show consistency by other means (as $\lambda^\star_n$ might not converge to infinity in this case).

\textbf{Case (i):} The confounded case $\E[B\xi_Y]\not =0$. It suffices to show that 
$$
\lambda^\star_n := \inf\{\lambda\geq 0 : T_n(\hat \theta^n
_{\lambda,\eta_{0},\mu})\leq q(\alpha)\}\underset{n\to\infty}{\stackrel{P}{\longrightarrow}} \infty.
$$
To that end, note that for fixed $\lambda \geq 0$ we have
that
\begin{align} \label{eq:ThetaLambdaConsistentEstimator}
\hat{\theta}^n_{\lambda,\eta_0,\mu} & \underset{n\to\infty}{\stackrel{P}{\longrightarrow}} \theta_\lambda,
\end{align}
where
\begin{align} \notag
\theta_\lambda \, &:= \left( \E\left[ BB^\top \right] + \lambda \E\left[ BC^\top   \right]\E\left[ CC^\top   \right]^{-1}\E\left[ CB^\top   \right] \right)^{-1} \\ 
& \quad  \times \left( \E \left[ B Y  \right]  + \lambda \E \left[ BC^\top \right] \E\left[ CC^\top   \right]^{-1} \E\left[ C Y   \right] \right).  \label{eq:ThetaLambdaFullRepresentation}
\end{align}
Recall that \eqref{eq:ExpansionOfECY_0} states that $
\E\left[ C Y \right] = \E\left[ C B^\top \right]\theta^0$.
Using \eqref{eq:ExpansionOfECY_0} and that $Y=B^\top \theta^0 + \xi_Y$ $\bP_{M}$-almost surely, we have that the latter factor of \eqref{eq:ThetaLambdaFullRepresentation} is given by
\begin{align*}
&\E \left[ B Y  \right]  + \lambda \E \left[ BC^\top \right] \E\left[ CC^\top   \right]^{-1} \E\left[ C Y   \right] \\
& \quad = \E \left[ B B^\top   \right]\theta^0 + \E \left[ B \xi_Y   \right] 
 + \lambda \E \left[ BC^\top \right] \E\left[ CC^\top   \right]^{-1} \E\left[ C B^\top    \right] \theta^0 \\
&\quad = \left( \E \left[ B B^\top   \right] + \lambda \E \left[ BC^\top \right] \E\left[ CC^\top   \right]^{-1} \E\left[ C B^\top    \right] \right)\theta^0  
 +  \E \left[ B \xi_Y   \right].
\end{align*}
Inserting this into  \eqref{eq:ThetaLambdaFullRepresentation} we arrive at the following representation of $\theta_\lambda$
\begin{align} \label{eq:ThetaLambdaInTermsOfTrueTheta}
&\theta_\lambda  = \theta^0 
+  \left( \E\left[ BB^\top \right] + \lambda \E\left[ BC^\top   \right]\E\left[ CC^\top   \right]^{-1}\E\left[ CB^\top   \right] \right)^{-1}
 \E \left[ B \xi_Y  \right].
\end{align}
Since $\E \left[ B \xi_Y  \right]\not =0$ by assumption, the above yields that
\begin{align} \label{eq:ThetaTrueNotEqualThetaLambda}
\forall \lambda \geq 0 : \quad \quad \theta^0 \not = \theta_\lambda.
\end{align}
Now we prove that $\lambda^\star_n$ diverges to infinity in probability as $n$ tends to infinity. That is, for any $\lambda \geq 0$ we will prove that
\begin{align*}
\lim_{n \to \infty }\bP (\lambda^\star_n \leq \lambda ) =0. 
\end{align*}
We fix an arbitrary $\lambda\geq 0$. By
\eqref{eq:ThetaTrueNotEqualThetaLambda} we have that
$\theta^0 \not = \theta_{\lambda}$. This implies that there exists an
$\ep>0$ such that $\theta^0\not \in \overline{B(\theta_{\lambda},\ep)}$, where $\overline{B(\theta_{\lambda},\ep)}$ is the 
closed ball in
$\R^k$ with center $\theta_{\lambda}$ and radius $\ep$. 
By the consistency result \eqref{eq:ThetaLambdaConsistentEstimator}, we know that 
the sequence of events $(A_n)_{n\in \N}$, for every $n \in \N$, given by 
$$A_n:= (|\hat \theta_{\lambda,\eta_0,\mu}^n -\theta_{\lambda}|\leq \ep) = (\hat \theta_{\lambda,\eta_0,\mu}^n \in\overline{B(\theta_{\lambda},\ep)}),$$ 
satisfies $\bP(A_n)\to 1$ as $n\to \infty$. 
By assumption \ref{ass:MonotonicityAndContinuityOfTest} we have that
\begin{align*}
\tilde \lambda\mapsto T_n(\theta^n_{\tilde \lambda, \eta_0,\mu} ), \qquad \text{and} \qquad \theta \mapsto T_n(\theta ),
\end{align*}
are weakly decreasing and continuous, respectively.  
Together with the continuity of $\tilde \lambda \mapsto \hat \theta_{\tilde \lambda,\eta_0,\mu }^n$,
this implies that also the mapping $\tilde \lambda \mapsto T_n(\hat \theta_{\tilde \lambda,\eta_0,\mu }^n)$
is continuous. It now follows from Assumption~\ref{ass:LambdaStarAlmostSurelyFinite} 
(stating that $\lambda^\star_n$ is almost surely finite) that for all $n \in \N$, 
$\bP(T_{n}( \hat \theta^{n}_{\lambda^\star_{n},\eta_0,\mu} ) \leq q(\alpha))=1$. Furthermore, since $\tilde \lambda\mapsto T_n(\theta^n_{\tilde \lambda, \eta_0,\mu} )$
is weakly decreasing, it follows that
\begin{align*}
 \hspace{-2.5cm}\bP (\lambda^\star_{n} \leq \lambda )
= &\, \bP( \{\lambda^\star_{n} \leq \lambda \} \cap \{T_{n}( \hat \theta^{n}_{\lambda^\star_{n}, \eta_0,\mu} ) \leq q(\alpha)\} )  \\
\leq &\, \bP( \{\lambda^\star_{n} \leq \lambda \} \cap \{T_{n}( \hat \theta^{n}_{\lambda, \eta_0,\mu} ) \leq q(\alpha) \}) \\
 = &\, \bP( \{\lambda^\star_{n} \leq \lambda \}\cap \{T_{n}( \hat \theta^{n}_{\lambda, \eta_0,\mu} ) \leq q(\alpha)\} \cap A_{n}) \\
   &\qquad + \, \, \bP( \{\lambda^\star_{n} \leq \lambda \} \cap \{T_{n}( \hat \theta^{n}_{\lambda, \eta_0,\mu} ) \leq q(\alpha)\} \cap A_{n}^c) \\
\leq &\,  \bP( \{\lambda^\star_{n} \leq \lambda \} \cap  \{ T_{n}( \hat \theta^{n}_{\lambda, \eta_0,\mu} ) \leq q(\alpha)\} 
 \cap \{ |\hat \theta_{\lambda,\eta_0,\mu}^n -\theta_{\lambda}|\leq \ep\} )
 + \bP(A_n^c).
\end{align*}
It now suffices to show that the first term converges to zero, since $\bP(A_{n}^c)\to 0$ as $n\to \infty$. 
We have
\begin{align*}
&\, \bP( \{\lambda^\star_{n} \leq \lambda \} \cap \{ T_{n}( \hat \theta^{n}_{\lambda, \eta_0,\mu} ) \leq q(\alpha)\}
\cap \{ |\hat \theta_{\lambda,\eta_0,\mu}^n -\theta_{\lambda}|\leq\ep\} ) \\
  &\qquad \leq \bP \Big( \{ \lambda^\star_{n} \leq \lambda \} \cap \Big\{ \inf_{\theta\in \overline{B(\theta_{\lambda},\ep)}}T_{n}( \theta ) \leq q(\alpha) \Big\}
  \cap \{ |\hat \theta_{\lambda,\eta_0,\mu}^n -\theta_{\lambda}|\leq\ep\} \Big)  \\
&\qquad \leq \bP \Big(  \inf_{\theta\in \overline{B(\theta_{\lambda},\ep)}}T_{n}( \theta ) \leq q(\alpha) \Big)\\ 
&\qquad \stackrel{P}{\to}0, 
\end{align*}
as $n\to \infty$, since $\overline{B(\theta_{\lambda},\ep)}$ is a compact set not containing $\theta^0$. Here, we used that the test statistic $(T_n)$ is assumed to have compact uniform power \ref{ass:ConsistentTestStatistic}.
Hence, $\lim_{n\to \infty} \bP (\lambda^\star_{n} \leq \lambda ) = 0 $ for any $\lambda \geq 0$,
proving that $\lambda^\star_n$ diverges to infinity in probability, which ensures consistency. 

\textbf{Case (ii):} the unconfounded case $\E[B(X)\xi_Y]=0$. Recall that
\begin{align} \notag
\hat{\theta}^n_{\lambda,\eta_0,\mu} \,&= \argmin_{\theta \in \R^{k}}  
\norm{\B{Y} - \B{B} \theta }_2^2 + \lambda \norm{\B{P}_\delta(\B{Y} - \B{B} \theta)}_2^2 + \gamma \theta^\top \B{K} \theta  \\ \label{eq:ThetaLambdaMinimizesObjectiveFunction}
&=\argmin_{\theta \in \R^{k}}   l_{\text{OLS}}^n(\theta) + \lambda l_{\text{TSLS}}^n(\theta) + \gamma l_{\text{PEN}}(\theta) ,
\end{align}
where we defined $l_{\text{OLS}}^n(\theta):= n^{-1}\norm{\B{Y} - \B{B} \theta }_2^2$, $l_{\text{TSLS}}^n(\theta) := n^{-1}\norm{\B{P}_\delta(\B{Y} - \B{B} \theta)}_2^2$, and $l_{\text{PEN}}(\theta) := n^{-1} \theta^\top \B{K} \theta$. 
For any $0\leq \lambda_1 < \lambda_2$ we have
\begin{align*}
&l_{\text{OLS}}^n(\hat{\theta}^n_{\lambda_1,\eta_0,\mu}) + \lambda_1 l_{\text{TSLS}}^n(\hat{\theta}^n_{\lambda_1,\eta_0,\mu}) + \gamma l_{\text{PEN}}(\hat{\theta}^n_{\lambda_1,\eta_0,\mu})   \\
&\quad \leq l_{\text{OLS}}^n(\hat{\theta}^n_{\lambda_2,\eta_0,\mu}) + \lambda_1 l_{\text{TSLS}}^n(\hat{\theta}^n_{\lambda_2,\eta_0,\mu})  + \gamma l_{\text{PEN}}(\hat{\theta}^n_{\lambda_2,\eta_0,\mu}) \\
&\quad = l_{\text{OLS}}^n(\hat{\theta}^n_{\lambda_2,\eta_0,\mu}) + \lambda_2 l_{\text{TSLS}}^n(\hat{\theta}^n_{\lambda_2,\eta_0,\mu}) + \gamma l_{\text{PEN}}(\hat{\theta}^n_{\lambda_2,\eta_0,\mu}) 
+ (\lambda_1-\lambda_2) l_{\text{TSLS}}^n(\hat{\theta}^n_{\lambda_2,\eta_0,\mu}) \\
&\quad \leq l_{\text{OLS}}^n(\hat{\theta}^n_{\lambda_1,\eta_0,\mu}) + \lambda_2 l_{\text{TSLS}}^n(\hat{\theta}^n_{\lambda_1,\eta_0,\mu}) + \gamma l_{\text{PEN}}(\hat{\theta}^n_{\lambda_1,\eta_0,\mu})
 + (\lambda_1-\lambda_2) l_{\text{TSLS}}^n(\hat{\theta}^n_{\lambda_2,\eta_0,\mu}),
\end{align*}
where we used \eqref{eq:ThetaLambdaMinimizesObjectiveFunction}. Rearranging this inequality and dividing by $(\lambda_1 - \lambda_2)$ yields 
\begin{align*}
 l_{\text{TSLS}}^n(\hat{\theta}^n_{\lambda_1,\eta_0,\mu}) \geq   l_{\text{TSLS}}^n(\hat{\theta}^n_{\lambda_2,\eta_0,\mu}),
\end{align*}
proving that $\lambda \mapsto l_{\text{TSLS}}^n(\hat{\theta}^n_{\lambda,\eta_0,\mu})$ is weakly decreasing. Thus, since $\lambda^\star_n \geq 0$ almost surely, we have that
\begin{align}
&l_{\text{TSLS}}^n(\hat{\theta}^n_{\lambda^\star_n,\eta_0,\mu})
\leq  \, l_{\text{TSLS}}^n(\hat{\theta}^n_{0,\eta_0,\mu})
 = \, n^{-1} (\B{Y} - \B{B} \hat{\theta}^n_{0,\eta_0,\mu})^{\top } \B{P}_\delta\B{P}_\delta(\B{Y} - \B{B} \hat{\theta}^n_{0,\eta_0,\mu}). \label{eq:unconfounded_IVinOLSboundedByConvergenceToZero}
\end{align}
Furthermore, recall from \eqref{eq:ThetaLambdaConsistentEstimator} that
\begin{align} \label{eq:unconfoundedOLSconvergence}
\hat{\theta}^n_{0,\eta_0,\mu}  \underset{n\to\infty}{\stackrel{P}{\longrightarrow}} \theta_0 = \theta^0,
\end{align}
where the last equality follows from \eqref{eq:ThetaLambdaInTermsOfTrueTheta} using that we are in the unconfounded case $\E[B(X)\xi_Y]=0$. By expanding and deriving convergence statements for each term, we get
\begin{align} \notag
&(\B{Y} - \B{B} \hat{\theta}^n_{0,\eta_0,\mu})^{\top } \B{P}_\delta\B{P}_\delta(\B{Y} - \B{B} \hat{\theta}^n_{0,\eta_0,\mu}) \\ \notag
& \underset{n\to\infty}{\stackrel{P}{\longrightarrow}}  (\E[ YC^\top ] - \theta_0\E[B C^\top ]) \E[C^\top C]^{-1}
 (
\E[CY]
- \E[CB^\top] \theta_0 ) \\
&  = 0, \label{eq:unconfoundedTSLSinOLSConvpZero}
\end{align}
where we used Slutsky's theorem, the weak law of large numbers, \eqref{eq:unconfoundedOLSconvergence} and \eqref{eq:ExpansionOfECY_0}. Thus, by \eqref{eq:unconfounded_IVinOLSboundedByConvergenceToZero} and \eqref{eq:unconfoundedTSLSinOLSConvpZero} it holds that
\begin{align*}
l_{\text{TSLS}}^n(\hat{\theta}^n_{\lambda^\star_n,\eta_0,\mu}) = n^{-1}\| \B{P}_\delta(\B{Y} - \B{B} \hat{\theta}^n_{\lambda^\star_n,\eta_0,\mu})  \|_2^2 \underset{n\to\infty}{\stackrel{P}{\longrightarrow}} 0.
\end{align*}
For any $z\in \R^n$ we have that 
\begin{align*}
\| \B{P}_\delta z  \|_2^2 &  = 
z^\top  \fC ( \fC^\top \fC + \delta \fM )^{-1}
\fC^\top\fC ( \fC^\top \fC + \delta \fM )^{-1}
\fC^\top z \\
& =
z^\top  \fC ( \fC^\top \fC + \delta \fM )^{-1}
(\fC^\top\fC)^{1/2}(\fC^\top\fC)^{1/2}
 ( \fC^\top \fC + \delta \fM )^{-1}
\fC^\top z \\ 
&= \| (\fC^\top\fC)^{1/2} ( \fC^\top \fC + \delta \fM )^{-1}
\fC^\top z  \|_2^2,
\end{align*}
hence
\begin{align}  \notag
\norm{H_n-G_n \hat{\theta}^n_{\lambda^\star_n,\eta_0,\mu} }_2^2 
&= \| n^{-1/2}(\fC^\top\fC)^{1/2} ( \fC^\top \fC + \delta \fM )^{-1}
\fC^\top (\B{Y} - \B{B} \hat{\theta}^n_{\lambda^\star_n,\eta_0,\mu})\|_2^2 \\ 
&\stackrel{P}{\to } 0, \label{eq:consitUconfoundedDifferenceConvPtoZero}
\end{align}
where for each $n \in \N$, $G_n\in \R^{k\times k}$ and $H_n \in \R^{k\times 1}$ are defined as
\begin{align*}
G_n &:= n^{-1/2}(\fC^\top\fC)^{1/2} ( \fC^\top \fC + \delta \fM )^{-1} \fC^\top \fB, \text{ and } \\
H_n &:= n^{-1/2}(\fC^\top\fC)^{1/2} ( \fC^\top \fC + \delta \fM )^{-1} \fC^\top \fY.
\end{align*}
Using the weak law of large numbers, the continuous mapping theorem and Slutsky's theorem,
it follows that, as $n \to \infty$,
\begin{align*}
G_n \stackrel{P}{\to}  G 	&:= E[CC^\top]^{1/2} E[CC^\top]^{-1} E[CB^\top], \text{ and }\\
H_n \stackrel{P}{\to}  H 	&:= E[CC^\top]^{1/2} E[CC^\top]^{-1} E[CY] \\
										& = E[CC^\top]^{1/2} E[CC^\top]^{-1} E[CB^\top ]\theta^0 \\
										&= G\theta^0,
\end{align*}
where the second to last equality follows from \eqref{eq:ExpansionOfECY_0}. 
Together with \eqref{eq:consitUconfoundedDifferenceConvPtoZero}, we now have that
\begin{align*}
&\, \norm{G_n \hat{\theta}^n_{\lambda^\star_n,\eta_0,\mu}  - G \theta^0}_2^2
\leq  \norm{G_n \hat{\theta}^n_{\lambda^\star_n,\eta_0,\mu}  - H_n}_2^2 + 
\norm{H_n  - G \theta^0}_2^2 
 \underset{n\to\infty}{\stackrel{P}{\longrightarrow}}  0.
\end{align*}
Furthermore, by the rank assumptions in \ref{ass:RankCondition} we have that $G_n\in \R^{k\times k}$ is of full rank (with probability tending to one), hence
\begin{align*}
\|\hat{\theta}^n_{\lambda^\star_n,\eta_0,\mu} -\theta^0\|_2^2 &= \|G_n^{-1}G_n( \hat{\theta}^n_{\lambda^\star_n,\eta_0,\mu} -\theta^0)\|_2^2 \\
& \leq \|G_n^{-1}\|_{\text{op}}^2 \|G_n( \hat{\theta}^n_{\lambda^\star_n,\eta_0,\mu} -\theta^0) \|_2^2 \\
&\stackrel{P}{\to}\|G^{-1}\|_{\text{op}}^2 \cdot 0 \\
&=0,
\end{align*}
as $n\to \infty$, proving the proposition.

\end{proof}

\bibliographystyle{abbrvnat}
\bibliography{ref}

\end{document}